\documentclass[11pt,prd,superscriptaddress,nofootinbib]{revtex4}
\usepackage[english]{babel}
\usepackage{graphicx}
\usepackage{mathtext}
\usepackage{indentfirst}
\usepackage{epsfig,amsmath,amsfonts}

\newcommand{\beq}[1]{\begin{eqnarray}\label{#1}}
\newcommand\eeq {\end{eqnarray}}
\newcommand\bqa {\begin{eqnarray}}
\newcommand\eqa {\end{eqnarray}}
\newcommand\pr {\partial}

\newcommand\nn {\nonumber}

\newcommand{\eq}[1]{(\ref{#1})}
\newcommand{\bear}{\begin{array}}
\newcommand{\enar}{\end{array}}

\begin{document}

\hfill ITEP-TH-32/13

\title{\Large \bf Lecture notes on interacting quantum fields in de Sitter space}


\maketitle

\centerline{\bf E.T. Akhmedov\footnote{E--mail: akhmedov@itep.ru}}

\centerline{Institute for Theoretical and Experimental Physics}

\centerline{117218, ul. B.Cheremushkinskaya, 25, Moscow}

\centerline{and}

\centerline{Moscow Institute of Physics and Technology}

\centerline{141700,Institutskii per. 9, Dolgoprudny, Moscow region}

\centerline{and}

\centerline{Mathematical Faculty of the}

\centerline{National Research University Higher School of Economics}

\centerline{117312, Vavilova 7, Moscow}

\vspace{5mm}

\centerline{\bf Abstract}


We discuss peculiarities of quantum fields in de Sitter space
on the example of the self-interacting massive real scalar, minimally coupled to the gravity background. Non-conformal quantum field theories in de Sitter space show very special infrared behavior,
which is not shared by quantum fields neither in flat nor in anti-de-Sitter space: in de Sitter space loops are not suppressed in comparison with tree level contributions because there are strong infrared corrections. That is true even for massive fields. Our main concern is the interrelation between these infrared effects, the invariance of the quantum field theory under the de Sitter isometry and the (in)stability of de Sitter invariant states (and of dS space itself) under nonsymmetric perturbations.


\vspace{10mm}


\newpage

\tableofcontents

\newpage

\section{Introduction}

$D$-dimensional de Sitter (dS) space solves the general relativity (GR) equations of motion with positive vacuum energy \cite{Schrodinger:2010zz}:

\bqa\label{first}
G_{\alpha\beta} = - \frac{(D-1)(D-2)}{2}\, H^2 \, g_{\alpha\beta}, \quad \alpha, \beta = 0, \dots, D-1.
\eqa
Here $G_{\alpha\beta}$ is the Einstein tensor and $H$ is the Hubble constant. The signature of the metric, $g_{\alpha\beta}$, is $(-,+,\dots,+)$. This space has a big isometry group, $SO(D,1)$, which is the analog of Poincar\'{e} invariance of Minkowski space.

Quantum effects produce an extra contribution, $\langle T_{\alpha\beta} \rangle$, to the right hand side of the GR equations of motion. By $\langle T_{\alpha\beta} \rangle$ we denote the quantum average of the energy-momentum tensor of whatever quantum fields are present on the dS background. What is the influence of this quantum average on the background geometry? Our final goal is to answer this question. However, in these lectures we concentrate on just a part of this program: We fix dS background and check whether the assumption of negligible or small effects of $\langle T_{\alpha\beta}\rangle$ is selfconsistent.

This quantum average, $\langle T_{\alpha\beta} \rangle$, contains the standard divergence due to zero-point fluctuations. It is proportional to $g_{\alpha\beta}$ and should be absorbed into the ultraviolet (UV) renormalization of the cosmological constant. Furthermore, quantum fields on the dS background are in a nonstationary situation. Hence, nontrivial finite contributions to $\langle T_{\alpha\beta} \rangle$, such as, e.g., fluxes, can be present and are of interest for us.

\subsection{Motivation}

It is natural to use the big isometry group of dS space in the formulation of quantum field theory (QFT) on this background, i.e., it is appropriate to look for a dS-invariant state, if any, and quantize excitations of such a state. (This is what one does when one quantizes fields on Minkowski background: one explores the Poincar\'{e} group.) But then, if the symmetry is respected at all stages of quantization, correlation functions depend only on dS-invariant geodesic distances rather than on each of their argument separately. Hence, in the free theory one can use the two-point correlation function to find that all contributions to $\langle T_{\alpha\beta} \rangle$ are proportional to the metric, $g_{\alpha\beta}$. That is just a consequence of the symmetry in question. Furthermore, with the use of the higher-point correlation functions one can extend this calculation to the interacting fields with the same conclusion, if the dS isometry is also respected on loop level.

Thus, in the case of perfectly dS-symmetric situation, quantum effects just renormalize the cosmological constant and dS space remains intact. Perhaps that is true unless exact correlators, as functions of geodesic distances, show an explosive behavior. The main goal of these notes, however, is to question the stability of dS space and to investigate if the dS isometry is broken or the dS-invariant state is unstable with respect to nonsymmetric perturbations.
It is probably worth stressing here that in these notes we study just the behavior of correlation functions in dS space and avoid using the notion of particle, unless this notion is meaningful and useful for the interpretation of the obtained equations.

At tree level the situation is as follows: For massless scalar fields, which are minimally coupled to the dS background, there is no Fock dS-invariant state \cite{Allen:1985ux}, \cite{Allen:1987tz}. Also the very possibility to define a dS-invariant state for gravity is still under discussion (see \cite{Miao:2013isa}, \cite{Tsamis:1993ub}--\cite{Tsamis:1996qm}, \cite{Garriga:2007zk}, \cite{Higuchi:2011vw}, \cite{Morrison:2013rqa} and \cite{PerezNadal:2008ju}--\cite{Roura:1999fq}). All these interesting and important issues will not be touched in these lectures. We will concentrate on the study of the real massive minimally coupled scalar field, $\phi$. Then the situation is conceptually simpler because in this case there is a one-parameter family of so-called $\alpha$-vacua, which respect the dS isometry at tree level \cite{Allen:1985ux},  \cite{Mottola:1984ar}.

First, we would like to understand whether or not the dS isometry is respected by loop contributions, if one quantizes over such a dS-invariant state: We will see that in some circumstances the dS isometry is indeed broken. Second, in those circumstances, in which the dS isometry is respected, we would like to understand whether or not the corresponding state is stable under nonsymmetric perturbations. (Note that the vacuum is still dS-invariant. We just consider a non trivial density matrix.)
We find it physically inappropriate to consider the stability of any system in a state in which all its symmetries are preserved. We will see that the exact dS-invariant state does exist, but it is unstable under sufficiently strong nonsymmetric perturbations.

A few points are worth stressing. First, Minkowski space is stable under nonsymmetric particle density perturbations over the Poincar\'{e} invariant vacuum. That is just a consequence of the energy conservation and of the H-theorem, neither of which is straightforwardly applicable in dS space. Second, in the presence of nonsymmetric density perturbations, even tree level two-point correlation functions depend on each of their arguments separately rather than on dS-invariant distances between them. In view of what we have said here one can conclude that argumentation, which is based on analytical properties of the correlators as functions of geodesic distances \cite{MarolfMorrison}--\cite{Marolf:2010nz}, \cite{Hollands:2010pr}, \cite{Hollands:2011we}, is not sufficient to support the stability of dS space. 

For the other IR issues in dS space please see, e.g., \cite{Antoniadis:2006wq}--\cite{Xue:2011hm}.

\subsection{General physical explanation of our main statements}

To explain our statements let us briefly describe the dS geometry. dS space can be totaly covered by the so-called global metric:

$$ds^2 = - dt^2 + \frac{\cosh^2(Ht)}{H^2} \, d\Omega^2_{D-1},$$
with $d\Omega^2_{D-1}$ being the line element on the unit $(D-1)$-dimensional sphere. At the same time the inflationary (aka planar, aka spatially flat) metric is

$$ds^2_+ = -d\tau_+^2 + e^{2H\tau_+} \, d\vec{x}_+^2 = \frac{1}{(H\eta_+)^2} \, \left[-d\eta_+^2 + d\vec{x}_+^2\right], \quad H\eta_+ = e^{-H\tau_+}.$$
It covers only geodesically incomplete half of the whole space. The latter is referred to as the expanding Poincar\'{e} patch (EPP). The other half is referred to as the contracting Poincar\'{e} patch (CPP) and is covered by the metric

$$ds^2_- = - d\tau_-^2 + e^{-2H\tau_-} \, d\vec{x}_-^2 = \frac{1}{(H\eta_-)^2} \left[-d\eta_-^2 + d\vec{x}_-^2\right],\quad H\eta_- = e^{H\tau_-}.$$
The boundary between these patches ($\eta_\pm = +\infty$), which is simultaneously the initial Cauchy surface of the EPP and the final one of the CPP, is light-like. One can obtain the CPP metric from the EPP one by reflecting the direction of the conformal time, $\eta_+ \in (+\infty, 0) \longrightarrow \eta_- \in (0, +\infty)$.

The EPP and CPP have a peculiarity in their geometry. The spatial part of their metric has the conformal factor $1/\eta^2_\pm$. Due to its presence every wave experiences strong blue shift towards the past (future) infinity of the EPP (CPP). I.e., these regions of the Poincar\'{e} patches correspond to the UV limit.

In loop integrals on the EPP background the vertex integration goes over the half of dS space. Hence, naively the dS isometry should be broken because there are generators of this symmetry which can move the EPP within the whole dS space. However, following the original work \cite{Polyakov:2012uc}, in these lectures we will show that the dS isometry can indeed be respected in the loops, but {\it only} if one starts {\it exactly} with the so-called Bunch--Davies (BD) state at the past infinity of the EPP. BD is such a state that there are no positive energy excitations at the past (future) infinity of the EPP (CPP) \cite{Bunch:1978yq}, \cite{Chernikov:1968zm}. In the region of dS space near the boundary between the EPP and CPP one can define what one means by particle and what one means by positive energy. This is possible because, as we have explained above, every momentum experiences infinite blue shift towards the past (future) of the EPP (CPP). In fact, high energy harmonics are not sensitive to the comparatively small curvature of the background space and behave as if they are in flat space.

After a Bogolyubov rotation from the BD harmonics to other modes, corresponding to other dS-invariant states from the aforementioned $\alpha$-family, one mixes the positive and negative energy states. That spoils the UV behavior of the correlation functions. As a result, any vacuum different from the BD one violates the dS isometry in the loops, even though it respects the symmetry at tree level.

Should one conclude then that the problem of the influence of massive scalar fields on the dS geometry is solved? It seems that in the EPP one just has to quantize fields over the BD state. But the solution of this problem is not yet complete because of large infrared (IR) effects: IR corrections may become destructive in the presence of nonsymmetric density perturbations because they are not suppressed in comparison with tree level contributions.

As we will explain, all QFTs in dS space, which are {\it not} conformally invariant\footnote{I.e., such QFTs, which can feel the difference between the flat space and  conformally flat dS one.}, share the same characteristic property, which is not present in massive QFTs in flat or anti-dS spaces \cite{AkhmedovSadofyev}. Obviously the UV limit of any meaningful QFT over the dS or anti-dS backgrounds should be the same as in flat space, hence, all the differences appear in the IR limit. Large IR effects, due to their large scale nature, are sensitive to the boundary and initial conditions in various patches of the entire space. Hence, they should be separately considered in the EPP, CPP and in global dS space.

Also, the character of these IR contributions in dS space crucially depends on the relation between the mass and the Hubble constant: If the mass of the scalar field is big, $m/H > (D-1)/2$, then the corresponding harmonic functions {\it oscillate} and decay to zero, as $\eta \to 0$. Scalar fields with such masses are composing the so-called principal series of theories. At the same time, harmonic functions of the scalars with small masses, $0 < m/H \leq (D-1)/2$, homogeneously decay to zero, as $\eta \to 0$. The corresponding theories are composing the so-called complementary series.

We first describe the situation in the EPP and then continue with the CPP and global dS space.
Due to the spatial homogeneity of the EPP and also of the initial states that we consider, it is natural to perform the partial Fourier transformation along $\vec{x}_+$ directions. Then, the two-point correlation function of interest for us acquires the form $D(p|\eta_+, \eta'_+) = \frac12 \, \left\langle\left\{ \phi(\eta_+, \vec{p}), \phi(\eta'_+,-\vec{p})\right\}\right\rangle$, where $\{\cdot, \cdot\}$ is the anticommutator. In the nonstationary situation it is the appropriate object to study and is referred to as the Keldysh propagator. As we review below, in the limit when the system approaches future infinity, $p\, \eta = p\,\sqrt{\eta_+ \eta'_+} \to 0$, the correlation function in question receives large corrections for any mass of the field.

E.g., for the $\lambda \, \phi^3$ scalar field theory from the principal series the first loop contains terms which are proportional to $\lambda^2 \eta^{D-1} \log(p\eta)$ \cite{PolyakovKrotov}, \cite{AkhmedovKEQ}. The same linear logarithmic corrections are also present in the second loop of $\lambda \phi^4$ theory \cite{AkhmedovPopovSlepukhin}. (The only difference between the $\phi^3$ and $\phi^4$ theories is in the mass-dependent coefficients of these IR contributions.) At the same time, the fields from the complementary series receive powerlike corrections, which are proportional to $\lambda^2 \eta^{D-1} \, (p\eta)^{-2\nu}$, where parameter $\nu$ depends on the mass and $0< \nu <(D-1)/2$ \cite{AkhmedovPopovSlepukhin}.

Thus, on the one hand, because of the factor $\eta^{D-1}$ in front of every contribution, which is due to the expansion of the spatial sections of the EPP, such loop corrections do not make the Keldysh propagator singular\footnote{Note that the only possible exception is given by the massless minimally coupled scalar field theory  \cite{Dolgov:1994cq}, \cite{Onemli:2002hr}, for which $\nu = (D-1)/2$.} in the limit $\eta\to 0$. But, on the other hand, even if $\lambda^2$ is very small the corrections in question become comparable to tree level contributions, as $p\eta\to 0$. Hence, one has to sum the unsuppressed leading IR contributions from all loops.

Of course one should not worry about the stability of the EPP, if the dS isometry is respected. But what if we consider some finite nonsymmetric density perturbation over the BD state at the past infinity of the EPP?
Due to the rapid expansion of the EPP it is usually believed that such density perturbations would quickly fade away and, thus, one should not care about their negligible influence on the background geometry. However, the situation is quite counterintuitive because the loop contributions go into the factor multiplying $\eta^{D-1}$ and marginally  depend on the expansion and/or contraction of spatial sections.

Let us clarify this observation here. The common wisdom is that excitations in the EPP can reproduce themselves only very slowly. (It is believed that the reproduction should be at most linear in time, e.g., due to the constant particle creation caused by the background field.) At the same time, the expansion of the spatial sections is exponential and, hence, will rapidly win over such a reproduction.

However, let us look more carefully at what is actually happening. Suppose we are in a situation where one can give a meaning to the notion of particle in the future infinity of the EPP. (We will see that this is possible under some conditions.) Then, consider, e.g., a particle dust in the EPP. Its density per physical volume is indeed rapidly decreasing. However, the density per comoving volume remains constant. (The comoving density is actually one of the quantities contributing to the factor multiplying $\eta^{D-1}$ in the two-point correlation function under consideration.) Suppose now there is some constant particle production process, i.e., it is the density per comoving volume which is linearly growing. (That is what we will actually see.) Then, sooner or later it will become very large and one will not be able to neglect the nonlinear particle self-reproduction processes. We will see that the non-linearities are proportional to the density per comoving volume rather than to that per physical one. (It may sound as very weird, but note that we are talking about waves whose size is growing with time and even redshifted outside the cosmological horizon in the progress towards future infinity.) As we will see, the nonlinear self-reproduction can actually cause the destruction and win over the expansion.

Although we just perform the calculation of loop contributions to correlation functions,
the summation of unsuppressed IR loop corrections allows the particle interpretation and is related to the above described particle kinetics. That happens to be true at least for the fields from the principal series. We are not yet able to perform the summation of the IR divergences for the complementary series because their physical meaning is not yet clear to us. But on general physical grounds we expect that the IR effects in this case will be even stronger.

To sum unsuppressed IR loop corrections one has to find an IR solution  of the system of Dyson--Schwinger (DS) equations for the vertices, propagators and self-energies. We show, however, that in the limit under study only the equation for the Keldysh propagator is relevant. Furthermore, for the principal series it reduces to a kinetic equation of Boltzmann type, where plane waves are substituted by exact dS harmonics. (As is known in condensed matter theory, loop effects may become classical. That is related to the fact that loop corrections are not suppressed in comparison with tree level contributions (see, e.g., \cite{LL}, \cite{Kamenev}).)


All elements of the kinetic equation that we obtain have a clear physical interpretation and describe various particle decay and creation processes in the future infinity of the EPP \cite{AkhmedovKEQ}, \cite{AkhmedovBurda}, \cite{AkhmedovGlobal}. Moreover, we can find solutions of this kinetic equation for various initial conditions which are set up by the initial (tree level) Keldysh propagator. One of the solutions shows an explosive behavior of the two-point correlation function.

What about IR effects in the CPP and global dS?  If we consider an exactly spatially homogeneous initial state, the loop calculation in the CPP follows straightforwardly from the one in the EPP. Note, however, that unlike the EPP, the spatially homogeneous state in the CPP is unstable with respect to the inhomogeneous perturbations. But it is still instructive to study loop effects in such an ideal situation.

The metric in the CPP is identical to the one in the EPP if one reverses the direction of the conformal time. But then, for the same reason as we have observed large IR contributions in the EPP, there are IR {\it divergences} in the CPP: For the scalar fields from the principal series with the $\lambda \phi^3$ self-interaction the corrections are proportional to $\lambda^2 \eta^{D-1} \log(\eta/\eta_0)$, if $p\eta \ll 1$. At the same time, if $p\eta \gg 1$, the corrections are proportional to $\lambda^2 \eta^{D-1} \, \log(p\eta_0)$. Here $\eta \equiv \sqrt{\eta_-\eta'_-}$, $\eta_0$ is the moment of time at past infinity, $p\eta_0 \to 0$, after which self-interactions are adiabatically turned on. The same linear logarithmic contributions, but with different mass-dependent coefficients, appear also in the second loop in the theory with $\lambda \phi^4$ self-interaction \cite{AkhmedovPopovSlepukhin}. For the scalars from the complementary series the divergences are powerlike.

If it were not for the presence of the IR cutoff, $\eta_0$, the loop integrals in the CPP would be explicitly divergent: $\eta_0$ cannot be taken to past infinity. Thus, one has to have an initial Cauchy surface at some finite $\eta_0$. But such a surface can be moved within the dS space by an isometry transformation. Then, holding $\eta_0$ fixed breaks the dS isometry and correlation functions start to depend separately on each of their time arguments.
Furthermore, the summation of the leading IR contributions in the CPP is performed similarly to the EPP case. In fact, the kinetic equation in the CPP is obtained from the one in the EPP just by the time reversal.



To study the situation in global dS space one has to keep in mind that it is the union of the CPP and EPP.
Hence, loops in global dS also have explicit IR divergences that break the dS isometry \cite{PolyakovKrotov}, \cite{AkhmedovGlobal}. In this case it is also possible to derive the kinetic equation for the fields from the principal series. But unlike the EPP and CPP case, this kinetic equation does not possess an obvious quasi-particle description in the IR limit and we do not expect to have a stationary solution in global dS, unless one considers a small enough part of it.

To conclude, the statements that we are going to advocate in these lectures are as follows. First, we will observe that the only way to respect the dS isometry in loop contributions is to start at the past infinity of the EPP exactly with the BD state. Any other dS-invariant vacuum, i.e., that state which respects the dS isometry at tree level, does break it at loop level. Second, we will see that any invariant initial state, including the so-called Euclidian one, in global dS space violates the isometry in loops, due to IR divergences of loop integrals. Similarly, due to IR divergences, any invariant initial state in the CPP also violates the dS isometry in loop contributions. And finally, after the summation of unsuppressed IR contributions for the principal series in all loops, we will proceed to show that even the BD state in the EPP is unstable under sufficiently strong nonsymmetric perturbations. For the complementary series we expect even more destructive IR effects, but we are not yet able to perform the loop summation.

The new content of these notes is mostly based on our previous work \cite{AkhmedovKEQ}, \cite{AkhmedovPopovSlepukhin}, \cite{AkhmedovBurda}, \cite{AkhmedovGlobal}.

\section{de Sitter geometry}

\begin{figure}
\begin{center}
\includegraphics[scale=0.4]{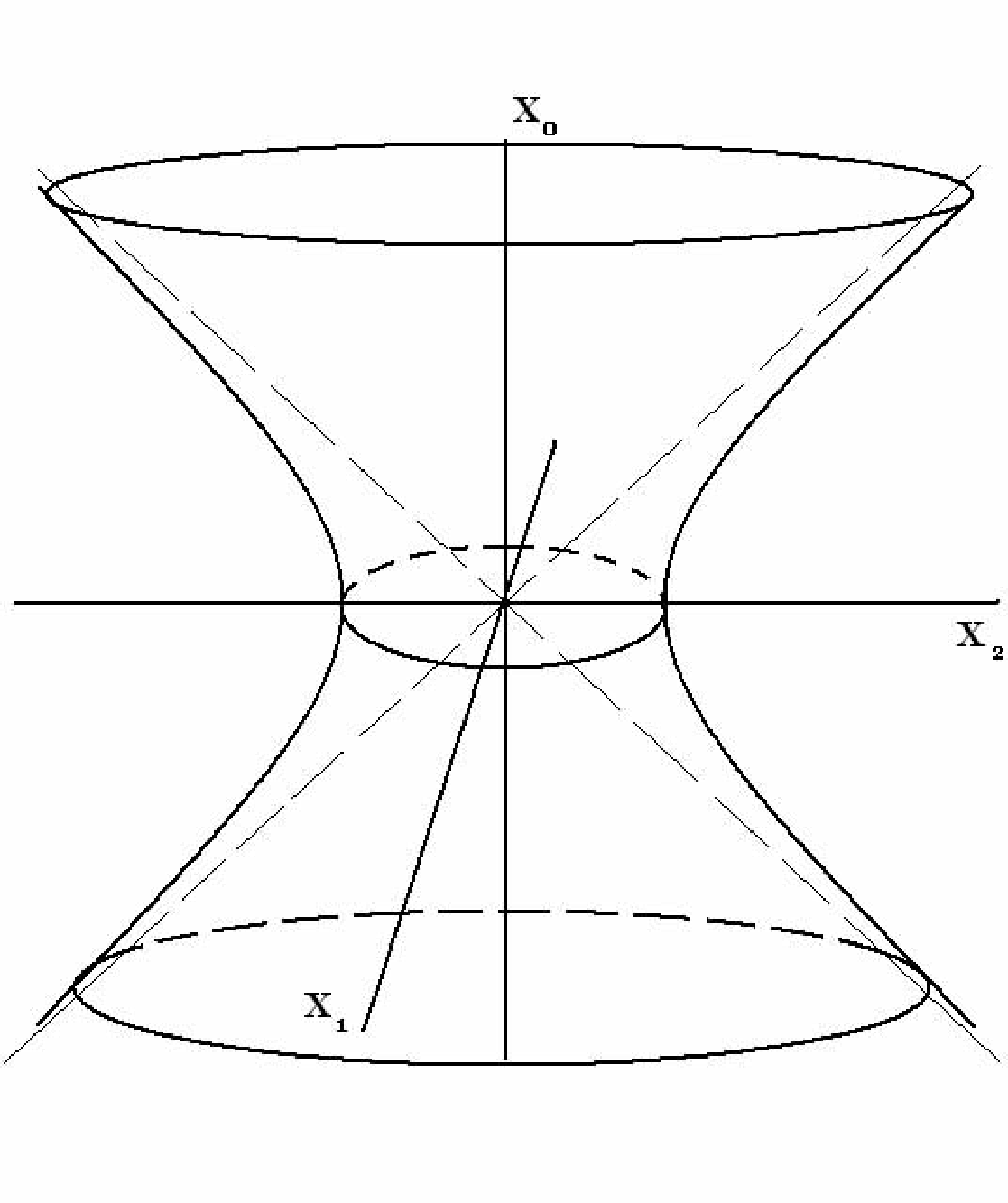}\caption{Each constant $X_0$ slice of this two-dimensional dS space is a circle of radius $(H^{-2} + X_0^2)$. }\label{fig1}
\end{center}
\end{figure}

$D$-dimensional dS space can be realized as hyperboloid,

\bqa\label{hyp}
- \left(X^0\right)^2 + \left(X^1\right)^2 + \dots + \left(X^D\right)^2 \equiv g_{AB} \, X^A \, X^B = H^{-2}, \quad A,B = 0, \dots, D,
\eqa
placed into the ambient $(D+1)$-dimensional Minkowski space, with metric $ds^2 = g_{AB} \, dX^A \, dX^B$.
One way to see that a metric on such a hyperboloid solves equation (\ref{first}) is to observe that it can be obtained from the sphere via the analytical continuation $X^{D+1} \to i\, X^0$. For illustrative reasons we depict two-dimensional dS space on fig. \ref{fig1}.

Another way to see that the hyperboloid in question has a constant curvature is to observe that eq. (\ref{hyp}) is invariant under $SO(D,1)$ Lorentz transformations of the ambient space. The stabilizer of any point obeying (\ref{hyp}) is $SO(D-1,1)$ group. Hence, dS is homogeneous, $SO(D,1)/SO(D-1,1)$, space and any its point is equivalent to another one. Furthermore, all directions at every point are equivalent. (Compare this with the sphere, which is $SO(D+1)/SO(D)$.) Thus, $SO(D,1)$ Lorentz group of the ambient Minkowski space is the dS isometry group.

The geodesic distance, $L_{12}$, between two points, $X_1^A$ and $X_2^A$, on the hyperboloid can be conveniently expressed via the so-called hyperbolic distance, $Z_{12}$, as follows:

\bqa
\frac{\cos \left(H \,L_{12}\right)}{H^2} \equiv \frac{Z_{12}}{H^2} \equiv g_{AB}\, X^A_1 \, X^B_2, \quad {\rm where} \quad g_{AB} \, X^A_{1,2} \, X^B_{1,2} = H^{-2}.
\eqa
To better understand the meaning of this expression, it is instructive to compare it to the geodesic distance, $l_{12}$, on a sphere of radius $R$:

$$R^2 \, \cos\left(\frac{l_{12}}{R}\right) \equiv R^2 \, z_{12} \equiv \left(\vec{X}_1\, \vec{X}_2\right),\quad
{\rm where} \quad \vec{X}^2_{1,2} = R^2.$$
While the spherical distance, $z_{12}$, is always less than unity, the hyperbolic one, $Z_{12}$, can acquire any value because of the Minkowskian signature of the metric.

All geodesics on the hyperboloid of Fig. (\ref{fig1}) are curves that are cut out on it by planes going through the origin of the ambient Minkowski space-time. (Compare this with the case of the sphere.) Hence, space-like geodesics are ellipses, time-like ones are hyperbolas and light-like are straight generatrix lines of the hyperboloid.

For every point $X^A$ on dS space, $g_{AB} \, X^A \, X^B = H^{-2}$, there is the antipodal one $\overline{X}^A = - X^A$, which is just its reflection with respect to the origin in the ambient Minkowski space. Note that then $Z_{1\overline{2}} = - Z_{12}$.

\subsection{Global de Sitter metric}

To define a metric on dS space, which is induced from the ambient space, one has to find a solution of eq. (\ref{hyp}). One possibility is as follows

\bqa
X^0 = \frac{\sinh (Ht)}{H}, \quad X^i = \frac{n_i \, \cosh(Ht)}{H}, \quad i = 1, \dots, D
\eqa
where $n_i$ is a unit, $n_i^2 = 1$, $D$-dimensional vector. One can choose:

\bqa
n_1 = \cos \theta_1, \quad -\frac{\pi}{2} \leq \theta_1 \leq \frac{\pi}{2} \nonumber \\
n_2 = \sin \theta_1 \, \cos \theta_2, \quad -\frac{\pi}{2} \leq \theta_2 \leq \frac{\pi}{2} \nonumber \\
\dots \\
n_{D-2} = \sin\theta_1 \, \sin \theta_2 \dots \sin \theta_{D-3} \, \cos\theta_{D-2}, \quad -\frac{\pi}{2} \leq \theta_{D-2} \leq \frac{\pi}{2} \nonumber \\
n_{D-1} = \sin\theta_1 \, \sin \theta_2 \dots \sin \theta_{D-2} \, \cos\theta_{D-1}, \quad -\pi \leq \theta_{D-1} \leq \pi \nonumber \\
n_{D} = \sin\theta_1 \, \sin \theta_2 \dots \sin \theta_{D-2} \, \sin\theta_{D-1}. \nonumber
\eqa
Then, the induced metric is:

\bqa\label{global}
ds^2 = - dt^2 + \frac{\cosh^2(Ht)}{H^2} \, d\Omega^2_{D-1},
\eqa
where

\bqa
d\Omega^2_{D-1} = \sum_{j=1}^{D-1} \left(\prod_{i=1}^{j-1} \sin^2\theta_i\right) \, d\theta_j^2
\eqa
is the line element on the unit $(D-1)$-dimensional sphere. The metric (\ref{global}) covers dS space totaly and is referred to as global. Its constant $t$ slices are compact $(D-1)$-dimensional spheres. Note that one can obtain from (\ref{global}) the metric of the $D$-dimensional sphere after the analytical continuation, $H \, t \to i \left(\theta_D - \frac{\pi}{2}\right)$.

The hyperbolic distance in these coordinates is given by:

\bqa
Z_{12} = - \sinh(H t_1) \, \sinh(H t_2) + \cosh(H t_1)\, \cosh(H t_2) \, \cos(\omega),
\eqa
where $\cos(\omega) = (\vec{n}_1, \vec{n}_2)$.

\begin{figure}
\begin{center}
\includegraphics[scale=0.4]{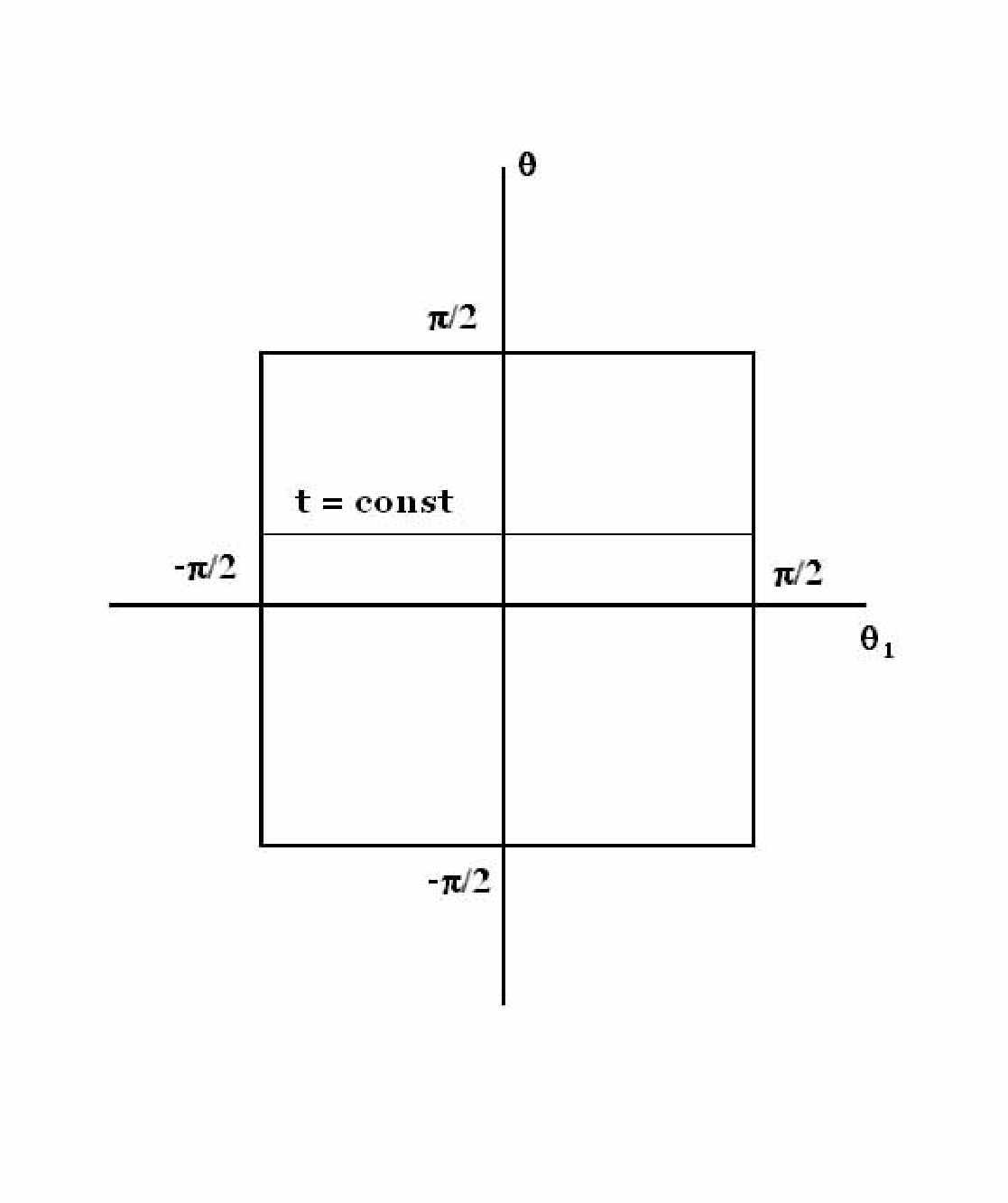}\caption{The standard quadratic Penrose diagram of $D$-dimensional dS space, when $D > 2$. The straight thin line is the constant $t$ and/or $\theta$ slice.}\label{fig2}
\end{center}
\end{figure}

\subsection{Penrose diagram}

To understand the causal structure of dS space it is convenient to transform the global coordinates as follows:

$$
\cosh^2(H t) = \frac{1}{\cos^2\theta}, \quad -\frac{\pi}{2} \leq \theta \leq \frac{\pi}{2}
$$
and to obtain the  metric

\bqa
ds^2 = \frac{1}{H^2 \, \cos^2\theta} \, \left[- d\theta^2 + d\Omega^2_{D-1}\right],
\eqa
which is conformal to that of the Einstein static universe, $ds^2_{ESU} = - d\theta^2 + d\Omega^2_{D-1}$, with compact time $\theta$. The causal structure of the latter universe coincides with that of dS space because sign of $ds^2$ coincides with that of $ds^2_{ESU}$. Hence, one can drop the conformal factor and depict the compact space. Such a procedure is just a variant of the stereographic projection. In the modern language the result of the projection of a space-time is referred to as Penrose diagram.

If $D>2$, then to draw the diagram on the two-dimensional sheet we should choose, in addition to $\theta$, one of the angles $\theta_j$, $j = 1, \dots, D-1$. The usual choice is $\theta_1$ because the metric in question has the form $d\theta^2 + d\theta_1^2 + \sin^2(\theta_1) d\Omega^2_{D-2}$, i.e., it is flat in the $(\theta-\theta_1)$-plain. The Penrose diagram for dS space, whose dimension is grater than $2$, is depicted on fig. \ref{fig2}.

Note that when $D>2$ angle $\theta_1$ is taking values in the range $\left[-\frac{\pi}{2},\, \frac{\pi}{2}\right]$. At the same time, when $D=2$ we have that $\theta_1 \in \left[-\pi, \, \pi\right]$. When $D>2$, the problem with the choice of $\theta_1$ in the Penrose diagram is that then cylindrical topology, $S^{D-1} \times R$, of dS space is not transparent. At the same time, the complication with the choice of $\theta_{D-1} \in \left[-\pi,\, \pi\right]$, instead of $\theta_1$, appears form the fact that the metric in the $(\theta-\theta_{D-1})$-plain is not flat. For this reason we prefer to consider just the stereographic projection in the two-dimensional case because it is sufficient to describe the causal structure and also clearly shows the topology of dS space.

\begin{figure}
\begin{center}
\includegraphics[scale=0.4]{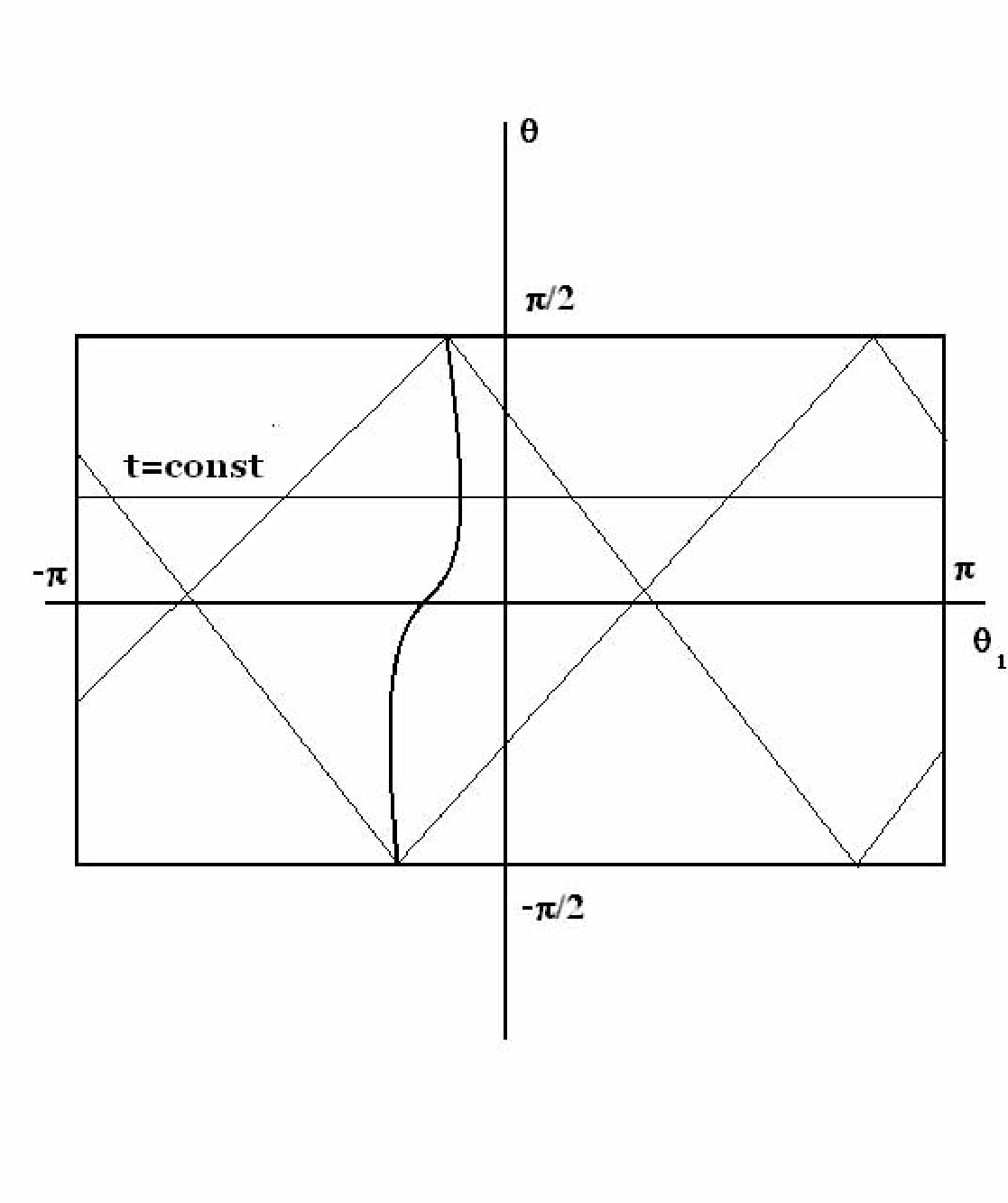}\caption{The rectangular Penrose diagram of the two-dimensional dS space. Note that the left and right sides of this rectangle are glued to each other. Thus, while on fig. \ref{fig2} the positions $\theta_1 = \pm \frac{\pi}{2}$ sit at the opposite poles of the spherical time slices, on the present figure the positions $\theta_1 = \pm \pi$ coincide.}\label{fig3}
\end{center}
\end{figure}

The Penrose diagram of the two-dimensional dS space is shown on fig. \ref{fig3}. This is the stereographic projection of the hyperboloid from fig. \ref{fig1}. What is depicted here is just a cylinder because the left and right sides of the rectangle are glued to each other. The fat solid curve is a world line of a massive particle. Thin straight lines, which compose $45^{\rm o}$ angle with both $\theta$ and $\theta_1$ axes, are light rays. From this picture one can see that every observer has a causal diamond within which he can exchange signals. Due to the expansion of dS space there are parts of it that are causally disconnected from the observer.

\subsection{Expanding and contracting Poincar\'{e} patches}

Another possible solution of (\ref{hyp}) is based on the choice:

\bqa
- \left(H\, X^0\right)^2 + \left(H\, X^D\right)^2 = 1 - \left(H\, x^i_+\right)^2 \, e^{2 \, H\, \tau_+}, \nonumber \\
\left(H\, X^1\right)^2 + \dots + \left(H\, X^{D-1}\right)^2 = \left(H\, x^i_+\right)^2 \, e^{2 \, H\, \tau_+}.
\eqa
Then, one can define

\bqa\label{choice}
H \, X^0 = \sinh\left(H \, \tau_+\right) + \frac{\left(H\, x^i_+\right)^2}{2}\, e^{H\,\tau_+}, \nonumber \\
H\, X^i = H x^i_+ \, e^{H \, \tau_+}, \quad i = 1, \dots, D-1, \nonumber \\
H\, X^D = - \cosh\left(H \, \tau_+\right) + \frac{\left(H \, x^i_+\right)}{2} \, e^{H\,\tau_+}.
\eqa
With such coordinates the induced metric is

\bqa\label{EPP1}
ds^2_+ = - d\tau_+^2 + e^{2\, H \, \tau_+} \, d\vec{x}_+^2.
\eqa
Note, however, that in (\ref{choice}) we have the following restriction:
$- X^0 + X^D = -\frac{1}{H} \, e^{H \, \tau_+} \leq 0$, i.e., metric (\ref{EPP1}) covers only half, $X^0 \geq X^D$, of the entire dS space. It is referred to as the expanding Poincar\'{e} patch (EPP). Another half of dS space, $X^0 \leq X^D$, is referred to as the contracting Poincar\'{e} patch (CPP) and is covered by the metric

\begin{figure}
\begin{center}
\includegraphics[scale=0.4]{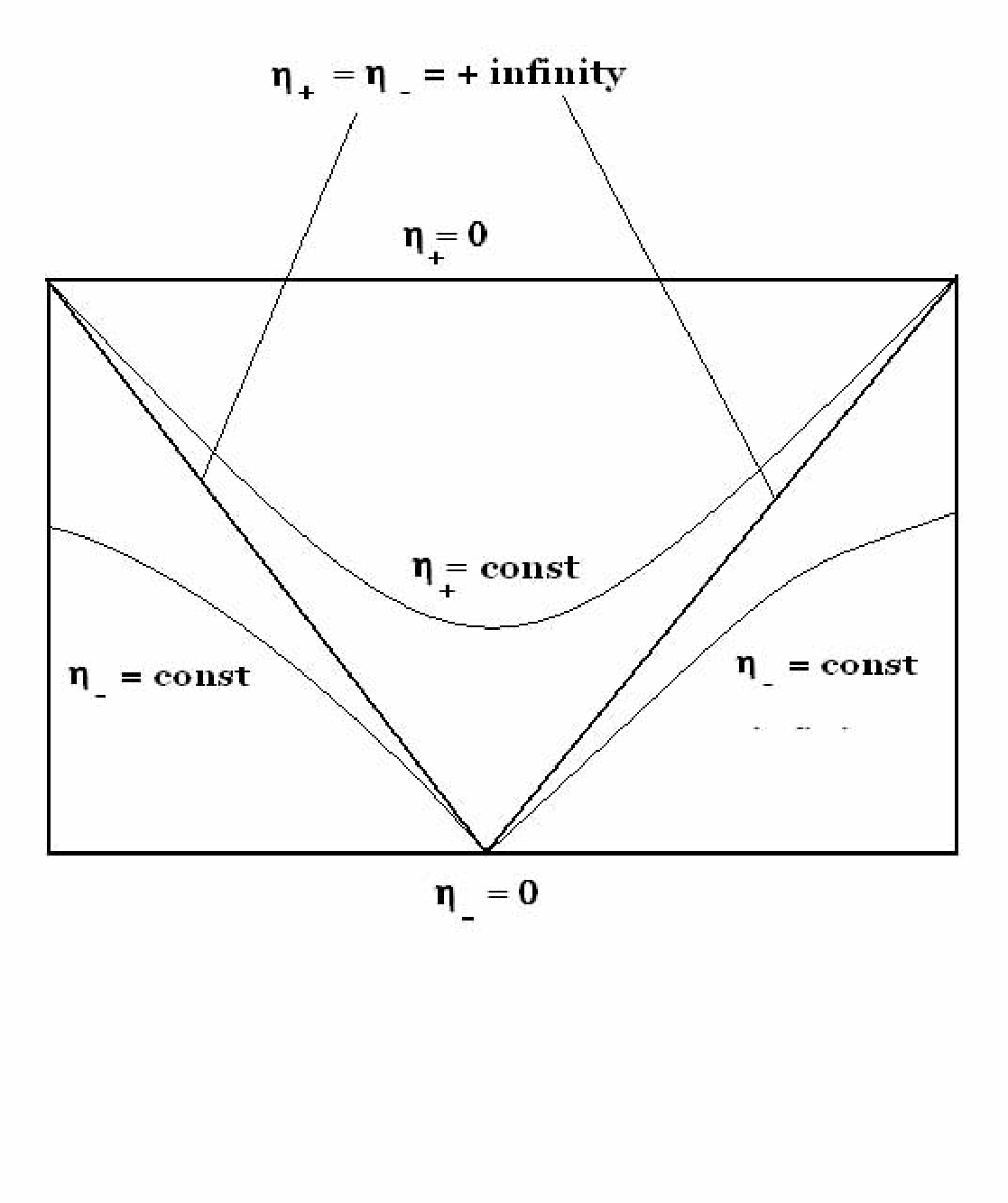}\caption{The boundary between the EPP and CPP is light-like and is situated at $\eta_\pm = + \infty$. We also show here the constant conformal time slices.}\label{fig4}
\end{center}
\end{figure}

\bqa\label{CPP1}
ds^2_- = - d\tau_-^2 + e^{- 2\, H \, \tau_-} \, d\vec{x}_-^2.
\eqa
In both patches it is convenient to change the proper time $\tau_\pm$ into the conformal one. Then, the EPP and CPP both possess the same metric:

\bqa\label{PP}
ds^2_\pm = \frac{1}{\left(H \, \eta_{\pm}\right)^2}\,\left[- d\eta_\pm^2 + d\vec{x}_\pm^2\right], \quad H\, \eta_\pm = e^{\mp H\, \tau_\pm}.
\eqa
However, while in the EPP the conformal time is changing form $\eta_+ = + \infty$ at past infinity ($\tau_+ = - \infty$) to $0$ at future infinity ($\tau_+ = + \infty$), in the CPP the conformal time is changing from $\eta_- = 0$ at past infinity ($\tau_- = -\infty$) to $+\infty$ at future infinity ($\tau_- = + \infty$).
Both the EPP and CPP are shown on fig. \ref{fig4}.

The hyperbolic distance in the EPP and CPP has the form:

\bqa\label{Z12}
Z_{12} = 1 + \frac{\left(\eta_1 - \eta_2\right)^2 - \left|\vec{x}_1 - \vec{x}_2\right|^2}{2 \, \eta_1 \, \eta_2}
\eqa
It is worth mentioning here that it is possible to cover simultaneously the EPP and CPP with the use of the metric (\ref{PP}), if one makes the changes $\vec{x}_\pm \to \vec{x}$ and $\eta_\pm \to \eta \in (-\infty, +\infty)$. Then, while at the negative values of the conformal time, $\eta = - \eta_+ < 0$, it covers the EPP, at its positive values, $\eta = \eta_- > 0$, this metric covers the CPP. The inconvenience of such a choice of global metric is due to that the boundary between the EPP and CPP simultaneously corresponds to $\eta = \pm \infty$.

\subsection{Other de Sitter metrics and patches}

There is another commonly used metric on dS space:

\bqa
ds^2 = - \left[1 - \left(H \, r\right)^2\right]\, dT^2 + \frac{dr^2}{\left[1 - \left(H \, r\right)^2\right]} + r^2 \, d\Omega_{D-2}^2, \quad 0 \leq r \leq 1/H.
\eqa
This metric covers only quoter of the entire space and is referred to as static. We are not going to consider quantum fields on this background. The reason for that is as follows. The metric in question contains nontrivial time component, $g_{00}(r) \neq 1$, and, hence, is seen by noninertial observers, unless the position of the observer is at $r=0$. As we explain below, we would like to understand the physics in dS space as seen by such observers which are not affected by any other force except the gravitational one. Hence, we restrict our considerations to the global, EPP and CPP metrics which are seen by inertial observers.

Note here that the transformation from the proper time, $\tau_\pm$, to the conformal one, $\eta_\pm$, (which makes $g_{00}$ time-dependent) is nothing but the change of the clock rate rather than a transition to the noninertial motion.

\subsection{Spatial volume in de Sitter space}

For our future considerations it is important to define here the physical and comoving spatial volumes in global dS space and in its Poincar\'{e} patches. The spatial sections in all aforementioned metrics contain conformal factors, $\frac{\cosh^2\left(H\, t\right)}{H^2}$ or $\frac{1}{\left(H \, \eta_\pm\right)^2}$. Then, there is the volume form, $d^{D-1}V$, with respect to the spatial metric, which is multiplying the corresponding conformal factor. This form remains constant during the time evolution of the spatial sections and is referred to as comoving volume.

It is important to observe that if one considers a dust in dS space, then its density per comoving volume remains constant independently of whether spatial sections are expanding or contracting.

At the same time if one takes into account the conformal factor, i.e., the expansion (contraction) of the EPP (CPP), then he has to deal with the physical volume, $\frac{d^{D-1}V_\pm}{\left(H \, \eta_\pm\right)^{D-1}}$. In global dS space the physical volume is $\frac{\cosh^{D-1}\left(H\, t\right) \, d^{D-1}V_{\rm sphere}}{H^{D-1}}$. Of course the density of the dust with respect to such a volume is changing in time.

\section{Free scalar fields in de Sitter space}

We start our discussion with free massive real scalar fields which are coupled to the dS background in the minimal way. From now on we set the curvature of dS space to one, $H=1$, and assume it to be fixed. In the following sections we will question this assumption in the presence of quantum effects in interacting theories.

\subsection{Free waves in Poincar\'{e} patches}

The action of the free theory under consideration is as follows:

\bqa\label{freeac}
S = \int d^Dx \sqrt{|g|}\, \left[g^{\alpha\beta} \, \pr_\alpha \phi \, \pr_\beta \phi + m^2 \, \phi^2\right].
\eqa
In the EPP or CPP the Klein--Gordon (KG) equation is:

\bqa\label{KG}
\left[- \eta^2 \pr_\eta^2 + (D-2)\,\eta \, \pr_\eta + \eta^2 \Delta - m^2 \right] \, \phi(\eta, \vec{x}) = 0,
\eqa
where $\Delta$ is the $(D-1)$-dimensional flat Laplacian.
At this point we do not distinguish the EPP from the CPP and, hence, drop
the ``$\pm$'' indexes of $\eta$ and $\vec{x}$.
The harmonics, which solve this equation, can be represented as $\phi_p (\eta, \vec{x}) = g_p(\eta) \, e^{\mp i\, \vec{p}\, \vec{x}}$. If one assumes the ansatz $g_p(\eta) = \eta^{\frac{D-1}{2}} \, h(p\eta)$, where $p = \left|\vec{p}\right|$, then (\ref{KG}) reduces to the Bessel equation for $h(p\eta)$. The index of this equation is $i\mu = i \sqrt{m^2 - \left(\frac{D-1}{2}\right)^2}$.

Generic solution of the Bessel equation with such an index behaves as follows:

\bqa\label{beha}
h(x) = \left\{\begin{array} {c} A \, \frac{e^{i\, x}}{\sqrt{x}} + B \, \frac{e^{- i\, x}}{\sqrt{x}}, \quad x \to \infty \\
C \, x^{i\mu} + D \, x^{-i\mu}, \quad x \to 0\end{array} \right.
\eqa
Here $A,B,C,D$ are some complex constants. Taking into account that $(p\eta)^{\pm i \mu} \sim e^{\pm i \, \mu \, \tau}$, one can interpret $x^{\pm i\, \mu}$, if $\mu$ is real, as a single wave in the future (past) of the EPP (CPP).

If the field is heavy, $m > (D-1)/2$, then, it belongs to the so-called principal series. The corresponding $g_p$ harmonics oscillate and decay to zero as $\eta^{(D-1)/2 \pm i \mu}$, when $\eta \to 0$. At the same time, for the light field from the complementary series, $m \leq (D-1)/2$, the harmonic functions homogeneously decay to zero as $\eta^{(D-1)/2 \pm \sqrt{(D-1)^2/4 - m^2}}$, when $\eta \to 0$. The only exception is the massless field, $m=0$, for which harmonics approach a non-zero constant in future infinity.

Bessel functions of the first, $h(x) = \sqrt{\frac{\pi}{\sinh (\pi \mu)}} \, J_{i\mu}(x)$, or second, $h(x) \propto Y_{i\mu}(x)$, kinds correspond to those modes for which either $C$ or $D$ in (\ref{beha}) is vanishing. But then both $A$ and $B$ are not zero. Thus, Bessel harmonics represent single free waves only in the future (past) infinity of the EPP (CPP). Correspondingly they are referred to as out- (in-)harmonics in the EPP (CPP).

Performing a Bogolyubov transformation, one can consider also other possibilities for $h(x)$. For example, Hankel functions of the first, $h(x) = \frac{\sqrt{\pi}}{2} \, e^{-\frac{\pi\, \mu}{2}} \, H^{(1)}_{i\mu}(x)$, or second, $h(x) = \frac{\sqrt{\pi}}{2} \, e^{\frac{\pi\, \mu}{2}} \, H^{(2)}_{i\mu}(x)$, kinds are such that either $A$ or $B$ in (\ref{beha}) is vanishing. But then both $C$ and $D$ are not zero. Thus, Hankels, which are referred to as Bunch--Davies (BD) modes \cite{Bunch:1978yq}, represent single free waves only in the past (future) infinity of the EPP (CPP).

For the beginning we do not specify our choice for $h(p\eta)$ because none of them behaves as single wave simultaneously at past and future infinity. Then, quantum field can be mode expanded in the usual way:

\bqa\label{mode}
\phi(\eta, \vec{x}) = \int d^{D-1}\vec{p}\left[a_{\vec{p}} \, g_p(\eta)\, e^{- i\, \vec{p} \, \vec{x}} + h.c. \right], \quad g_p(\eta) = \eta^{\frac{D-1}{2}} \, h(p\eta).
\eqa
Corresponding annihilation, $a_{\vec{p}}$, and creation, $a^+_{\vec{p}}$, operators obey the proper Heisenberg commutation relations. They follow from the commutation relations of $\phi$ with its conjugate momentum and are the corollary of the time-independence of the Wronskian, $\eta^{2-D}\, \left(g_p \, \dot{g}^*_p - \dot{g}_p \, g^*_p\right) = \pm i$, which follows from the equations of motion. This observation allows to fix the proper normalization of the harmonic functions.

For the illustrative reasons let us consider the free Hamiltonian in $D=4$. In an arbitrary dimension the formulas are similar. The Hamiltonian can be found from (\ref{freeac}), using the machinery presented, e.g., in Ref. \cite{Misner:1974qy}. The energy-momentum tensor is $T_{\alpha\beta} = \pr_\alpha \phi \, \pr_\beta \phi - g_{\alpha\beta} L$, where $L$ is the Lagrangian density. Then, the free Hamiltonian (before the normal ordering) is $H_0(\eta) = \frac{1}{\eta^2} \, \int d^3x \, T_{00}(\eta, \vec{x})$ and can be expressed via the creation and annihilation operators as:

\bqa\label{freeH}
H_0(\eta) = \int d^3p \left[A_p(\eta) \, a_{\vec{p}}^+ \, a_{\vec{p}} + B_p(\eta) \, a_{\vec{p}} \, a_{-\vec{p}} + h.c.\right], \nonumber \\
A_p(\eta) = \frac{1}{2\eta^2}\, \left\{\left|\dot{g}_p\right|^2 + \left[p^2 + \frac{m^2}{\eta^2}\right]\, \left|g_p\right|^2\right\}, \nonumber \\
B_p(\eta) = \frac{1}{2\eta^2}\, \left\{\dot{g}_p^2 + \left[p^2 + \frac{m^2}{\eta^2}\right]\, g_p^2\right\}, \quad \dot{g}_p \equiv \frac{dg_p}{d\eta}.
\eqa
The main characteristic feature of this Hamiltonian is that one cannot diagonalize it once and forever. That is because there is no solution of the KG equation which coincides with the function that solves equation $B_p(\eta) = 0$. (In flat space the simultaneous solution of the corresponding KG equation and of $B_p = 0$ is the plane wave.) Moreover, one cannot use such $g_p(\eta)$ which solve $B_p(\eta) = 0$ equation in place of the mode functions in (\ref{mode}) because then the corresponding creation and annihilation operators will not obey the appropriate Heisenberg algebra.

However, $B_p(\eta)$ can be set to zero as $\eta \to +\infty$. This can be done if one chooses the BD modes because they behave as single {\it plane} waves when $\eta\to +\infty$. Then, we have a clear meaning of the positive energy and of the particle in this region of space-time because the Hamiltonian is diagonal. Recalling that the past (future), $\eta \to + \infty$, of the EPP (CPP) corresponds to the UV limit of the physical momentum, $p\eta$, one can see that other harmonics have wrong UV behavior. In fact, after a Bogolyubov rotation of the BD modes to other harmonics one mixes positive and negative energy excitations.

There are no modes that allow to set $B_p(\eta)$ to zero as $\eta\to 0$. That is because the gravitational field is never switched off in this limit. In fact, $B_p(\eta)$ does not asymptotically approach a constant and one has to re-diagonalize the free Hamiltonian at each new value of $\eta$, as $\eta\to 0$. (Note that in the above UV limit the gravitational field is effectively switched off because high energy harmonics are not sensitive to the comparatively small curvature of the background space.) Hence, {\it naively} even the Bessel functions, out- (in-)modes, do not provide a proper quasi-particle description in the future (past) infinity of the EPP (CPP).

\subsection{Digression on particle interpretation in de Sitter space}

It is probably worth stressing here that throughout these notes we will avoid using the notion of particle, unless this notion is meaningful. We will concentrate on the behavior of the correlation functions. However, let us make here some comments about the particle interpretation.

In curved space-times one usually avoids the use of the notion of particle because it is believed to be an observer dependent phenomenon. In fact, different observers may detect different particle fluxes. However, one should separate the Unruh effect \cite{Unruh:1976db} from what we would like to call as the real particle production. In Minkowski space, both inertial and noninertial observers see the same state --- Minkowski (Poincar\'{e} invariant) vacuum. However, while an inertial observer sees it as the empty space, a noninertial one sees it as the thermal state. That is due to the specific correlation of the vacuum fluctuations along its worldline \cite{AkhmedovUnruh}, \cite{Akhmedov:2007xu}. Note that there is no nontrivial gravitational field present in the circumstances under consideration because in flat space the Riemanian tensor is exactly zero.

The real particle creation is due to a change of the ground state under the influence of quantum effects in a non-trivial background field. Then a flux is seen by all sorts of observers. That is exactly what happens in the strong electric field, in dS space and in the collapsing background. Rephrasing that, we would like to say here that, while in Minkowski space there is one type of observers that does not see any particle flux, in dS space there is no such an observer that sees nothing. On general grounds, we expect that the least flux is seen by inertial observers --- they do not see the extra Unruh type of flux, so to say. (Note here that the least possible flux in a given space-time is an observer independent/invariant notion, i.e. is a characteristic feature of the given space--time.)
Apart from that, if we see strong backreaction in a noninertial frame it is not clear whether it should be attributed to the background gravitational field or to the extra non-gravitational force acting on the corresponding observer.
While in an inertial frame the situation is unambiguous.

In any case, independently of whatever name we use for different quantities that are calculated below, all of them are just components of correlation functions. And at the end of the day the objects that we calculate are just correlation functions. However, the notion of particle sometimes is convenient for the physical interpretation of various equations that will appear below.

So, what do we mean by particle? In general if the free Hamiltonian of a theory is diagonal then one indeed can have a particle interpretation. Furthermore, if for some reason the anomalous quantum average $\langle a_p \, a_{-p}\rangle$ is strongly suppressed in comparison with $\langle a^+_p a_p\rangle$ then one also can give a meaning to particle like excitations, as we will see below.

Then, in principle at every given moment of time, $\eta$, one can make an instantaneous Bogolyubov rotation as follows \cite{GribMamaevMostepanenko}:

\bqa
b_{\vec{p}}(\eta) = \alpha_p(\eta) \, a_{\vec{p}} + \beta_p(\eta) \, a_{-\vec{p}}^+, \quad
b^+_{\vec{p}}(\eta) = \alpha^*_p(\eta) \, a^+_{\vec{p}} + \beta^*_p(\eta)\, a_{-\vec{p}},
\eqa
with

\bqa
\alpha_p(\eta) = \sqrt{\frac{A_p + \Omega_p}{2\, \Omega_p}} \quad
{\rm and} \quad \beta_p(\eta) = \frac{B^*_p}{A_p + \Omega_p}\, \alpha_p(\eta),
\eqa
where $\Omega_p(\eta) = \sqrt{A_p^2 - \left|B_p\right|^2}$. The Hamiltonian (\ref{freeH}) becomes diagonal

\bqa
H_0(\eta) = \int d^3\vec{p} \, \Omega_p(\eta) \, \left[b^+_{\vec{p}}(\eta) \, b_{\vec{p}}(\eta) + h.c.\right].
\eqa
The rotated harmonics are

\bqa
\bar{g}_p(\eta) = \alpha_p^* \, g_p - \beta^*_p\, g^*_p = \frac{i\, \eta}{\left[p^2 + \frac{m^2}{\eta^2}\right]^{\frac14}}\, \frac{\dot{g}_p - i\, \sqrt{p^2 + \frac{m^2}{\eta^2}}\, g_p}{\left|\dot{g}_p - i\, \sqrt{p^2 + \frac{m^2}{\eta^2}}\, g_p\right|}.
\eqa
This allows to have a particle interpretation around any given moment of time $\eta$.

Note that the new creation and annihilation operators, $b_{\vec{p}}$ and $b^+_{\vec{p}}$, depend on time $\eta$. But $a_{\vec{p}}$ and $a_{\vec{p}}^+$ are time independent in the free theory --- all their time dependence is, then, gone into the harmonics $g_p(\eta)$. They start to depend on time if one turns on interactions.

One reason to avoid using $b$'s is that if one knows all expressions in terms of $a$'s then it is not hard to restor their form in terms of $b$'s. Another reason is that as we will see, if $m > (D-1)/2$, $a$'s do provide a proper quasi-particle description in the IR limit. In fact, it will happen that for some choice of harmonics, $\langle a_p \, a_{-p}\rangle$ is suppressed in comparison with $\langle a^+_p a_p\rangle$, as the system approaches future infinity. Hence, one does not really need to make the rotation to $b$'s.

\subsection{Free waves in global de Sitter space}

In global dS space the KG equation is as follows:

\bqa\label{25}
\left[- \pr_t^2 + (D-2) \, \tanh(t) \, \pr_t + \frac{\Delta_{D-1}(\Omega)}{\cosh^2(t)} - m^2\right] \phi(t, \Omega) = 0.
\eqa
Here $\Delta_{D-1}(\Omega)$ is the $(D-1)$-dimensional spherical Laplacian. The solution of this equation can be represented as $\phi_{j,\vec{m}}(t,\Omega) = g_j(t)\, Y_{j,\vec{m}}(\Omega)$, where $\Delta_{D-1}(\Omega) \, Y_{j,\vec{m}}(\Omega) = - j\, (j + D - 2)\, Y_{j,\vec{m}}(\Omega)$, and $\, Y_{j,\vec{m}}(\Omega)$ are $(D-1)$-dimensional spherical harmonics, $\vec{m}$ is the multi-index enumerating them in the dimension grater than two.

The equation for $g_j(t)$, following from (\ref{25}), can be reduced to the hypergeometric one. One possible its solution is \cite{BoussoStrominger}:

\bqa
g^{(in)}_j(t) = \frac{2^{j + \frac{D}{2} - 1}}{\sqrt{\mu}} \cosh^j(t) \, e^{\left(j + \frac{D-1}{2} \pm i\, \mu\right)\, t} \, F\left(j+ \frac{D-1}{2}, j + \frac{D-1}{2} \mp i \, \mu; 1 \mp i \, \mu; - e^{2\, t}\right).
\eqa
Where, from now on, $F(a,b;c;x)$ is the hypergeometric function of the $(2,1)$ type and $\mu$ was defined in the previous subsection. These harmonics behave as single waves at the past infinity, $t \to -\infty$, of global dS space:

$$
g^{(in)}_j (t) \sim e^{\frac{D-1}{2}\, t} \, e^{\mp i \, \mu \, t}.
$$
That is the reason why they are referred to as in-harmonics. At the same time, as $t\to +\infty$, these modes behave as

$$
g^{(in)}_j(t) \sim e^{\frac{D-1}{2}\, t} \, \left(C_1 \, e^{\mp i \, \mu \, t} + C_2 \, e^{\pm i \, \mu \, t}\right),
$$
where $C_{1,2}$ are both non-zero complex constants in even dimensional dS space. In odd dimensional dS spaces $C_2=0$ \cite{BoussoStrominger}. The out-harmonics in global dS are as follows $g^{(out)}_j (t) = \left[g^{(in)}_j(-t)\right]^*$. Their name is justified by the observation that they behave as single waves at future infinity.

Another peculiar type of harmonics in global dS space is given by the so-called Euclidian ones \cite{Allen:1985ux}, \cite{Mottola:1984ar}, \cite{BoussoStrominger}. They are defined as:

\bqa
g^{(E)}_j(t) = \frac{2^{j + \frac{D}{2} - 1} \, i^{- j + \frac{D-1}{2}}}{\sqrt{\mu}} \cosh^j(t) \, e^{\left(j + \frac{D-1}{2} \pm i\, \mu\right)\, t}\times \nonumber \\ \times F\left(j+ \frac{D-1}{2}, j + \frac{D-1}{2} \pm i \, \mu; 2\, j + D - 1; 1 + e^{2\, t}\right)
\eqa
and are regular on the lower hemisphere after the analytical continuation, $t \to i\,\left(\theta - \pi/2\right)$.
Then, complete wave functions of the Euclidian modes obey $\phi^{(E)}_{j,\vec{m}} \left(\overline{X}\right) = \left[\phi^{(E)}_{j,\vec{m}}\left(X\right)\right]^*$, where $\overline{X}$ is the antipodal point of $X$.

All the aforementioned mode functions in global dS space belong to the one-parameter family of the so-called $\alpha$-harmonics \cite{Allen:1985ux}, \cite{BoussoStrominger}:

\bqa\label{bogrot}
\phi_{j,\vec{m}}^{(\alpha)}(X) = \frac{1}{\sqrt{1 - e^{\alpha + \alpha^*}}} \, \left[\phi^{(E)}_{j,\vec{m}}(X) + e^\alpha \, \overline{\phi_{j,\vec{m}}^{(E)}}(X)\right],
\eqa
where $\alpha$ is a complex number.

Note the coincidence, after the identification $\eta_\mp = e^{\pm t}$, of the above behavior of the global dS harmonic functions with that of the modes at the past and future of the CPP and EPP. In fact, under such an identification the metric of global dS space at its past and future infinity can be well approximated by those of the CPP and EPP, correspondingly. Hence, there is a one-to-one correspondence between harmonic functions in the EPP (CPP) and $\alpha$-modes in global dS.

The free Hamiltonian of the four-dimensional theory can be written as:

\bqa
H_0(t) = \frac12 \, \sum_{j,\vec{m}} \left[A_{j}(t)\, a_{j,\vec{m}}^+\, a_{j,\vec{m}} + B_{j}(t) \, a_{j,\vec{m}}\, a_{j,-\vec{m}} + h.c. \right], \nonumber \\
A_{j}(t) = \frac{\cosh^3(t)}{2}\, \left\{\left|\dot{g}_j\right|^2 + \left[\frac{L\left(L+2\right)}{\cosh^2(t)} + m^2\right]\, \left|g_j\right|^2\right\}, \nonumber \\
B_{j}(t) = \frac{\cosh^3(t)}{2}\, \left\{\dot{g}_j^2 + \left[\frac{L\left(L+2\right)}{\cosh^2(t)} + m^2\right]\, g_j^2\right\} \times \left({\rm a \,\, phase}\right).
\eqa
There is no choice of harmonics which allows to diagonilize this Hamiltonian neither in the past nor in the future infinity because in global dS space the background field is never switched off. In fact, $B_j(t)$ does not approach a constant neither in the past nor in the future infinity of global dS space.

\subsection{Green functions in de Sitter space}

We continue with the construction of two-point correlation functions. The Wightman function, $\langle \phi(X_1) \, \phi(X_2)\rangle$, is a solution of the homogeneous KG equation: $\left(\Box - m^2\right) \, G\left(X_1, X_2\right) = 0$. Because of the dS isometry invariance, it should be a function of the invariant distance between $X_1$ and $X_2$. The KG operator, $\left(\Box - m^2\right)$, when acting on a function of $Z = Z_{12}$ can be reduced to \cite{Mottola:1984ar}, \cite{Allen:1985ux}:

\bqa\label{KGW}
\left[\left(Z^2 - 1\right)\, \pr^2_Z + D\, Z \, \pr_Z + m^2\right] \, G(Z) = 0.
\eqa
This equation coincides with that for the Wightman function on the $D$-dimensional sphere. We just have to keep in mind that in the latter case $Z$ is the spherical distance rather than the hyperbolic one. The same equation is also valid in anti-dS and Euclidian anti-dS (Lobachevsky) space. One just has to change the sign of $m^2$ term, due to the change of the sign of the curvature $H^2$, and keep in mind that $Z$ is the hyperbolic distance in the corresponding space.

Eq. (\ref{KGW}) has three singular points $Z = \pm 1, \infty$ in the complex $Z$-plain. Hence, it is not hard to recognize in it the hypergeometric equation. After the transformation to the new variable, $z = (1+Z)/2$, one puts the singular points into their standard positions, $z = 0, 1, \infty$. Then, the generic solution of this equation is:

\bqa\label{alpha}
G_W\left(Z\right) = A_1 \, F\left(\frac{D-1}{2} + i\, \mu, \frac{D-1}{2} - i\, \mu; \frac{D}{2}; \frac{1 + Z}{2}\right) + \nonumber \\ + A_2 \, F\left(\frac{D-1}{2} + i\, \mu, \frac{D-1}{2} - i\, \mu; \frac{D}{2}; \frac{1 - Z}{2}\right).
\eqa
Here $A_{1,2}$ are some complex constants and $\mu$ was defined above. The two hypergeometric functions in (\ref{alpha}) behave, when $Z \to \pm 1$, as follows:

\bqa\label{pole}
F\left(\frac{D-1}{2} + i\, \mu, \frac{D-1}{2} - i\, \mu; \frac{D}{2}; \frac{1 \pm Z}{2}\right) \sim \frac{1}{\left(1 \mp Z\right)^{\frac{D}{2} - 1}}.
\eqa
Also they have the branching point at $Z \to \infty$:

$$G_W(Z) \sim B_+ \, Z^{-\frac{D-1}{2} + i\, \mu} + B_- \, Z^{-\frac{D-1}{2} - i \, \mu}.$$
Here $B_\pm$ are some complex constants which depend on $A_{1,2}$. Thus, $G_W(Z)$ is an analytical function on the complex $Z$-plain with two branch cuts going from $Z = \pm 1$ to infinity.

What is the physical meaning of the singularities that $G_W(Z)$ has? Is there any state for which $G_W(Z)$ looks as in (\ref{alpha})? We are going to address these questions now.

\subsubsection{Two-point correlation functions in global de Sitter space}

We first scetch the situation in global dS and then continue with a bit more extensive discussion of the EPP (CPP) case. We choose some solution of the corresponding KG equation and, thus, specify $\alpha$-modes. Then, we define the $\alpha$-vacuum as the state which is annihilated by the corresponding annihilation operators: $a^{(\alpha)}_{j,\vec{m}} \, |\alpha\rangle = 0$. For every choice of $\alpha$ there is the corresponding $\alpha$-vacuum \cite{Allen:1985ux}. Then, one can construct the two point Wightman function as

$$G_\alpha\left(X_1, X_2\right) \equiv \left\langle \alpha\left| \phi(X_1) \, \phi(X_2) \right|\alpha \right\rangle,$$
where he has to substitute the $\alpha$-harmonic expansion of $\phi(X)$. It is straightforward to show (using, e.g., \cite{GradshteinRizhik}) that {\it away from its singularity points} this function looks as (\ref{alpha}) with such $A_{1,2}$ which depend on $\alpha$. At the same time $Z=Z_{12}$ is the hyperbolic distance, which is expressed via the global coordinates of $X_1$ and $X_2$.

For the Euclidian vacuum we have that $A_2 = 0$ and the singularity point of the corresponding Wightman function $G_E(Z_{12})$ is at $Z_{12} = 1$. From (\ref{pole}) one can see that it is the standard UV behavior, when $X_1$ is sitting on the light-cone whose apex is at $X_2$.
The demand that this UV singularity should be the same as in flat space allows to fix simultaneously the $A_1$ coefficient in (\ref{alpha}) and the $\epsilon$-prescription (the resolution of the singularity):

\bqa\label{WBD}
G_{E}(Z) = G\left(Z  - i\, \epsilon \, sign \Delta t\right), \quad {\rm where} \nonumber \\
G(Z) \equiv \frac{\Gamma\left(\frac{D-1}{2} + i\, \mu\right)\, \Gamma\left(\frac{D-1}{2} - i\, \mu\right)}{\left(4\, \pi\right)^{\frac{D}{2}} \, \Gamma\left(\frac{D}{2}\right)} \, F\left(\frac{D-1}{2} + i\, \mu, \frac{D-1}{2} - i\, \mu; \frac{D}{2}; \frac{1 + Z}{2}\right).
\eqa
The same value of $A_1$ also follows from the proper normalization of the Euclidian harmonics.

The $\epsilon$-prescription for the $\alpha$-harmonics follows after the Bogolyubov rotation (\ref{bogrot}). The result is as follows \cite{BoussoStrominger}:

\bqa\label{alphaG}
G_{\alpha}(Z) = \frac{1}{1 - e^{\alpha + \alpha^*}} \, \left[G\left(Z - i\, \epsilon \, sign \Delta t\right) + e^{\alpha + \alpha^*} \, G\left(Z + i\, \epsilon \, sign \Delta t\right) + \right. \nonumber \\ \left.  + e^{\alpha^*} \, G\left(- Z + i\, \epsilon \, sign \Delta t\right) + e^{\alpha} \, G\left(- Z - i\, \epsilon \, sign \Delta t\right)\right],
\eqa
where $G(Z)$ is defined in (\ref{WBD}). If one puts $\epsilon = 0$ in this expression, he can reproduce (\ref{alpha}) with generic $A_{1,2}$.

At the same time $\epsilon$-prescription for the T-ordered (Feynman) propagator in the Euclidian vacuum is $G_{T}(Z) = G\left(Z  - i\, \epsilon\right)$. This correlation function is just the analytical continuation of the propagator on the sphere in the complex $Z$-plain.

\subsubsection{Two-point correlation functions in the Poincar\'{e} patches}

To define the Wightman function in the EPP or CPP, one also has to pick up a solution of the Bessel equation and define the corresponding vacuum, $a_{\vec{p}} \, |vac \rangle = 0$. Then, doing the same as it was done above, one obtains the Fourier expansion of the correlation function:

\bqa\label{Fourier}
G_W(X_1, X_2) = \int d^{D-1}\vec{p} \, e^{i\, \vec{p}\, \left(\vec{x}_1 - \vec{x}_2\right)}\, \left(\eta_1\, \eta_2\right)^{\frac{D-1}{2}}\, h(p\eta_1)\, h^*(p\eta_2).
\eqa
Using \cite{GradshteinRizhik}, one can calculate this integral for different choices of $h(p\eta)$. 6.672.1-4 of \cite{GradshteinRizhik} can be used for the calculation in the case of 2D, and generalizations to higher dimensions are straightforward. There are two conclusions that follow. First one is that $G_W$ depends on the invariant hyperbolic distance $Z_{12}$ expressed via $\eta_{1,2}$ and $\vec{x}_{1,2}$ as (\ref{Z12}). And the second one is that different solutions of the Bessel equation, $h(p\eta)$, are in one--to--one correspondence with the concrete values of $A_{1,2}$ in (\ref{alpha}) \cite{Allen:1985ux}.

In particular, for the BD state we have that $A_2 = 0$. Thus, this state in the EPP and CPP leads to the same propagator as the Euclidian state in global dS space. Then, for the BD state the singularity of $G_W(Z)$ is at $Z=1$ and corresponds to the UV limit of the physical momentum, $p\eta\to\infty$. Hence, similarly to the flat space case, for the integral in (\ref{Fourier}) to be properly defined at the singularity, there should be an appropriate shift as follows: $\eta_1 - \eta_2 \to \eta_1 - \eta_2 \pm i\, \epsilon$. The sign of this shift depends on which one among $\eta_1$ or $\eta_2$ is grater than the other. Thus, the Wightman function in the BD state is also defined by (\ref{WBD}). The difference is that now $Z_{12}$ is expressed via the EPP (CPP) coordinates of $X_1$ and $X_2$ and $\Delta t$ should be substituted by $\mp \Delta \eta_{\pm}$. Here the ``$-$'' sign for the EPP is due to the reverse order of the time flow.

After a Bogolyubov rotation to other modes in the EPP (CPP) $A_2$ becomes nonzero and $A_1$ is changed. Thus, the residue of the singularity at $Z_{12} = 1$ is changed and also appears another singularity at $Z_{12} = -1$. Recalling that $Z_{12} = - Z_{1\overline{2}}$, one can conclude that another singularity corresponds to the situation when $X_1$ is sitting on the light-cone with the apex at $\overline{X}_2$ --- antipodal point of $X_2$. This singularity is causally disconnected from the one at $Z_{12} = 1$. In fact, a light ray passing through any point $X$ on the hyperboloid on fig. \ref{fig1} is a generatrix of this hyperboloid. Two generatrix lines crossing at $X_2$ never intersect those which are crossing at $\overline{X}_2$. This can be seen from the Penrose diagram.

Thus, for the other harmonics in the EPP and CPP the Wightman functions are also given by (\ref{alphaG}), where instead of the Euclidian harmonics in (\ref{bogrot}) one should use the BD ones.

\subsection{Digression on an alternative quantization}

There is a different way of field quantization in dS space\footnote{I would like to thank V.Losyakov and A.Morozov for the discussions on this issue.}. While in flat space this procedure leads to the same result, in dS space it provides an alternative quantization.

Consider for the illustrative reasons two-dimensional scalar field in global dS space:

\bqa
S = \frac12 \, \int dt \int_0^{2\pi} d\theta \, \cosh(t) \, \left[-\dot{\phi}^2 + m^2 \phi^2 + \frac{1}{\cosh^2(t)}\, \left(\pr_\theta \phi\right)^2\right].
\eqa
Let us Fourier expand the field in the spatial direction $\phi(t,\theta) = \sum_{p=-\infty}^{+\infty} g_p(t) \, e^{i\,p\,\theta}$, where $g_p$ obeys the condition $g_p = g_{-p}^*$ because $\phi$ is real. In terms of $g_p$ the Lagrangian is:

\bqa
L = \pi \, \sum_{p=0}^{+\infty} \cosh(t) \, \left[\dot{g}_p \, \dot{g}^*_p - \left(m^2 + \frac{p^2}{\cosh^2(t)}\right)\, g_p \, g_p^*\right].
\eqa
If one defines $g_p = \frac{1}{\sqrt{4\pi}}\, \left(q_p + i Q_p\right)$, $M(t) = \cosh(t)$ and $\omega_p^2(t) = m^2 + \frac{p^2}{\cosh^2(t)}$, then he can write the corresponding Hamiltonian as follows:

\bqa
H_0 = \sum_{p=0}^{+\infty} \left[\frac{1}{2\, M(t)} \left(p_p^2 + P_p^2\right) + \frac{M(t)\, \omega_p^2(t)}{2}\, \left(q_p^2 + Q_p^2\right)\right].
\eqa
Here $p_p$ and $P_p$ are momenta conjugate to $q_p$ and $Q_p$, correspondingly. Then, via the definition of the operators

$$a_p = \frac{1}{\sqrt{2}} \left(p_p - i q_p\right) \quad {\rm and} \quad \quad A_p = \frac{1}{\sqrt{2}} \left(P_p - i Q_p\right),$$
one can rewrite the Hamiltonian as:

\bqa
H_0 = \frac12 \, \sum_{p=0}^{+\infty} \left\{\left[\frac{1}{M(t)} + M(t) \, \omega_p^2(t)\right]\, \left[a_p^+ \, a_p + A_p^+ \, A_p\right] + \right. \nonumber \\ \left. \left[\frac{1}{M(t)} - M(t) \, \omega_p^2(t)\right]\, \left[a_p^2 + A_p^2\right] + h.c.\right\}.
\eqa
It is straightforward to show, however, that this way of quantization leads to propagators that are not dS-invariant because $a$'s and $A$'s here depend on time. This makes such a procedure inappropriate for our considerations. In fact, our goal is to understand if the dS isometry can be respected at all stages of quantization. Hence, we would not like to break it by the choice of a non-invariant vacuum state. However, otherwise this way of quantization is perfectly sensible and is also worth studying.

\section{Loops in de Sitter space QFT}

In this section we study one-loop contributions to propagators and vertices in the Poincar\'{e} patches and in global dS space.

\subsection{Brief introduction to the Schwinger--Keldysh diagrammatic technique}

Because free Hamiltonians in global dS and in the EPP (CPP) depend on time, the system under consideration is in a nonstationary state and one has to apply Schwinger--Keldysh (SK) (aka in-in, aka nonstationary) diagrammatic technique. The systematic introduction to it can be found in \cite{LL} or \cite{Kamenev}. To set the notations we will sketch here the general physical motivation for this technique.

Suppose one would like to calculate the expectation value of an operator ${\cal O}$ at some moment of time $t$:

\bqa\label{ave}
\left\langle {\cal O} \right\rangle(t) \equiv \left\langle \Psi\left| \overline{T} e^{i\,\int_{t_0}^t dt' H(t')}\,{\cal O} \, T e^{-i\,\int_{t_0}^t dt' H(t')}\right| \Psi\right\rangle,
\eqa
where $H(t) = H_0(t) + H^{int}(t)$ is the full Hamiltonian of a theory; while $T$ denotes the time-ordering, $\overline{T}$ is the reverse time-ordering; $t_0$ is an initial moment of time; $\left|\Psi\right\rangle$ is an initial state. The initial value of  $\left\langle {\cal O} \right\rangle(t_0)$ is supposed to be given in the setup of the problem. The expression (\ref{ave}) is valid both in the Heisenberg picture, when the evolution operators are attributed to ${\cal O}$, and in the Schrodinger one, when they are attributed to the bra and ket states. The generalization of our considerations to multiple operators under the average is straightforward.

After the transformation to the interaction picture, we get \cite{LL}:

\bqa\label{step}
\left\langle {\cal O} \right\rangle(t) = \left\langle \Psi\left| S^+(t, t_0)\, {\cal O}_0(t) \, S(t,t_0) \right| \Psi\right\rangle = \left\langle \Psi\left| S^+(t, t_0)\, T\left[ {\cal O}_0(t) \, S(t,t_0)\right]\right| \Psi\right\rangle = \nonumber \\ = \left\langle \Psi\left| S^+(t, t_0) S^+(+\infty, t)\, S(+\infty, t)\, T\left[ {\cal O}_0(t) \, S(t,t_0)\right]\right| \Psi\right\rangle = \nonumber \\ = \left\langle \Psi\left| S^+(+\infty, t_0)\, T\left[ {\cal O}_0(t) \, S(+\infty,t_0)\right]\right| \Psi\right\rangle,
\eqa
where $S(t,t_0) = T e^{-i\,\int_{t_0}^t dt' H^{int}_0(t')}$; ${\cal O}_0(t)$ and $H^{int}_0(t)$ are the same operators as were defined above, but written in the interaction picture.

To perform the first step in (\ref{step}) we have used the Baker-Hausdorff formula:

\bqa \label{exprrr} e^{A + B} = {\rm T}\, \exp\left\{\int^1_0 dt
\, e^{-t\, B} \, A e^{t\, B}\right\} \, e^B \eqa
which follows from the logarithmic $t$ derivative of the operator
$G(t)=e^{t \, (A + B)}e^{- t \, B}$:

\bqa G(t)^{-1}d_t G(t) =
e^{-t \, B} \, A \, e^{t \, B} \nonumber. \eqa
To perform the step on the second line of (\ref{step}) we had inserted the following resolution of the unit operator: $1 = S^+(+\infty, t)\, S(+\infty, t)$. That allows one to extend the original evolution (from $t_0$ to $t$ and back) to that which goes from $t_0$ to future infinity and back. We put the operator ${\cal O}_0(t)$ on the forward going part of the time contour.

To understand the meaning of the technique in question, let us slightly change the problem. We adiabatically turn on the interaction term, $H^{int}$, after $t_0$, i.e., $\left|\Psi \right\rangle$ does not evolve before $t_0$. Then, one can rewrite the expectation value (\ref{step}) as follows:

\bqa\label{calo}
\left\langle {\cal O} \right\rangle_{t_0}(t) = \left\langle \Psi\left| S^+(+\infty, -\infty)\, T\left[ {\cal O}_0(t) \, S(+\infty,-\infty)\right]\right| \Psi\right\rangle.
\eqa
A good question is if one can take $t_0$ to past infinity, $t_0 \to -\infty$, i.e. to get rid of the dependence of $\left\langle {\cal O} \right\rangle_{t_0}(t)$ on $t_0$. The seminal example when one can do so is as follows: The free Hamiltonian, $H_0$, does not depend on time and $\left|\Psi \right\rangle$ coincides with its ground state $\left|vac \right\rangle$, $H_0 \, \left|vac \right\rangle = 0$. One also assumes that the interaction term is adiabatically switched off at future infinity --- after the time $t$.

If $\left|vac \right\rangle$ is the true vacuum state of the free theory, then, by adiabatic turning on and then switching off the interactions, one cannot disturb such a state, i.e., $\left\langle vac \left| S^+(+\infty, -\infty)\right| excited \,\, state\right\rangle = 0$, while $\left|\left\langle vac \left| S^+(+\infty, -\infty)\right| vac \right\rangle\right| = 1$. Hence,

\bqa\label{43}
\left\langle {\cal O} \right\rangle(t) = \sum_{state} \left\langle vac \left| S^+(+\infty, -\infty)\right| state \right\rangle \, \left\langle state \left| T\left[ {\cal O}_0(t) \, S(+\infty,-\infty)\right]\right| vac \right\rangle = \nonumber \\ = \left\langle vac \left| S^+(+\infty, -\infty)\right| vac \right\rangle \, \left\langle vac \left| T\left[ {\cal O}_0(t) \, S(+\infty,-\infty)\right]\right| vac \right\rangle = \nonumber \\ =
\frac{\left\langle vac \left| T\left[ {\cal O}_0(t) \, S(+\infty,-\infty)\right]\right| vac \right\rangle}{\left\langle vac \left| S(+\infty, -\infty)\right| vac \right\rangle} .
\eqa
To perform the first step in (\ref{43}), we have inserted the resolution of unity $1 = \sum_{state} \left| state \right\rangle \, \left\langle state \right|$, where the sum is going over the complete basis of eigen-states of $H_0$. To perform the second step, we have used that $\left|vac \right\rangle$ is the only state from the sum which gives a non-zero contribution.

Thus, the dependence on $t_0$ disappears and we arrive at the expressions which contain only T-ordering (and no any $\overline{\rm T}$-orderings), i.e., we obtain the standard Feynman diagrammatic technique. Note that one can also apply the SK technique in the stationary situation because then the $\overline{\rm T}$-ordered expressions just cancel out vacuum diagrams.

However, if $\left|\Psi\right\rangle$ is not a ground state and/or $H_0$ depends on time, one cannot use the above machinery and has to deal directly with \eq{calo} or (\ref{step}). If one knows the matrix element

$${\cal A}_{12} = \left\langle \Psi_1\left| S^+(t_1, t_2) \right| \Psi_2\right\rangle $$
for arbitrary $t_{1,2}$ and generic states $\left|\Psi_{1,2}\right\rangle$ (which do not have to belong to the same Fock space), then one can calculate (\ref{step}) with the use of such a generalized Feynman technique.
Unfortunately, usually there are no algorithmic tools to calculate such matrix elements as ${\cal A}_{12}$ or even to deal with their unusual divergences. In this case the efficient method is the so-called SK technique, where one has to perturbatively expand both $S$ and $S^+$ under the quantum average. Many comparatively simple and interesting examples of the application of this technique are presented in \cite{Kamenev}.

We continue with the concrete example of real massive scalars with $\lambda \, \phi^3$ self-interaction. We have chosen the theory with such an unstable potential just to simplify the equations because effects, that we consider below, are not affected by such an instability. The situation in the stable $\lambda \, \phi^4$ theory is described in \cite{AkhmedovPopovSlepukhin} and is similar to the case under consideration (see the last section).

The functional integral for the theory in question can be derived in the standard manner. The functional integral form of (\ref{step}) is:

\bqa\label{propave}
\left\langle {\cal O} \right\rangle(t) = \int {\cal D} \varphi^+(x)\, {\cal D} \varphi^-(x) \, \left\langle \varphi^+ \left| \rho(t_0) \right| \varphi^-\right\rangle \, \int_{\varphi^+}^{\varphi^-} {\cal D} \phi^+(t,x)\, {\cal D} \phi^-(t,x) \, {\cal O}(t) \times \nonumber \\ \times  \exp \left\{\frac{i}{2}\, \int_{t_0}^{+\infty} dt \int d^{D-1}x \, \sqrt{|g|}\left[\left(\pr_\mu \phi_{-}\right)^2 + m^2\, \phi^2_{-} + \frac{\lambda}{3}\, \phi_{-}^3 - \left(\pr_\mu \phi_{+}\right)^2 - m^2 \, \phi^2_{+} - \frac{\lambda}{3}\, \phi_{+}^3\right]  \right\},
\eqa
where $\left\langle \varphi^+ \left| \rho(t_0) \right| \varphi^-\right\rangle$ is the matrix element of the initial density matrix, which in our case is $\rho(t_0) = \left| \Psi \right \rangle \, \left\langle \Psi \right|$. While $\phi^+$ is defined on the direct side of the time contour, $\phi^-$ belongs to its reverse side; $\varphi^\pm$ are initial/final values of $\phi^\pm$ at $t_0$, correspondingly. All these complications are due to the simultaneous presence of $S$ and $S^+$ in (\ref{step}).

Also it is worth stressing here that in the nonstationary situation one usually cannot take $t_0\to -\infty$, unless the system has a stationary state, towards which it has to evolve inevitably. We will encounter such a strongly nonstationary situation at one loop in the CPP and in global dS. At the same time, in the EPP we will encounter an unusual stationary situation --- there $t_0$ can be taken to past infinity.

From the functional integral (\ref{propave}) one can deduce that in the SK technique every vertex carries the ``$\pm$'' index, depending on whether it belongs to the ``$+$'' or ``$-$'' sides of the time contour.
As a result, if every particle is described by the propagator matrix,

\bqa
\hat{G}_0\left(X_1, X_2\right) = \left(\begin{array} {c c} G^0_{++}\left(X_1, X_2\right) &  G^0_{+-}\left(X_1, X_2\right)\\
G^0_{-+}\left(X_1,X_2\right) & G^0_{--}\left(X_1, X_2\right)
\end{array} \right),
\eqa
all loop expressions can be written in a matrix form (see, e.g., \cite{vanderMeulen:2007ah} and the next subsection for the details). The mixed ``$\pm$'' propagators appear because of the presence of the non-trivial initial density matrix \cite{Kamenev}.

With the use of the Wightman function all of the constituents of the propagator matrix can be written as follows

\bqa
G^0_{-+}\left(X_1,X_2\right) = i\,\langle \phi_-(X_1) \phi_+(X_2) \rangle = i\,\langle \hat{\phi}(X_1) \hat{\phi}(X_2) \rangle, \nonumber \\  G^0_{+-}\left(X_1,X_2\right) = i\,\langle \phi_+(X_2) \phi_-(X_1) \rangle = i\,\langle \hat{\phi}(X_2) \hat{\phi}(X_1) \rangle, \nonumber \\ G^0_{++}\left(X_1,X_2\right) = \langle T \,\hat{\phi}(X_1) \hat{\phi}(X_2) \rangle = \theta(t_1 - t_2) \, G^0_{-+}\left(X_1,X_2\right) + \theta(t_2 - t_1)\, G^0_{+-}\left(X_1,X_2\right), \nonumber \\
G^0_{--}\left(X_1,X_2\right) = \langle \overline{T} \,\hat{\phi}(X_1) \hat{\phi}(X_2) \rangle = \theta(t_1 - t_2) \, G^0_{+-}\left(X_1,X_2\right) + \theta(t_2 - t_1)\, G^0_{-+}\left(X_1,X_2\right).
\eqa
They obey one relation $G^0_{+-} + G^0_{-+} = G^0_{++} + G^0_{--}$.
To reduce the number of propagators it is convenient to perform the Keldysh rotation \cite{LL}:

\bqa\label{47}
\left(\begin{array} {c} \phi_{cl} \\
\phi_q \end{array} \right) = \left(\begin{array} {c} \frac{1}{2}\, \left[\phi_{+} + \phi_{-}\right] \\
\phi_{+} - \phi_{-} \end{array} \right) = \hat{R} \, \left(\begin{array} {c} \phi^+ \\
\phi^- \end{array} \right), \quad {\rm where} \quad \hat{R} = \left(\begin{array} {c c} \frac12 &  \frac12 \\
1 & -1
\end{array} \right).
\eqa
Then, the action becomes:

\bqa\label{Kelac}
S = \int_{t_0}^{+\infty} dt \int d^Dx \, \sqrt{|g|}\left[\pr_\mu \phi_{cl}\, \pr^\mu \phi_{q} + m^2 \phi_{cl} \, \phi_q + \frac{\lambda}{3!}\, \left(\phi_q\,\phi_{cl}^2 + \frac14 \, \phi_q^3 \right)\right].
\eqa
At the same time the propagator matrix gets converted into:

\bqa
\hat{D}_0 \left(X_1, X_2\right) \equiv \hat{R}\, \hat{G}_0\left(X_1, X_2\right)\, \hat{R}^{\rm T} = \left(\begin{array} {c c} i\, D_0^K\left(X_1, X_2\right) &  D_0^R\left(X_1, X_2\right)\\
D_0^A\left(X_1,X_2\right) & 0 \end{array} \right),
\eqa
where

\begin{figure}[t]
\begin{center} \includegraphics[scale=0.4]{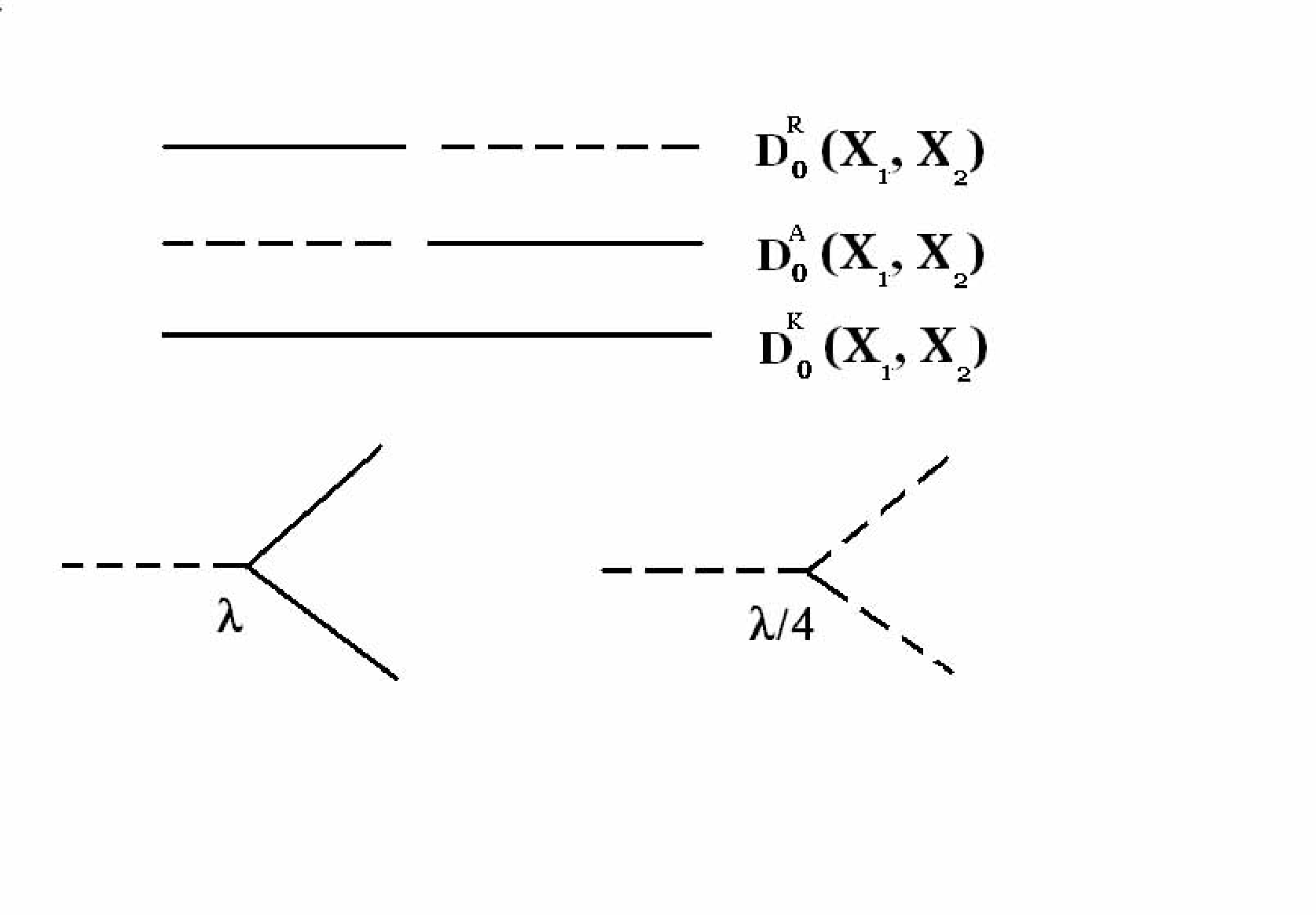}
\caption{While the solid line corresponds to $\phi_{cl}$, the dashed one --- to $\phi_q$.}\label{Diag}
\end{center}
\end{figure}

\bqa
D^{\frac{R}{A}}_0\left(X_1, X_2\right) = G^0_{++}\left(X_1, X_2\right) - G^0_{\pm \mp}\left(X_1, X_2\right) = \nonumber \\ = \theta\left(\pm \Delta t_{12}\right)\, \left[G_{\mp \pm}\left(X_1, X_2\right) - G_{\pm \mp}\left(X_1, X_2\right)\right] = \nonumber \\ = \pm \theta\left(\pm \Delta t_{12}\right) \, \left\langle \left[\phi(X_1), \phi(X_2)\right]\right\rangle, \quad \Delta t_{12} = t_1 - t_2
\eqa
are the retarded and advanced propagators. They define the spectrum of excitations in the theory under consideration. At tree level they do not depend on the state with respect to which the averaging is done because the commutator, $\left[\cdot, \cdot\right]$, of $\phi$'s is just a c-number. The Keldysh propagator is:

\bqa
D^K_0\left(X_1, X_2\right) = - \frac{i}{2} \left[G^0_{-+}\left(X_1, X_2\right) + G^0_{+-}\left(X_1, X_2\right)\right] = - \frac{i}{2} \, \left\langle \left\{\phi(X_1), \phi(X_2)\right\}\right\rangle,
\eqa
where $\{\cdot, \cdot\}$ is the anti-commutator. Note that while the retarded propagator is given by $D^R \sim \langle \phi_{cl} \, \phi_{q} \rangle$ in the rotated variables, the Keldysh one is as follows $D^K \sim \langle \phi_{cl} \, \phi_{cl} \rangle$. Although it is not obvious from the action (\ref{Kelac}), the Keldysh propagator is not trivial because of the initial density matrix in (\ref{propave}) (see, e.g., \cite{Kamenev} for the explanations).
The Feynman rules for (\ref{Kelac}) are depicted on fig. \ref{Diag}.

As we will see below, the Keldysh propagator is sensitive to the time evolution of the background state and essentially is a classical quantity. That in particular, explains the adapted in condensed matter theory notations that $\phi_{cl}$ is the ``classical'' field, while $\phi_q$ is the ``quantum'' one \cite{Kamenev}.

One last point which should be stressed here is that the SK technique, unlike the Feynman one, is strictly causal. In terms of, e.g., Eq. (\ref{propave}) that means that all contribution to the expectation value in question comes from the causal past of time $t$. Rephrasing this, the result of the calculation of a quantity with the use of the SK technique is a solution of a Cauchy problem whose initial data are set up by the tree level value of the quantity under study.

\subsection{On de Sitter isometry invariance at loop level}

To start with, we show that loop corrections to the BD state in the EPP are dS isometry invariant\footnote{I would like to thank A.Polyakov for explaining this point to me.}.
For this problem we prefer to use the SK technique {\it before} the Keldysh rotation (\ref{47}).
Then, e.g., the one-loop correction to the propagator matrix can be written as:

\bqa\label{oneloop}
\hat{G}^1(Z_{XY}) = \lambda^2 \, \int [dW] \int [dU]\, \hat{G}^0(Z_{XW}) \, \hat{\Sigma}^0(Z_{WU}) \, \hat{G}^0(Z_{UY}) \eqa
where

\bqa
\hat{G}^{0,1}(Z) = \left(
  \begin{array}{cc}
    G^{0,1}_{++}(Z) & G^{0,1}_{+-}(Z) \\
    G^{0,1}_{-+}(Z) & G^{0,1}_{--}(Z) \\
  \end{array}
\right) \quad {\rm and} \quad \hat{\Sigma}^0(Z) = \left(
  \begin{array}{cc}
    \left[G^0_{++}(Z)\right]^2 & \left[G^0_{+-}(Z)\right]^2 \\
    \left[G^0_{-+}(Z)\right]^2 & \left[G^0_{--}(Z)\right]^2 \\
  \end{array}
\right),
\eqa
and the measure is written in terms of the embedding coordinates of the ambient Minkowski space

$$[dW] = d^{(D+1)} W \, \delta\left(W^A\,W_A - 1\right)\,\theta \left(W^0 - W^D\right).$$
It is equivalent to $\frac{d\eta_+}{\eta^D_+}\, d^{D-1}\vec{x}_+$ after the substitution of the EPP coordinates of $W^A$. This formula for $\hat{G}^1$ is valid for any dS-invariant $\alpha$-vacuum. Note that in (\ref{oneloop}) the dS isometry is naively broken by the presence of the Heavyside $\theta$-function, which restricts to the EPP.

As follows from the discussion in the previous section, for the BD state we have that \cite{Polyakov:2007mm}:

\bqa\label{plmi}
G^0_{++}\left[Z_{12}\right] = G\left[Z_{12} + i\, \epsilon\right], \quad G^0_{+-}\left[Z_{12}\right] = G\left[Z_{12} - i \, \epsilon \, sign(\eta_2 - \eta_1)\right], \nonumber \\ G^0_{--}\left[Z_{12}\right] = G\left[Z_{12} - i\, \epsilon\right], \quad G^0_{-+}\left[Z_{12}\right] = G\left[Z_{12} + i \epsilon \, sign(\eta_2 - \eta_1)\right].
\eqa
Here $G(Z)$ is defined in (\ref{WBD}). (Note the reverse time flow in the EPP.)

Several comments are in order at this point. First, it is not hard to check that for the BD state the UV divergence in (\ref{oneloop}) is the same as in flat space. For the other vacua this is not the case. Second, all the arguments of the present subsection are valid only if there are no IR divergences in the loop integrals. We will see in the next subsection that in the EPP there are no IR divergences in the field theory under consideration. There are only large IR contributions.

Let us examine now a variation of $\hat{G}^{1}$ under a transformation of $SO(D,1)$ isometry group which changes arguments of $\theta$-functions in $[dW]$ and $[dU]$. (Here we reproduce the arguments of \cite{Polyakov:2012uc}.) Let us perform an infinitesimal rotation around $X^0$ towards, e.g., $X^1$: $X^D \to X^D - \psi \, X^1$. Taylor expanding the integration measure up to the first order in $\psi$, we get:

\bqa
\delta_{\psi}\, \int[dW]\dots = \int d^{(D+1)} W \, \delta\left(W^A\, W_A - 1\right)\,\delta\left(W^0 - W^D\right)\, \psi\, W^1 \dots = \nonumber \\ = \int d\left(W^0 + W^D\right) \, d^{(D-1)}W \, \delta\left(W^A \, W_A - 1\right)\,\psi\, W^1 \dots \nonumber
\eqa
and similarly for $[dU]$ integration.

Consider, e.g., the situation with $d\left(W^0 + W^D\right)$ integral. Its integrand is a function of

$$Z_{XW} = -\frac12\, (X^0-X^D)\, (W^0+W^D) - \frac12\, (X^0+X^D)\, (W^0-W^D) + X^a\, W^a$$
and of $Z_{UW}$. Here $W^0 - W^D = 0$ because of the presence of $\delta\left(W^0 - W^D\right)$ in the integration measure for the variation $\delta_\psi \hat{G}^1$. Also $\left(X^0 - X^D\right) \ge 0$ because we are in the EPP. This means that $G\left[W^0 + W^D\right] \equiv G\left[Z\left(W^0 + W^D\right)\right]$, when considered as the function of $\left(W^0 + W^D\right)$, has the same analytical properties as those in the complex $Z_{XW}$-plain. Furthermore, because of $\delta\left(W^0 - W^D\right)$, $\eta_w$ in $sign(\eta_x - \eta_w)$ goes to past infinity. Hence, we have a definite sign of the $\epsilon$-prescription inside all propagators. The same is true for the functions of $Z_{UW}$.

As a result, due to the $\epsilon$-prescription for the BD state the integrand of $d\left(W^0 + W^D\right)$ is an analytical function on the complex $\left(W^0 + W^D\right)$-plain with the cut going from $1$ to infinity and slightly shifted to either the upper or lower plane (depending on whether it is $+$ or $-$ vertex in the contribution under consideration). Then, because propagators have a powerlike decay, as $\left(W^0 + W^D\right)\to \infty$, one can close the integration contour by an infinite semicircle in either lower or upper half of the complex $\left(W^0 + W^D\right)$-plain, correspondingly. As we just explained, the integrand is analytical function inside the contour, hence, the integral is zero.

The same arguments work for the $d(U_0 + U_D)$ integral and also for the infinitesimal rotations in the other directions around $X^0$. Furthermore, it is straightforward to extend these arguments to higher loops and higher-point correlation functions. Hence, in the case of the BD state the exact matrix propagator $\hat{G}(X_1,X_2)$ is a function of $Z_{12}$ only.

The arguments of this subsection do not work for the other $\alpha$-vacua because then tree level propagators have another cut going from $Z=-1$ to infinity and another $\epsilon$-prescription. That spoils the analytical properties of the corresponding Wightman function in the complex $Z$-plain. Furthermore, because of the IR divergences, which we will discuss below, these arguments also do not work in the CPP for any dS-invariant vacuum.

It is tempting to propose, however, that the dS isometry will be also respected in loops, if one would consider that half of the entire dS space, which corresponds to $t\geq 0$ in global coordinates, and the Euclidian vacuum as the initial state. In fact, then the propagators are also given by (\ref{plmi}) and the expression for the one-loop correction is the same as (\ref{oneloop}), with a bit different measure

$$[dW] = d^{D+1} W \, \delta\left(W^A\, W_A - 1\right) \, \theta\left(W^0\right).$$
However, one of the possible transformation of $SO(D,1)$ group, which moves the boundary of this subspace, is the infinitesimal Lorentz boost in the $W^D$ direction: $W^0 \to W^0 + \psi \, W^D$. Under such a boost
the measure is changed by

\bqa
\delta_{\psi}\, \int[dW]\dots = \int d^{(D+1)} W \, \delta\left(W^A\, W_A - 1\right)\,\delta\left(W^0\right)\, \psi\, W^D \dots = \nonumber \\ = \int d^{D}W \, \delta\left(\sum_{i=1}^D W_i \, W_i - 1\right)\,\psi\, W^D \dots \nonumber
\eqa
But the integrand here, considered as the function of any one among $W_i, i = 1, \dots, D$, ($G[W_i] = G[Z(W_i)]$) does not have the same analytical properties as those in the complex $Z_{WX}$-plain: The cut in the complex $W_i$-plain ($i$ is fixed) does not coincide with the one in the $Z_{WX}$-plain because, unlike $(X^0-X^D)$ in the EPP, $W_i$ does not have a definite sign. Hence, by cutting global dS space at its neck, one breaks the dS isometry with any initial state.

\subsection{One--loop correction in the expanding Poincar\'{e} patch}

In this subsection we calculate leading IR one-loop contributions to propagators $D^{R,A,K}\left(X_1, X_2\right)$ and vertices. Due to spatial homogeneity of the EPP itself and due to spatial homogeneity of background states that we are going to consider, we find it convenient to perform the Fourier transformation of all quantities along the spatial directions: $D^{K,R,A}_{0}\left(p\,|\eta_1, \eta_2\right) \equiv \int d^{D-1}x \, e^{i\, \vec{p}\, \vec{x}} D^{K,R,A}_0(\eta_1, \vec{x}; \eta_2, 0)$. To simplify the notations below, we drop the ``$\pm$'' indexes that distinguish coordinates of the EPP from the CPP.

Below we do not care about UV divergences, i.e., we assume some kind of UV renormalization and also assume that masses of the fields and coupling constants possess their physical renormalized values.
But it is probably worth stressing here that mixed expressions, with the partial Fourier transformation along only the spatial sections, are not sensitive to the UV divergences. In fact, to reveal the latter even in the flat space Feynman diagrammatic technique one has to either transform back to the configuration space or to make the remaining Fourier transformation over the time direction. Also to observe the UV divergences in the Keldysh technique in dS space one has to go back to the configuration, $(\eta, \vec{x})$, space.

We consider such an IR limit in which

$$p\eta \to 0, \quad {\rm where} \quad \eta = \sqrt{\eta_1 \eta_2} = e^{-(\tau_1 + \tau_2)/2}, \quad {\rm and} \quad \eta_1/\eta_2 = e^{- (\tau_1 - \tau_2)} = const.$$
This is the limit when both arguments of the propagators are taken to the future infinity, while the time distance between them is kept fixed. The presence of large IR corrections in such a limit means that there are growing with time contributions, as the system progresses towards future infinity.

\begin{figure}[t]
\begin{center} \includegraphics[scale=0.4]{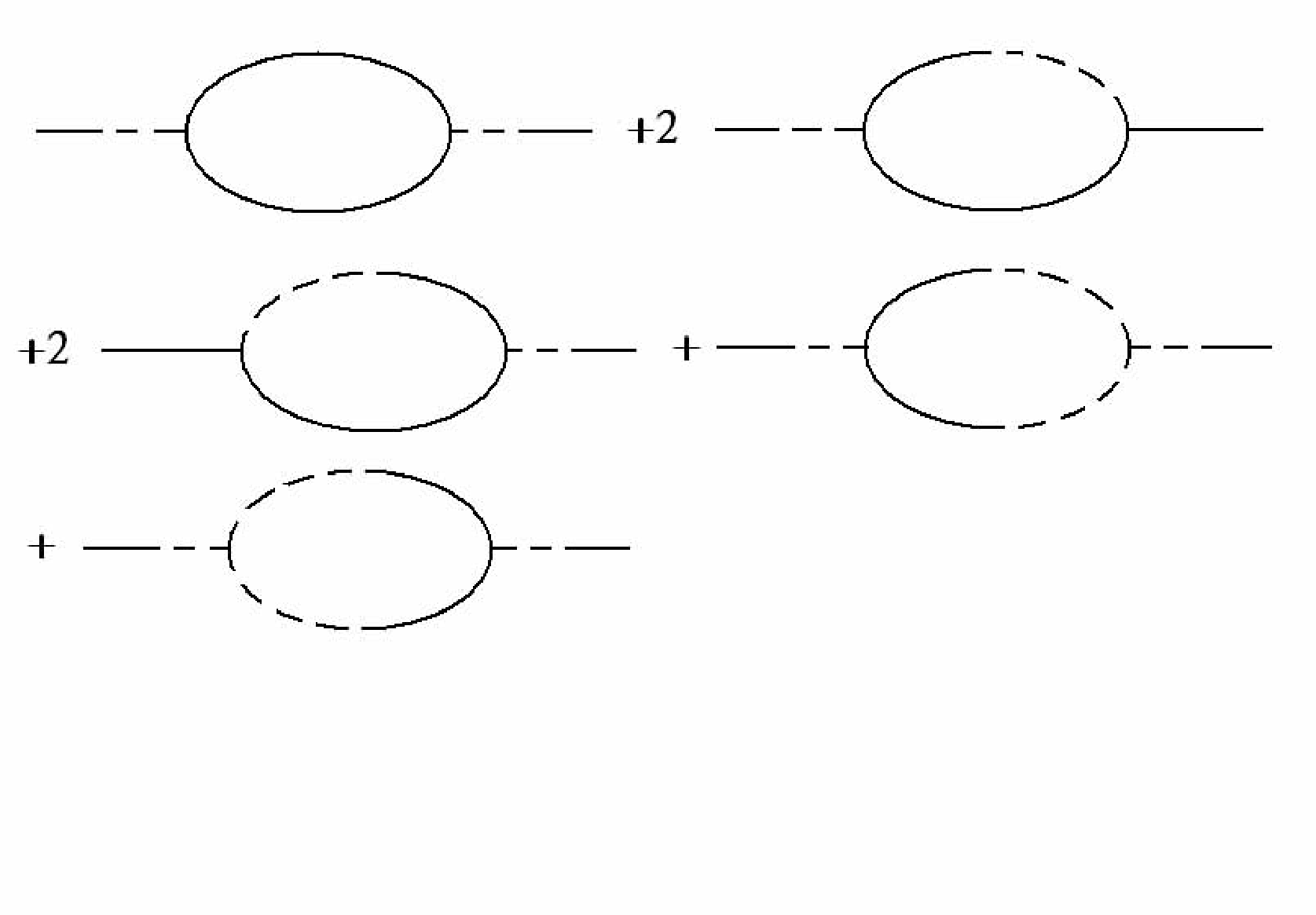}
\caption{One loop correction to the Keldysh propagator $D^K$.}\label{LoopK}
\end{center}
\end{figure}

Let us start with the Keldysh propagator. The retarded (advanced) propagator and the vertices will be described below. One-loop diagrams contributing to the Keldysh propagator are depicted on fig. \ref{LoopK}. (It is straightforward to see that tadpole diagrams do not contain large IR corrections.) The corresponding expression is:

\bqa\label{DSDK}
D_1^{K}\left(p\, |\eta_1,\eta_2\right) =  \lambda^2 \int \frac{d^{D-1}\vec{q}}{(2\pi)^{D-1}} \iint_{\eta_0}^0 \frac{d\eta_3 d\eta_4}{(\eta_3\eta_4)^D} \times \nonumber \\ \times \Biggl[ \Biggr. D_{0}^R\left(p\,|\eta_1,\eta_3\right) \, D_0^K\left(q|\eta_3,\eta_4\right) \, D_0^K\left(\left|\vec{p} - \vec{q}\right|\, |\eta_3,\eta_4\right) \, D_0^A\left(p\, |\eta_4,\eta_2\right) + \nonumber \\ + 2 \, D_{0}^R\left(p\,|\eta_1,\eta_3\right) \, D_0^R\left(q\,|\eta_3,\eta_4\right) \, D_0^K\left(\left|\vec{p} - \vec{q}\right|\, |\eta_3,\eta_4\right) \, D_0^K\left(p\,|\eta_4,\eta_2\right) + \nonumber \\ + 2 \, D_{0}^K\left(p \, |\eta_1,\eta_3\right) \, D_0^K\left(q\, |\eta_3,\eta_4\right) \, D_0^A\left(\left|\vec{p} - \vec{q}\right|\, |\eta_3,\eta_4\right) \, D_0^A\left(p\, |\eta_4,\eta_2\right) - \nonumber \\ - \frac{1}{4} \, D_{0}^R\left(p\, |\eta_1,\eta_3\right) \, D_0^R\left(q\, |\eta_3,\eta_4\right) \, D_0^R\left(\left|\vec{p} - \vec{q}\right|\, |\eta_3,\eta_4\right) \, D_0^A\left(p\, |\eta_4,\eta_2\right) - \nonumber \\ -\frac{1}{4} \, D_{0}^R\left(p\, |\eta_1,\eta_3\right) \, D_0^A\left(q\, |\eta_3,\eta_4\right) \, D_0^A\left(\left|\vec{p} - \vec{q}\right|\, |\eta_3,\eta_4\right) \, D_0^A\left(p\, |\eta_4,\eta_2\right) \Biggl. \Biggr].
\eqa
Here $\eta_0>\eta_{1,2}$ is an early, $\eta_0\to +\infty$, moment after which the self-interactions are adiabatically turned on; the Keldysh propagator is

$$D_{0}^{K}\left(p\,|\eta_1,\eta_2\right) = (\eta_1\eta_2)^{\frac{D-1}{2}}\,{\rm Re}\left[h(p\eta_1)h^*(p\eta_2)\right],$$
the retarded and advanced propagators are

$$D_{0}^{R}\left(p\,|\eta_1,\eta_2\right) = \theta\left(\eta_2-\eta_1\right) \,2\,(\eta_1\eta_2)^{\frac{D-1}{2}}\,{\rm Im}\left[h(p\eta_1)h^*(p\eta_2)\right],$$ and

$$D_{0}^{A}\left(p\,|\eta_1,\eta_2\right) = - \theta\left(\eta_1-\eta_2\right) \, 2\, (\eta_1\eta_2)^{\frac{D-1}{2}}\,{\rm Im}\left[h(p\eta_1)h^*(p\eta_2)\right],$$ correspondingly.
Note that in these lectures we define ${\rm Re} \left[h_1 \, h_2^*\right] = h_1 \, h_2^* + h_1^*\, h_2$ and ${\rm Im}\left[h_1\, h_2^*\right] = \frac{1}{i}\left[h_1\, h_2^* - h_1^*\, h_2\right]$.

We will show below, that the leading IR contribution to $D_1^K\left(p\, |\eta_1,\eta_2\right)$ is hidden within the following expression

\bqa\label{EPP}
D_{1}^{K}\left(p\, |\eta_1,\eta_2\right) \approx (\eta_1\,\eta_2)^{\frac{D-1}{2}}\, \left[h(p\eta_1)\, h^*(p\eta_2) \, n_p(\eta) + h(p\eta_1)\, h(p\eta_2) \,\kappa_p(\eta)^{\phantom{\frac12}} + c.c. \right], \nonumber \\ {\rm where} \quad
n_p(\eta) \approx - \frac{\lambda^2}{(2\pi)^{2(D-1)}} \, \int d^{D-1}q_1 \int d^{D-1}q_2 \iint_{\eta_0}^\eta d\eta_3 \, d\eta_4 \, (\eta_3\, \eta_4)^{\frac{D-3}{2}}\, \delta\left(\vec{p} + \vec{q}_1 + \vec{q}_2\right) \times \nonumber \\ \times h^*\left(p\eta_3\right) \, h\left(p\eta_4\right) \, h^*(q_1\,\eta_3)\,  h(q_1\eta_4) \, h^*(q_2\eta_3)\, h(q_2\eta_4), \nonumber \\ {\rm and} \quad
\kappa_p(\eta) \approx \frac{2\,\lambda^2}{(2\pi)^{2(D-1)}} \, \int d^{D-1}q_1 \, \int d^{D-1}q_2 \int_{\eta_0}^\eta d\eta_3 \int_{\eta_0}^{\eta_3} d\eta_4 \, (\eta_3\, \eta_4)^{\frac{D-3}{2}}\, \delta\left(\vec{p} + \vec{q}_1 + \vec{q}_2\right) \times \nonumber \\ \times h^*\left(p\eta_3\right) \, h^*\left(p\eta_4\right) \, h^*(q_1\eta_3)\,  h(q_1\eta_4) \, h^*(q_2\eta_3)\, h(q_2\eta_4).
\eqa
In deriving this expression from (\ref{DSDK}) we have used that in the IR limit in question one can neglect the difference between $\eta_1$ and $\eta_2$ and substitute the average conformal time, $\eta = \sqrt{\eta_1\eta_2}$, instead of both $\eta_1$ and $\eta_2$, in the upper limits of the integrations over $\eta_3$ and $\eta_4$. I.e., in (\ref{EPP}) we dropped off subleading terms of the order of $\lambda^2 \, \log \left(\eta_1/\eta_2\right)$ or smaller, which may arise from integrals between $\eta_{1,2}$ and $\sqrt{\eta_1\eta_2}$.

Before performing the explicit calculation of (\ref{EPP}) for the concrete choice of $h(x)$, it is worth checking what is happening in the flat space limit. To take this limit in (\ref{EPP}) we just change $\eta \to t$, substitute $\eta^{(D-1)/2}\, h(p\eta)$ by the plane wave, $e^{-i \, \epsilon(p)\, t}/\sqrt{\epsilon(p)}$, and exchange $\sqrt{g} = 1/\eta^D$ for one. Then, as the upper limits of the time integrals in (\ref{EPP}) go to future infinity, while the lower ones are taken to past infinity, the time integrals are converted into the $\delta$-functions establishing the energy conservation: The integrands of $\iint dq_1 \, dq_2$ in the expressions for $n_p$ and $\kappa_p$ will contain the time-line integrals of $\delta\left[\epsilon(p) + \epsilon(q_1) + \epsilon(q_2)\right]$. The argument of this $\delta$-function is never zero. Hence, the whole expressions for $n_p$ and $\kappa_p$ are vanishing.

Thus, we see that in the flat space limit the expressions for $n_p$ and $\kappa_p$ do vanish, which is the consequence of the energy conservation. This gives us a hint for the physical interpretation of (\ref{EPP}). In fact, these expressions can arise in the presence of the particle creation process, when there is no energy conservation due to the background gravitational field. Also this observation shows that for the case of the EPP the contributions to $n_p$ and $\kappa_p$ from the past infinity, i.e., from the region where $p\eta\gg \mu$ and $q_{1,2}\eta \gg \mu$, are negligible: When $q\eta \gg \mu$ the function $g_p(\eta)$ can be well approximated by the plane wave with the minor pre-exponential factor. Furthermore, this allows one to safely take $\eta_0$ to past infinity, $\eta_0\to +\infty$. In fact, $d\eta_{3,4}$ integrals in (\ref{EPP}) are converging in the generalized sense on their lower limits of integration. The origin of these observations can be traced back to the presence of the infinite blue shift towards the boundary between the EPP and CPP.

\subsubsection{Bunch--Davies vacuum}

From now on we talk about the scalar fields from the principal series, $m > (D-1)/2$, unless otherwise stated.
We comment on the complementary series in the last section.

So, we would like to estimate leading contributions to $n_p$ and $\kappa_p$ in (\ref{EPP}) for the BD modes.
First, we take the integrals over $\vec{q}_2$ with the use of the $\delta$-functions and rename $\vec{q}_1 \to \vec{q}$. Second,
it is straightforward to check that the largest IR contribution to (\ref{EPP}) comes from the region where $q\gg p$ inside the integral over $dq$. Hence, we neglect $p$ in comparison with $q$ under the integrals in (\ref{EPP}). That changes, however, the lower limits of integration over $\eta_{3,4}$ from $\eta_0 \to +\infty$ to $\mu/p$. Note that by doing this we just neglect the contributions to $n_p$ and $\kappa_p$ from the high energy region $p\eta_{3,4} \gg \mu$.

Third, we make the following change of the integration variables in (\ref{EPP}): $q$ to $x_3 = q\eta_3$, $\eta_4$ to $x_4 = q\eta_4$ and leave $\eta_3$ unchanged. As a result, the integration measure is converted into

$$\iiint dq \, q^{D-2} \, d\eta_3 \, d\eta_4 \, \left(\eta_3 \, \eta_4\right)^{\frac{D-3}{2}} \dots = \iiint \frac{d\eta_3}{\eta_3} \, dx_3 \, dx_4 \, \left(x_3\, x_4\right)^{\frac{D-3}{2}} \dots.$$
Note that once the integral over $q_2$ was taken and $\vec{p}$ was neglected in comparison with $\vec{q}\equiv \vec{q}_1$, the integrand of $d^{D-1}q$ depends only on $\left|\vec{q}\right|$.

Now we can harmlessly extend the lower limits of integration over $x_{3,4}$ to infinity both in $n_p$ and $\kappa_p$. Furthermore, in the expression for $n_p$ we can extend the upper limits of integration over $x_{3,4}$ to zero. In the expression for $\kappa_p$ the latter change is done only for $x_3$. These changes are harmless because after them the integrals remain finite and pre-factors of the expressions, that we will find below, are just slightly changed by the contributions from the high energy region.

Fourth, the integrals for $n_p$ and $\kappa_p$ are saturated around $q\eta_{3,4} \sim \mu$, while $p\eta_{3,4} \ll q\eta_{3,4}$. Hence, we can expand

$$h(p\eta_{3,4}) \approx A_{+} \, \left(p\eta_{3,4}\right)^{i\mu} + A_{-} \, \left(p\eta_{3,4}\right)^{-i\mu}, \quad
{\rm where} \quad A_\pm = \frac{\sqrt{\pi} \, e^{\pm \frac{\pi\mu}{2}}}{2^{\pm i \mu + \frac12} \, \Gamma\left(1 \pm i \mu\right) \, \sinh\left(\pm \pi \mu\right)}.$$
And finally, we perform the integration over $\eta_3$ from $\mu/p$ to $\eta$ and keep in the expressions for $n_p$ and $\kappa_p$ only divergent, as $p\eta \to 0$, terms. The result is

\bqa\label{nkappa}
n_p(\eta) \approx \frac{\lambda^2 S_{D-2}}{(2\pi)^{D-1}}\, \log\left(\frac{\mu}{p \, \eta}\right)\, \iint^0_\infty dx_3 \, dx_4 \, \left(x_3 \, x_4\right)^{\frac{D-3}{2}}\times \nn \\ \times \left[\left|A_+\right|^2 \, \left(\frac{x_4}{x_3}\right)^{i\mu} + \left|A_-\right|^2 \, \left(\frac{x_3}{x_4}\right)^{i\mu}\right]\, h^2(x_3)\, \left[h^*(x_4)\right]^2, \nn \\
\kappa_p(\eta) \approx - \frac{2\,\lambda^2 S_{D-2}}{(2\pi)^{D-1}}\, \log\left(\frac{\mu}{p \, \eta}\right)\, A_+ \, A_- \, \int^0_\infty dx_3 \,\int_\infty^{x_3} dx_4 \, \left(x_3 \, x_4\right)^{\frac{D-3}{2}}\times \nn \\ \times \left[\left(\frac{x_4}{x_3}\right)^{i\mu} + \left(\frac{x_3}{x_4}\right)^{i\mu}\right]\, h^2(x_4)\, \left[h^*(x_3)\right]^2,
\eqa
where $S_{D-2}$ is the volume of the $(D-2)$-dimensional sphere.

Once the initial state is the BD vacuum, the correlation function (\ref{EPP}), (\ref{nkappa}) respects the dS isometry. Hence, we are in a dS-homogenous stationary state and particle interpretation of our formulas is not quite appropriate because $\langle T_{\alpha\beta}\rangle$ does not contain any fluxes\footnote{It is interesting to note, however, that a free floating detector will click in such a stationary state. This is just a consequence of the fact that in the stationary state the Wightman function depends only on the hyperbolic distance $Z$, which is equal to $Z = \cosh(\tau_1 -\tau_2)$ if the spatial position of the detector is not changing. The detector will click because according to (\ref{pole}) the Wightman function has poles on the imaginary axis in the complex proper time-plain \cite{AkhmedovUnruh}, \cite{Akhmedov:2007xu}.}. In fact, the inverse Fourier transformation of (\ref{EPP}), with $\eta$-dependent $n_p$ and $\kappa_p$ from (\ref{nkappa}), will be a function of the invariant geodesic distance between $(\eta_1, \vec{x}_1)$ and $(\eta_2,\vec{x}_2)$. One just has to bear in mind that in (\ref{nkappa}) we have kept inside $n_p$ and $\kappa_p$ the leading contributions in the limit under consideration. I.e., we have forgot that in principle $n_p$ and $\kappa_p$ depend on $\eta_1$ and $\eta_2$ separately rather than on their combination $\eta = \sqrt{\eta_1 \eta_2}$.

However, to understand the physical meaning of the obtained expressions, let us for a moment forget about the dS-invariance of (\ref{EPP}), (\ref{nkappa}). That makes sense for the following reasons: At the past infinity of the EPP $n_p$ and $\kappa_p$ have the clear interpretation as the particle density per comoving volume, $\langle a^+_p \, a_p \rangle$, and as the anomalous quantum average, $\langle a_p \, a_{-p}\rangle$, correspondingly. Then, if we take some non-zero initial value of $n_p$, the dS isometry is broken even at tree level (by the initial state). In fact, for non-zero, $\eta$-independent value of $n_p$ (even when $\kappa_p = 0$) the inverse Fourier transform of $D^K(p|\eta_1, \eta_2)$ depends on each its time argument separately rather than on the invariant geodesic distance. Moreover, below we will repeat the calculation in the situation when the dS isometry is broken by an initial perturbation. The result will be the same as (\ref{EPP}), but with a bit different expressions for $n_p$ and $\kappa_p$.

Thus, having in mind what we have just said, the physical interpretation of (\ref{EPP}), (\ref{nkappa}) is as follows.
If we start from the vacuum, then both $n_p$ and $\kappa_p$ are zero at past infinity. As we have seen above, in the absence of the gravitational field they also remain zero in future infinity. However, once the gravitational field is turned on, they are generated in loops and are comparable to one, even though they have been zero initially and $\lambda^2$ is much less than one. So the presence of these quantities signals the particle creation in the EPP. Moreover, below we give arguments favoring that $\kappa_p$ is the measure of the strength of the backreaction of quantum effects on the background state. Hence, its significant presence signals that the backreaction on the initial BD state is strong, i.e., in future infinity the system should relax to a different state from the BD one.

\begin{figure}[t]
\begin{center} \includegraphics[scale=0.3]{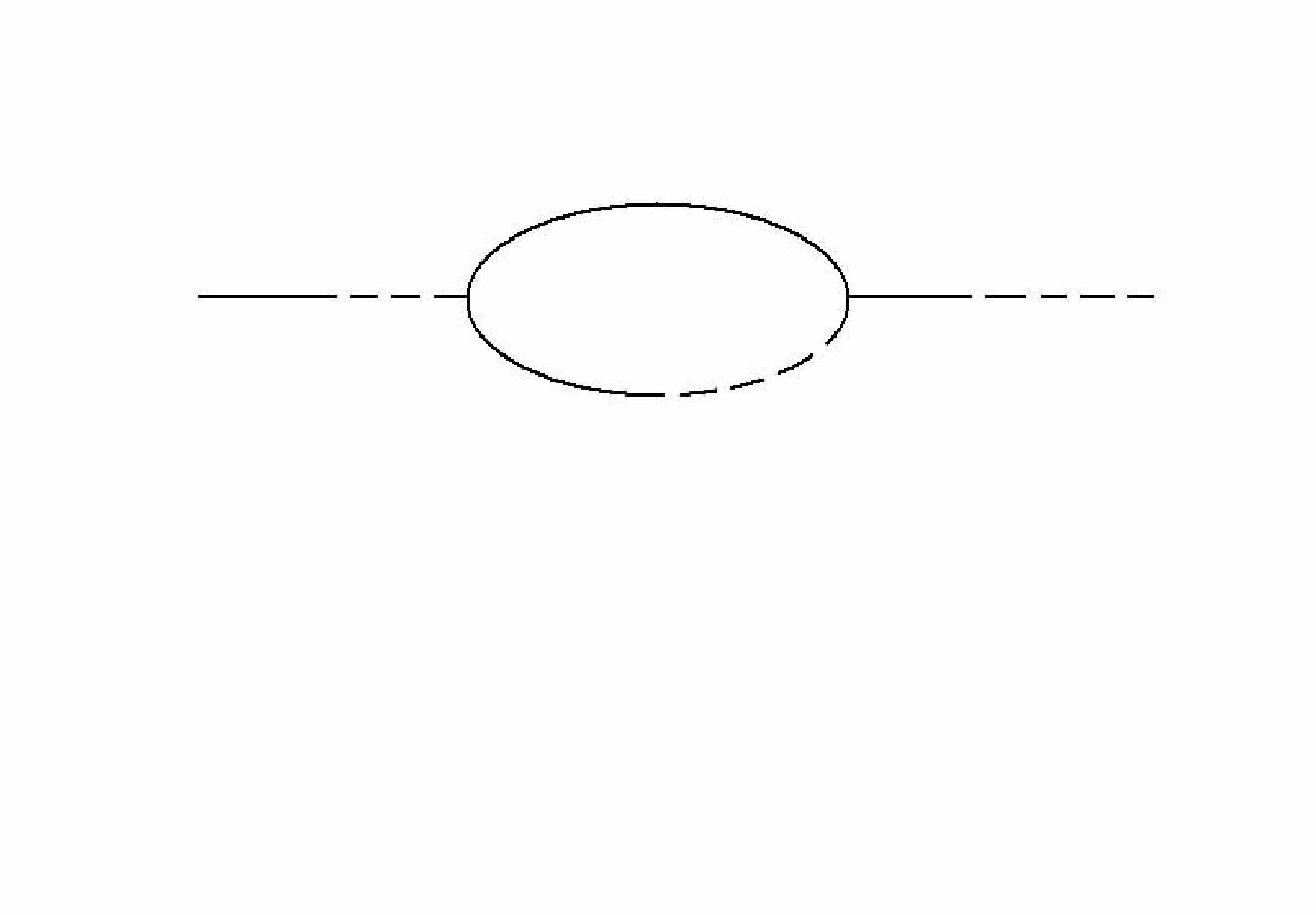}
\caption{One-loop correction to $D^R$. For $D^A$ the diagram is mirror symmetric to this one.}\label{loopR}
\end{center}
\end{figure}

\begin{figure}[t]
\begin{center} \includegraphics[scale=0.4]{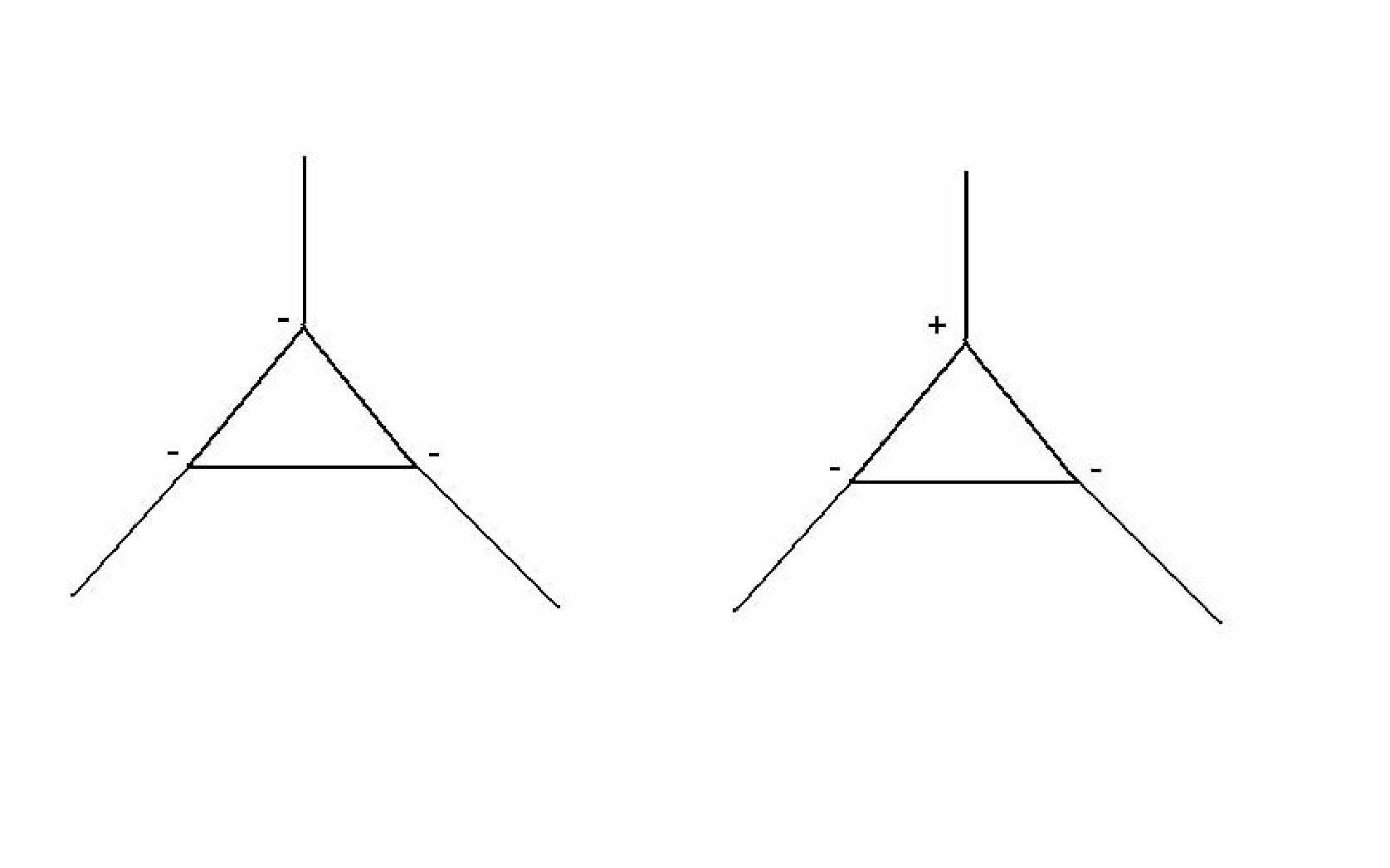}
\caption{Here we show some of the diagrams. The other diagrams are just complex conjugate of those that are presented here.}\label{vertex}
\end{center}
\end{figure}

\subsubsection{Contributions to the retarded and advanced propagators and to the vertices}

Once we have understood the origin of the large IR corrections to the Keldysh propagator, let us consider the situation in the case of the retarded (advanced) propagator and vertices. This will be important for the summation of the leading IR corrections from all loops.

The one-loop diagram, that contributes to $D^R$, is depicted on fig. \ref{loopR}. The corresponding expression is:

\bqa\label{oneloopR}
D^R_{1}\left(p\,|\eta_1, \eta_2\right) = \lambda^2 \, \int \frac{d^{D-1}\vec{q}}{\left(2\, \pi\right)^{D-1}} \, \iint_{+\infty}^0 \frac{d\eta_3\, d\eta_4}{\left(\eta_3\, \eta_4\right)^{D}} \times \nonumber \\ \times D^R_{0}\left(p\,|\eta_1, \eta_3\right)\, D^R_{0}\left(q\,|\eta_3, \eta_4\right)\, D^K_{0}\left(\left|\vec{q} - \vec{p}\right|\,|\eta_3, \eta_4\right)\, D^R_0\left(p|\eta_4, \eta_2\right).
\eqa
Here one can immediately see that due to the presence of $D^R$ inside the loop and in the external legs the limits of integration over $\eta_{3,4}$ are such that $\eta_1 > \eta_3 > \eta_4 > \eta_2$. As a result, the loop integral (\ref{oneloopR}) does not receive large corrections in such a limit when $\eta_1/\eta_2$ is held fixed. In the different limit, when say $\eta_1 \to +\infty$ and $\eta_2 \to 0$, this integral can receive at most a correction proportional to $\lambda^2 \, \log\left(\eta_1/\eta_2\right)$ \cite{Leblond}, but it is not of interest for us here. The situation with the advanced propagator is the same.

The diagrams that contribute to the vertex renormalization are depicted on fig. \ref{vertex}. We find it more convenient to consider loop contributions to vertices in the SK technique before the Keldysh rotation (\ref{47}).
Then the expressions for these diagrams are as follows\footnote{I would like to thank F.Popov and V.Slepukhin for the discussions on this point.}:

\bqa
\lambda^{---}(\eta_{1,2,3}; p_{1,2,3}) = \left(-i\,\lambda\right)^3 \, \left(\eta_1\, \eta_2\, \eta_3\right)^{D-1} \, \int \frac{d^{D-1}p}{\left(2\pi\right)^{D-1}} \times \nonumber \\ \times \left[\theta(\eta_1 - \eta_2)^{\phantom{\frac12}} h^*(p\eta_1)\, h(p\eta_2) + \theta(\eta_2 - \eta_1)^{\phantom{\frac12}} h^*(p\eta_2)\, h(p\eta_1)\right]\nonumber \\
\left[\theta(\eta_2 - \eta_3)^{\phantom{\frac12}} h^*\left(\left|\vec{p}_1 + \vec{p}\right|\eta_2\right)\, h\left(\left|\vec{p}_1 + \vec{p}\right|\eta_3\right) + \theta(\eta_3 - \eta_2)^{\phantom{\frac12}} h^*\left(\left|\vec{p}_1 + \vec{p}\right|\eta_3\right)\, h\left(\left|\vec{p}_1 + \vec{p}\right|\eta_2\right)\right] \nonumber \\
\left[\theta(\eta_3 - \eta_1)^{\phantom{\frac12}} h^*\left(\left|\vec{p}_2 - \vec{p}\right|\eta_3\right)\, h\left(\left|\vec{p}_2 - \vec{p}\right|\eta_1\right) + \theta(\eta_1 - \eta_3)^{\phantom{\frac12}} h^*\left(\left|\vec{p}_2 - \vec{p}\right|\eta_1\right)\, h\left(\left|\vec{p}_2 - \vec{p}\right|\eta_3\right)\right], \nonumber \\
{\rm and} \quad \lambda^{+--}(\eta_{1,2,3}; p_{1,2,3}) = - \left(-i\,\lambda\right)^3 \, \left(\eta_1\, \eta_2\, \eta_3\right)^{D-1} \, \int \frac{d^{D-1}p}{\left(2\pi\right)^{D-1}} \times \nonumber \\ \times \left[\theta(\eta_1 - \eta_2)^{\phantom{\frac12}} h^*(p\eta_1)\, h(p\eta_2) + \theta(\eta_2 - \eta_1)^{\phantom{\frac12}} h^*(p\eta_2)\, h(p\eta_1)\right]\nonumber \\
h^*\left(\left|\vec{p}_1 + \vec{p}\right|\eta_2\right)\, h\left(\left|\vec{p}_1 + \vec{p}\right|\eta_3\right) \,\, h\left(\left|\vec{p}_2 - \vec{p}\right|\eta_3\right)\, h^*\left(\left|\vec{p}_2 - \vec{p}\right|\eta_1\right),
\eqa
up to the factor of $\delta\left(\vec{p}_1 + \vec{p}_2 + \vec{p}_3\right)$.
Now it is straightforward to check that in the limit $p_i \eta_i \to 0$, $i=1,2,3$, there are no large IR contributions to these expressions. The fastest way to make this observation is to put all the external momenta $p_i$ to zero and to see that, then, $\lambda^{\pm\pm\pm}$ are all finite. At the same time, the one-loop correction to $D^K$ is divergent in this limit.

\subsubsection{Other $\alpha$-harmonics}

Other $\alpha$-modes can be relevant in the IR limit, although they show wrong behavior in the UV limit.
That is a quite frequent situation in condensed matter physics: One notable instance is the BCS theory for superconductivity, where, to observe the Cooper pairing, one has to perform the Bogolyubov rotation into harmonics, which, however, are inappropriate in the UV limit.
Because of that it is instructive to perform the one-loop calculation for them.

The leading IR contribution in this case also can be expressed in such a form as (\ref{EPP}). For the most values of $\alpha$, both $n_p$ and $\kappa_p$ receive large contributions which are similar to (\ref{nkappa}), but with different prefactors of $\log(p\eta)$. The only exception is given by the out-modes. In the latter case $h(x) \approx A\, x^{i\mu}$, as $x\to 0$, where $A$ is some complex constant following from the proper normalization of the harmonic functions and the expansion of the Bessel functions around zero.

To see the peculiarity of the out-modes and to calculate $n_p$ and $\kappa_p$ in this case, one has to perform the same manipulations as have been done during the one-loop calculation for the BD harmonics. However, if one substitutes the corresponding $h(p\eta)$ into (\ref{EPP}), he should take into account that they behave at past infinity, $x\to \infty$, as follows

\bqa
h(x) = \frac{1}{\sqrt{x}}\, \left[A_1 \, e^{i\,x} + A_2 \, e^{- i\,x}\right],
\eqa
where $A_{1,2}$ are some complex constants. Due to the interference terms between $e^{ix}$ and $e^{-ix}$, the  integrals over $x_{3,4}$ do not converge fast enough: they are saturated in the vicinity of $px_{3,4}/q \sim \mu$ rather than at $x_{3,4}\sim \mu$. (Note that according to our approximations $p/q\ll 1$.) Hence, naively one cannot Taylor expand $h\left(px_{3,4}/q\right) = h(p\eta_{3,4})$ around zero. However, if we are in $D=4$, we can subtract from and then add to $h^2(x)$, under these integrals, the value of the interference term, $\frac{A_1\, A_2}{|x|}$:

$$h^2(x) = h^2(x) - \frac{A_1\, A_2}{|x|} + \frac{A_1\,A_2}{|x|}.$$
Then, the $x_{3,4}$ integrals of $h^2(x) - \frac{A_1\,A_2}{|x|}$ are saturated around $x_{3,4}\sim \mu$ and, hence, one can Taylor expand $h\left(px_{3,4}/q\right)$ around zero inside the corresponding expressions. At the same time, the contributions from the additional integrals of $\frac{A_1\, A_2}{|x|}$ are suppressed in the IR limit. In fact, due to extra powers of $\eta_3$, the integrals over $d\eta_3$ are not divergent in the limit as $p\eta\to 0$. Thus, performing these manipulations we fetch out the leading IR correction. In higher dimensions the procedure is the same, but may demand a subtraction of higher powers of $1/|x|$. It is interesting to note that this kind of a problem does not appear in $\phi^4$ theory \cite{AkhmedovPopovSlepukhin}.

Thus, the contribution to the two-point function is as follows:

\bqa\label{62}
n_p(\eta) \approx \frac{\lambda^2 S_{D-2}}{(2\pi)^{D-1}}\, \left|A\right|^2 \,\log\left(\frac{\mu}{p \, \eta}\right)\, \iint^0_\infty dx_3 \, dx_4 \, \left(x_3 \, x_4\right)^{\frac{D-3}{2}}\, \left(\frac{x_4}{x_3}\right)^{i\mu} \, V(x_3)\, V^*(x_4), \nonumber \\ {\rm where} \quad
V(x) \equiv h^2(x) - \frac{A_1\, A_2}{|x|} - \dots,
\eqa
and $\kappa_p$ does not receive any large IR correction! As we will explain below, $\kappa_p$ is the measure of the backreaction strength. Hence, this observation is crucial for the proper quasi-particle interpretation in future infinity.

\subsection{One-loop correction in the contracting Poincar\'{e} patch}

Let us consider now one-loop corrections in the CPP. Similarly to the case of the EPP, it is not hard to show that $D^{R,A}$ and vertices do not receive large IR corrections. At the same time, in the limit $p\eta_0 \to 0$ the leading one-loop contribution to $D^K(p|\eta_1,\eta_2)$ can be written as (\ref{EPP}). The crucial difference, however, is that now $0 < \eta_0 < \eta_{1,2} < + \infty$. Also one has to exchange the positive and negative energy harmonics because of the time reversal.

To estimate the largest contribution to the integrals in $n_p$ and $\kappa_p$, one has to perform the same manipulations as we have been doing in the calculation on the EPP background.
The results are as follows. First, unlike the case of the EPP, now $\eta_0$ cannot be taken to past infinity, $\eta_0 \to 0$. Otherwise the integrals in $n_p$ and $\kappa_p$ will be explicitly divergent. Second, the answer for $n_p$ and $\kappa_p$ depends on the value of $p\eta = p\sqrt{\eta_1\eta_2}$:

For the BD modes, which are out-harmonics in the CPP, the IR divergence is present both in $n_p$ and $\kappa_p$. If $p\eta \ll \mu$, it is proportional to $\log\left(\eta/\eta_0\right) = \tau - \tau_0$. That is just the proper time elapsed from $\eta_0 = e^{\tau_0}$ to $\eta = e^\tau$. If, however, $p\eta \gg \mu$, then the divergence is proportional to $\log\left(\mu/p\eta_0\right)$. The coefficients of these divergences in both situations are the same as in (\ref{nkappa}).

For the Bessel or in-harmonics in the CPP, $n_p(\eta)$ diverges as $\log\left(\eta/\eta_0\right)$, if $p\eta \ll \mu$. At the same time, when $p\eta \gg \mu$, the divergence in $n_p$ is as $\log\left(\mu/p\eta_0\right)$. (The coefficients in both cases are the same as in (\ref{62}).) At the same time, in this case $\kappa_p(\eta)$ does not have any divergence or large IR contribution.

For any other type of $\alpha$-harmonics the IR behavior of $n_p$ and $\kappa_p$ is similar to that of the BD modes.

\subsection{One-loop correction in global de Sitter space}

We continue with the one-loop calculation in global dS. In this case the situation with the retarded (advanced) propagator and with vertices is the same as in the EPP and CPP. Thus, we consider here the Keldysh propagator.
Following \cite{PolyakovKrotov}, we would like to show that in global dS space the moment of turning on self-interactions cannot be taken to past infinity. Below we are going to work with even dimensional dS spaces. In odd-dimensional global dS spaces the situation is a bit different due to some cancelations \cite{Polyakov:2012uc}, but in essential sense it is similar to the even-dimensional case. The existence of the particle creation in the odd-dimensional case can be traced back to the fact that also in odd dimensions it is not possible to diagonalize the corresponding free Hamiltonian once and forever.

Unlike higher dimensional dS space, in the 2D case the one-loop calculation in global coordinates can be made quite explicit. So let us present it before continuing with the IR divergences in general dimensions.
Again, due to spatial homogeneity of 2D global dS space together with background states under consideration, we find it convenient to perform the Fourier transformation of all quantities along the spatial direction $\varphi \in [0,2\pi]$: $D^{K,R,A}(p|t_1, t_2) \equiv \int_0^{2\pi} d\varphi \, e^{i\, p \, \varphi} D^{K,R,A}(t_1, \varphi; t_2, 0)$.
Then, the one-loop contribution to $D^K$ is as follows:

\bqa\label{DSDK1}
D_{1}^{K}(p|t_1,t_2) = \lambda^2 \sum_{q=-\infty}^{+\infty} \iint_{t_0}^{+\infty} dt_3 dt_4 \cosh(t_3)\, \cosh(t_4) \times \nonumber \\ \times \Biggl[ \Biggr. D_{0}^R(p|t_1,t_3) \, D_{0}^K(q|t_3,t_4) \, D_{0}^K(|\vec{p}-\vec{q}|\,|t_3,t_4) \, D_{0}^A(p|t_4,t_2) + \nonumber \\ + 2 \, D_{0}^R(p|t_1,t_3) \, D_{0}^R(q|t_3,t_4) \, D_{0}^K(|\vec{p} - \vec{q}|\,|t_3,t_4) \, D_{0}^K(p|t_4,t_2) + \nonumber \\ + 2 \, D_{0}^K(p|t_1,t_3) \, D_{0}^K(q|t_3,t_4) \, D_{0}^A(|\vec{p}-\vec{q}|\,|t_3,t_4)\, D_{0}^A(p|t_4,t_2) - \nonumber \\ - \frac{1}{4} \, D_{0}^R(p|t_1,t_3) \, D_{0}^R(q|t_3,t_4) \, D_{0}^R(|\vec{p} - \vec{q}|\,|t_3,t_4) \, D_{0}^A(p|t_4,t_2) - \nonumber \\ -\frac{1}{4} \, D_{0}^R(p|t_1,t_3) \, D_{0}^A(q|t_3,t_4) \, D_{0}^A(|\vec{p} - \vec{q}|\,|t_3,t_4) \, D_{0}^A(p|t_4,t_2) \Biggl. \Biggr].
\eqa
Here the Keldysh propagator is

$$D_{0}^{K}(p|t_1,t_2) = {\rm Re}\left[g_p(t_1)g^*_p(t_2)\right],$$
the retarded and advanced propagators are

$$D_{0}^{R}(p|t_1,t_2) = \theta\left(t_1-t_2\right) \,2\,{\rm Im}\left[g_p(t_1)g^*_p(t_2)\right], \quad {\rm and} \quad D_0^{A}(p|t_1,t_2) = - \theta\left(t_2-t_1\right) \, 2\,{\rm Im}\left[g_p(t_1)g^*_p(t_2)\right],$$
correspondingly; $g_p(t)$ is the temporal part of a solution of the KG equation, which we do not specify for the beginning; $t_0$ is the moment when self-interactions are turned on.

The above one-loop expression can be rewritten as:

\bqa\label{63}
D_{1}^{K}(p|t_1,t_2) = -\frac{\lambda^2}{2} \, g_p(t_1)\, g^*_p(t_2) \, \sum_q \iint_{t_0}^{+\infty} dt_3 dt_4 \cosh(t_3)\, \cosh(t_4) g_p^*(t_3)\, g_p(t_4)\times \nn \\ \times \Biggl\{ \Biggr. - \theta(t_1 - t_3)\, \theta(t_2 - t_4) \, \Biggl[g_q(t_3) \, g_q^*(t_4) \, g_{|p-q|}(t_3)\,g_{|p-q|}^*(t_4) + c.c.\Biggr] + \nn \\ \Biggl[ \theta(t_1 - t_3)\, \theta(t_3 - t_4) + \theta(t_2 - t_4)\, \theta(t_4 - t_3)\Biggr] \, \Biggl[g_q(t_3) \, g_q^*(t_4) \, g_{|p-q|}(t_3)\,g_{|p-q|}^*(t_4) - c.c.\Biggr] \Biggl. \Biggr\} - \nn \\
-\frac{\lambda^2}{2} \, g_p(t_1)\, g_p(t_2) \, \sum_q \iint_{t_0}^{+\infty} dt_3 dt_4 \cosh(t_3)\, \cosh(t_4) g_p^*(t_3)\, g^*_p(t_4) \times \nn \\ \times \Biggl\{ \Biggr. \theta(t_1 - t_3)\, \theta(t_2 - t_4) \, \Biggl[g_q(t_3) \, g_q^*(t_4) \, g_{|p-q|}(t_3)\,g_{|p-q|}^*(t_4) + c.c.\Biggr] + \nn \\ \Biggl[ \theta(t_1 - t_3)\, \theta(t_3 - t_4) - \theta(t_2 - t_4)\, \theta(t_4 - t_3)\Biggr] \, \Biggl[g_q(t_3) \, g_q^*(t_4) \, g_{|p-q|}(t_3)\,g_{|p-q|}^*(t_4) - c.c.\Biggr] \Biggl. \Biggr\} + c.c.
\eqa
We need to find the leading contribution from this expression in the limit $t\equiv \frac{t_1 + t_2}{2}\to +\infty$, $t_1 - t_2 = const$ and $t_0 \to -\infty$. (We take $t_0\to -\infty$ just to check if this limit is smooth.)
As in the case of the EPP and CPP, we can substitute $t_{1,2}$ by $t$ in the arguments of the $\theta$-functions.

One can estimate directly the expression in Eq. (\ref{63}). But instead of doing this we apply a different procedure which allows us to find the leading corrections in any dimension.
If $t\to+\infty$ and $t_0 \to -\infty$, then the leading IR contributions to $D_{1}^K$ appear from the regions of integration over $t_{3,4}$ in the vicinity of $t$ and $t_0$. As we have explained above, in these regions the global dS metric can be well approximated by those of the EPP and CPP. It is this observation which allows one to estimate the one-loop correction in an arbitrary dimension.
In fact, then $g_p(t)$ can be approximated by

\bqa
g_p(t) \approx \left\{
                  \begin{array}{cc}
                    \eta_+^{\frac{D-1}{2}} \, h_+(p\eta_+), & \eta_+ = e^{-t}, \quad t \to +\infty \\
                    \eta_-^{\frac{D-1}{2}} \, h_-(p\eta_-), & \eta_- = e^t, \quad t\to -\infty \\
                  \end{array}
                \right.
\eqa
for non-zero $p$. (In an arbitrary dimension $p$ coincides with the index of the temporal part of the harmonics.) Then, as $p\eta\to 0$ and $p\eta_0\to 0$, where $\eta = e^{-t}$ and $\eta_0 = e^{t_0}$, the leading IR contribution to $D^K$ is hidden within \cite{PolyakovKrotov}, \cite{AkhmedovGlobal}:

\bqa\label{Leadingoneloop}
D_{1}^{K}(p|t_1,t_2) = g_p(t_1)\, g^*_p(t_2) \, n_p(t) + g_p(t_1)\, g_p(t_2) \,\kappa_p(t) + c.c., \nn \\
{\rm where} \quad n_p(t) \approx - \frac{\lambda^2 \, S_{D-2}}{(2\pi)^{D-1}} \, \int d^{D-1}q \iint_\infty^\eta d\eta_3 \, d\eta_4 \, (\eta_3\, \eta_4)^{\frac{D-3}{2}} \times \nn \\ \times h_+^*\left(p\eta_3\right) \, h_+\left(p\eta_4\right) \, h_+^*(q\,\eta_3)\,  h_+(q\eta_4) \, h^*_+\left(\left|\vec{q} - \vec{p}\right|\eta_3\right)\, h_+\left(\left|\vec{p} - \vec{q}\right|\eta_4\right) - \nn \\
- \frac{\lambda^2 \, S_{D-2}}{(2\pi)^{D-1}} \, \int d^{D-1}q \iint^\infty_{\eta_0} d\eta_3 \,\eta_4\, (\eta_3\, \eta_4)^{\frac{D-3}{2}} \times \nn \\ \times h_-^*\left(p\eta_3\right) \, h_-\left(p\eta_4\right) \, h_-^*(q\eta_3)\,  h_-(q\eta_4)\, h_-^*\left(\left|\vec{q} - \vec{p}\right|\eta_3\right) \, h_-\left(\left|\vec{q} - \vec{p}\right|\eta_4\right), \nn \\ {\rm and} \quad
\kappa_p(t) \approx \frac{2\,\lambda^2 \, S_{D-2}}{(2\pi)^{D-1}} \, \int d^{D-1}q \int_\infty^\eta d\eta_3 \int_{\infty}^{\eta_3} d\eta_4 \, (\eta_3\, \eta_4)^{\frac{D-3}{2}} \times \nn \\ \times h_+^*\left(p\eta_3\right) \, h_+^*\left(p\eta_4\right) \, h_+^*(q\eta_3)\,  h_+(q\eta_4) \, h_+^*\left(\left|\vec{q} - \vec{p}\right|\eta_3\right)\, h_+\left(\left|\vec{q} - \vec{p}\right|\eta_4\right) + \nn \\
+ \frac{2\,\lambda^2 \, S_{D-2}}{(2\pi)^{D-1}} \, \int d^{D-1}q \int^\infty_{\eta_0} d\eta_3 \int^{\eta_3}_{\eta_0} d\eta_4 \, (\eta_3\, \eta_4)^{\frac{D-3}{2}} \times \nn \\ \times h_-^*\left(p\eta_3\right) \, h_-^*\left(p\eta_4\right) \, h_-^*(q\eta_3) \,  h_-(q\eta_4)\, h^*_-\left(\left|\vec{q} - \vec{p}\right|\eta_3\right) \, h_-\left(\left|\vec{q} - \vec{p}\right|\eta_4\right).
\eqa
Thus, IR corrections in global dS space are just sums of the contributions from the contracting (past infinity, i.e. around $t_0$) and expanding (future infinity, i.e. around $t$) Poincar\'{e} patches of the whole space.

We continue with the description of the character of the IR divergences for different choices of the $\alpha$-modes. That may help to specify the convenient choice of the harmonics for the proper definition of quasi--particles in the IR limit.

We start with the Euclidian harmonics. In this case, $h_-^*(x) = h_+(x) \propto {\cal H}^{(1)}_{i\mu}(x)$. Using that $h_+(x) \approx A_+ \, x^{i\mu} + A_-\, x^{-i\mu}$, as $x\to 0$, and performing the same manipulations as in the EPP and CPP, we obtain:

\bqa
n_p(t) \approx \frac{\lambda^2 S_{D-2}}{(2\pi)^{D-1}}\, \log\left(\frac{\mu^2}{p^2 \, \eta \, \eta_0}\right)\, \iint_0^\infty dx_3 \, dx_4 \, \left(x_3 \, x_4\right)^{\frac{D-3}{2}}\times \nn \\ \times \left[\left|A_+\right|^2 \, \left(\frac{x_4}{x_3}\right)^{i\mu} + \left|A_-\right|^2 \, \left(\frac{x_3}{x_4}\right)^{i\mu}\right]\, h^2(x_3)\, \left[h^*(x_4)\right]^2, \nn \\
\kappa_p(t) \approx - \frac{2\,\lambda^2 S_{D-2}}{(2\pi)^{D-1}}\, \log\left(\frac{\mu^2}{p^2 \, \eta \, \eta_0}\right)\, A_+ \, A_- \, \int^0_\infty dx_3 \,\int_\infty^{x_3} dx_4 \, \left(x_3 \, x_4\right)^{\frac{D-3}{2}}\times \nn \\ \times \left[\left(\frac{x_4}{x_3}\right)^{i\mu} + \left(\frac{x_3}{x_4}\right)^{i\mu}\right]\, h^2(x_4)\, \left[h^*(x_3)\right]^2.
\eqa
Note that to calculate the expression for $\kappa_p$ we have used the identity $A_+^* \, A_-^* = A_+ \, A_-$. Thus, for the Euclidian vacuum the leading IR divergence is proportional to $\log\left(1/\eta\eta_0\right) = (t-t_0)$ and is present both in $n_p$ and $\kappa_p$.

At the same time, for the in-harmonics $h_-(x)$ is proportional to the Bessel function $Y_{i\mu}(x)$. Hence, $h_-(x) \approx A^*\, x^{-i\mu}$, as $x\to 0$ (the past infinity of the CPP). But $h_+(x) \approx C_+ \, x^{i\mu} + C_-\, x^{-i\mu}$, as $x\to 0$ (the future infinity of the EPP). Here $C_\pm$ are some complex constants related to $A_\pm$ via the corresponding Bogolyubov $\alpha$-rotation.

Then, the leading IR contribution in this case looks as:

\bqa
n_p(t) \approx \frac{\lambda^2 S_{D-2}}{(2\pi)^{D-1}}\, \log\left(\frac{\mu}{p \, \eta}\right)\, \iint^0_\infty dx_3 \, dx_4 \, \left(x_3 \, x_4\right)^{\frac{D-3}{2}}\times \nn \\ \times \left[\left|C_+\right|^2 \, \left(\frac{x_4}{x_3}\right)^{i\mu} + \left|C_-\right|^2 \, \left(\frac{x_3}{x_4}\right)^{i\mu}\right]\, V_+(x_3)\, V^*_+(x_4) + \nn \\ + \frac{\lambda^2 S_{D-2}}{(2\pi)^{D-1}}\, \log\left(\frac{\mu}{p \, \eta_0}\right)\, \iint_0^\infty dx_3 \, dx_4 \, \left(x_3 \, x_4\right)^{\frac{D-3}{2}}\, \left|A\right|^2 \, \left(\frac{x_4}{x_3}\right)^{i\mu} \, V^*_-(x_3)\, V_-(x_4), \nn \\
\kappa_p(t) \approx - \frac{2\,\lambda^2 S_{D-2}}{(2\pi)^{D-1}}\, \log\left(\frac{\mu}{p \, \eta}\right)\, C^*_+ \, C^*_- \, \int^0_\infty dx_3 \,\int_\infty^{x_3} dx_4 \times \nn \\ \times \left(x_3 \, x_4\right)^{\frac{D-3}{2}}\, \left[\left(\frac{x_4}{x_3}\right)^{i\mu} + \left(\frac{x_3}{x_4}\right)^{i\mu}\right]\, V_+(x_3)\, V_+^*(x_4).
\eqa
where $V_\pm(x) = h_\pm^2 (x) - \frac{A_1 A_2}{|x|} - \dots$ and was defined above. (Dots here stand for the higher powers of $1/|x|$.) Note that for the in-harmonics $n_p(t)$ depends separately on $t$ and $t_0$ rather than on their difference $(t-t_0)$ and $\kappa_p(t)$ does not diverge as $t_0\to-\infty$.

Similar conclusions are true for the out-harmonics. The crucial difference, however, is that in the latter case $\kappa_p$ does diverge as $t_0\to -\infty$, but is finite as $t\to+\infty$.
For the other $\alpha$--harmonics the leading IR contributions to $n_p$ and $\kappa_p$ also depend on $t$ and $t_0$ separately and diverge simultaneously when $t\to +\infty$ and $t_0 \to -\infty$.

The presence of the IR divergence means that one cannot take the regulator, $t_0$, to minus infinity. Thus, one has to fix an initial Cauchy surface at some finite $t_0$. (Such a surface does not necessarily have to be taken somewhere around past infinity. One can put it, e.g., around the neck of global dS space.) But holding $t_0$ fixed, violates the dS-isometry.

\section{Summation of leading infrared contribution in all loops}

Thus, we see that $\lambda^2 \log(\eta)$ and $\lambda^2 \log(\eta_0)$ contributions can become large, as $\eta \to 0$ and $\eta_0 \to 0$,  even if $\lambda^2$ is very small. I.e., loop corrections are not suppressed in comparison with classical tree level contributions to the propagators. That is true even if the dS isometry is respected in loops. Hence, to understand the physics in dS space, one has to sum unsuppressed IR corrections from all loops.

We start with the discussion of the situation in the EPP and then continue with the CPP and global dS space. In the EPP there is a separate interesting problem to sum dS-invariant IR contributions to the correlation functions of the exact BD state (see, e.g., \cite{Youssef:2013by}, \cite{Gautier:2013aoa}). As we pointed out in the Introduction, solution of this problem is not sufficient to make a definite conclusion about the stability of dS space: We find it as more physically sensible to question the stability of dS space under nonsymmetric perturbations.

Hence, we propose to consider an initial nonsymmetric perturbation on top of the BD state. But we cannot just put an initial comoving density at past infinity of the EPP, because then the physical density will be infinite. One has to put the density perturbation at an initial Cauchy surface which is different from the past infinity $+ \infty > \eta_0 > 0$. Moreover, due to the UV divergences the comoving momentum also should be cutoff at the UV scale. But, as was explained above $n(p\eta)$ and $\kappa(p\eta)$ are attributed to the comoving volume and, hence, do not change before $p\eta \sim \mu$. Their behaviour for $p\eta > \mu$ is not much different from that in flat space--time. Thus, cutting simultaneously comoving momentum and conformal time integrals effectively amounts to cut the physical momentum integrals at $\mu$. On the other hand, due to the symmetries of the EPP we can put an initial comoving density at an initial value of the physical momentum $(p\eta)_0 \sim \nu$ and cutoff all the integrals over the physical momentum at this value. 
Then, the dS isometry is broken by the presence of a density on top of the dS-invariant BD state. (In the case of the CPP and global dS space we do not even have to do this because the dS isometry is broken in loops for any initial state.) We would like to trace the destiny of such a density perturbation and to understand the effect of the large IR contributions on the initial state, as the system progresses towards future infinity.

\subsection{From the Dyson--Schwinger to the kinetic equation in the Poincar\'{e} patches}

To sum loop contributions one has to solve the system of Dyson-Schwinger (DS) equations for the propagators, self-energies and vertices. We would like to sum only powers of the leading contribution, $\lambda^2 \log(p\eta)$, and neglect suppressed terms, such as, e.g., $\lambda^4 \log(p\eta)$ or $\lambda^2 \log(\eta_1/\eta_2)$ and etc.. As we have seen above, the retarded and advanced propagators do not receive large IR contributions from the first loop. The same is true for the vertices. However, these quantities may receive large contributions from the second or higher loops. In fact, the latter may come from those corrections to the Keldysh propagator which appear at lower loops. But this just means that such contributions to $D^{R,A}$ and vertices will be suppressed by higher powers of $\lambda$. Thus, if we need to perform only the summation of the leading corrections, then we have to care only about the DS equation for the Keldysh propagator.

In the following, we assume that $D^{R,A}$ and vertices take their tree level values (perhaps renormalized by the UV contributions) and assume that all IR corrections go into IR exact $D^K$. Then, the relevant DS equation acquires the form:

\bqa\label{DSDK2}
D^{K}(p|\eta_1,\eta_2) =  D_{0}^K(p|\eta_1,\eta_2) + \nonumber \\ + \lambda^2 \int \frac{d^{D-1}\vec{q}}{(2\pi)^{D-1}} \iint_\infty^0 \frac{d\eta_3 d\eta_4}{(\eta_3\eta_4)^D} \, \Biggl[ \Biggr. D_{0}^R(p|\eta_1,\eta_3) \, D^K(q|\eta_3,\eta_4) \, D^K(|\vec{p} - \vec{q}|\,|\eta_3,\eta_4) \, D_0^A(p \,| \eta_4,\eta_2) + \nonumber \\ + 2 \, D_{0}^R(p|\eta_1,\eta_3) \, D_0^R(q|\eta_3,\eta_4) \, D^K(|\vec{p}-\vec{q}|\,|\eta_3,\eta_4) \, D^K(p|\eta_4,\eta_2) + \nonumber \\ + 2 \, D^K(p|\eta_1,\eta_3) \, D^K(q\eta_3,\eta_4) \, D_0^A(|\vec{p}-\vec{q}|\,|\eta_3,\eta_4)\, D_0^A(p|\eta_4,\eta_2) - \nonumber \\ - \frac{1}{4} \, D_{0}^R(p|\eta_1,\eta_3) \, D_0^R(q|\eta_3,\eta_4) \, D_0^R(|\vec{p}-\vec{q}|\,|\eta_3,\eta_4) \, D_0^A(p|\eta_4,\eta_2) - \nonumber \\ - \frac{1}{4} \, D_{0}^R(p|\eta_1,\eta_3) \, D_0^A(q|\eta_3,\eta_4) \, D_0^A(|\vec{p}-\vec{q}|\,|\eta_3,\eta_4) \, D_0^A(p|\eta_4,\eta_2) \Biggl. \Biggr],
\eqa
where $D_0^K(p|\eta_1,\eta_2)$ is the initial (tree level) value of the propagator.
We propose the following ansatz to solve this equation:

\bqa\label{ansatz}
D^{K}(p|\eta_1,\eta_2) = \left(\eta_1\eta_2\right)^{\frac{D-1}{2}}\, d^K(p\eta_1,p\eta_2), \quad {\rm where} \nonumber \\
d^K\bigl(p\eta_1,p\eta_2\bigr) =  \frac12 h\bigl(p\eta_1\bigr)\,h^*\bigl(p\eta_2\bigr) \biggl[1+2\,n_p\bigl(\eta\bigr)\biggr] + h\bigl(p\eta_1\bigr)h\bigl(p\eta_2\bigr)\,\kappa_p\bigl(\eta\bigr) + c.c..
\eqa
Here $\eta = \sqrt{\eta_1 \eta_2}$, $n_p(\eta)$ and $\kappa_p(\eta)$ are unknown functions to be defined by the
equations under derivation. Below we also use the notation $d^-(p\eta_1, p\eta_2) = 2\, {\rm Im} \left[h(p\eta_1)\, h^*(p\eta_2)\right]$ to simplify the equations.
The ansatz for the IR exact Keldysh propagator is inspired by the one-loop calculation and its physical interpretation.

For general values of $\eta_1$ and $\eta_2$ the ansatz (\ref{ansatz}) does not solve (\ref{DSDK2}). However, in the limit $p\eta_{1,2}\to 0$ and $\eta_1/\eta_2=const$ one can neglect the difference between $\eta_1$ and $\eta_2$ in the expressions that follow and substitute the average conformal time $\eta = \sqrt{\eta_1\eta_2}$ instead of both $\eta_1$ and $\eta_2$ for the limits of integrations over $\eta_3$ and $\eta_4$. Then, the ansatz in question solves the DS equation if $n_p$ and $\kappa_p$ obey:

\bqa\label{np}
n_p(\eta) \approx n_p^{(0)} - \lambda^2 \, \int \frac{d^{D-1}q}{(2\pi)^{D-1}} \, \iint_{\infty}^{\eta} d\eta_3 \, d\eta_4 \, (\eta_3\eta_4)^{\frac{D-3}{2}}  \times \nn\\
\times \Biggl\{ \Biggr. \Biggl[ d^K\biggl(q\eta_3,q\eta_4\biggr)\, d^K\biggl(\left|\vec{p} - \vec{q} \right| \, \eta_3, \left|\vec{p} - \vec{q} \right|\eta_4\biggr) + \frac14 \, d^-\biggl(q\eta_3,q\eta_4\biggr) \, d^-\biggl(\left|\vec{p} - \vec{q} \right|\eta_3, \left|\vec{p} - \vec{q}\right|\eta_4\biggr)
- \Biggr. \nn \\ - \Biggl. d^-\biggl(q\eta_3,q\eta_4\biggr)\, d^K\biggl(\left|\vec{p} - \vec{q} \right| \, \eta_3, \left|\vec{p} - \vec{q}\right| \, \eta_4\biggr) \, \biggl[1 + 2\,n\bigl(p\eta_{13}\bigr)\biggr] \Biggr] \, h^*\bigl(p\eta_3\bigr)\,h\bigl(p\eta_4\bigr)
+ \nn \\ + 4 \, \theta\left(\eta_4-\eta_3\right) \, d^K\biggl(\left|\vec{p} - \vec{q}\right|\eta_3, \left|\vec{p} - \vec{q} \right| \, \eta_4\biggr) \, {\rm Re} \biggl[d^-\biggl(q\eta_3,q\eta_4\biggr)\,h\bigl(p\eta_3\bigr)\,h\bigl(p\eta_4\bigr)\, \kappa\bigl(p\eta_{42}\bigr) \biggr] \Biggl. \Biggr\}
\eqa
and
\bqa\label{kappa}
\kappa_p(\eta) \approx \kappa_p^{(0)} - \lambda^2 \, \int \frac{d^{D-1}q}{(2\pi)^{D-1}} \, \iint_{\infty}^{\eta} d\eta_3 \, d\eta_4 \, (\eta_3\eta_4)^{\frac{D-3}{2}}  \times \nn\\
\times \Biggl\{ \Biggr. \Biggl[ d^K\biggl(q\eta_3,q\eta_4\biggr) \, d^K\biggl(\left|\vec{p} - \vec{q}\right| \, \eta_3, \left|\vec{p} - \vec{q}\right| \, \eta_4\biggr) + \frac14 \, d^-\biggl(q\eta_3,q\eta_4\biggr) \, d^-\biggl(\left|\vec{p} - \vec{q}\right| \, \eta_3, \left|\vec{p} - \vec{q}\right| \, \eta_4\biggr)
+ \Biggr. \nn \\ + \Biggl. d^-\biggl(q\eta_3,q\eta_4\biggr) \, d^K\biggl(\left|\vec{p} - \vec{q}\right| \, \eta_3, \left|\vec{p} - \vec{q}\right| \, \eta_4\biggr)\, \biggl[1 + 2\,n\bigl(p\eta_{13}\bigr)\biggr] \Biggr] \, h^*\bigl(p\eta_3\bigr) \, h^*\bigl(p\eta_4\bigr)
+ \nn \\ + 4 \, \theta\left(\eta_4-\eta_3\right) \, d^K\biggl(\left|\vec{p} - \vec{q}\right| \, \eta_3, \left| \vec{p} - \vec{q}\right| \, \eta_4\biggr) \, d^-\biggl(q\eta_3,q\eta_4\biggr) \, h^*\bigl(p\eta_3\bigr) \, h\bigl(p\eta_4\bigr) \, \kappa\bigl(p\eta_{42}\bigr) \Biggl. \Biggr\}
\eqa
where $n_p^{(0)}$ and $\kappa_p^{(0)}$ define the initial propagator $D^K_{0}(p|\eta_1, \eta_2)$ via Eq. (\ref{ansatz}).

In the derivation of (\ref{np}) and (\ref{kappa}) we have used the following relations $d^-(p\eta_1,p\eta_2) = - d^-(p\eta_2,p\eta_1) = - \biggl[d^-(p\eta_1,p\eta_2)\biggr]^*$ and $\int d^{D-1}\vec{q} f\left(q,\left|\vec{p} - \vec{q}\right|\right) = \int d^{D-1}\vec{q} f\left(\left|\vec{p} - \vec{q}\right|,q\right)$. Below we assume that $n_p(\eta)$ and $\kappa_p(\eta)$ are slow functions in comparison with $h(p\eta)$. Then, one can safely change arguments of all $n$'s and $\kappa$'s to the external time $\eta$. This is possible because of the usual separation of scales, which lays in the basis of the kinetic theory \cite{LL}. In our case this approximation is correct at least for the fields from the principal series, $m>(D-1)/2$, because their harmonics $h(p\eta)$ oscillate at future infinity.

Note that for the scalars from the complementary series, $m\le (D-1)/2$ the ansatz (\ref{ansatz}) also solves (\ref{DSDK2}). But in this case harmonics do {\it not} oscillate at future infinity and, hence, can be as slow as $n_p$ and $\kappa_p$. The main problem with the situation when $h(p\eta)$ is slow function is that then one cannot derive the kinetic equation of the usual form. More complicated integrodifferential equations are available whose solution and physical interpretation is not yet known to us.

To avoid such a situation when the equations themselves depend on their initial data, we would like to convert the integral equations (\ref{np}), (\ref{kappa}) into their integrodifferential form, i.e., into the form of the kinetic equation \cite{LL}. This is done via a kind of the renormalization group procedure as follows \cite{AkhmedovBurda}, \cite{AkhmedovGlobal}, \cite{AkhmedovPopovSlepukhin}. In the given settings, $n^{(0)}_p$ and $\kappa^{(0)}_p$ are the particle density and anomalous average at some moment after $\eta_* \sim \mu/p$. In fact, as we will explain below, before this moment, all the kinetic processes in the theory under consideration are suppressed because of the strong blue shift. Hence, before $\eta_* \sim \mu/p$, $n_p(\eta)$ and $\kappa_p(\eta)$ remain practically unchanged, i.e. can be set equal to their initial values, $n^{(0)}_p$ and $\kappa^{(0)}_p$, correspondingly. Then, from (\ref{np}), (\ref{kappa}) it is straightforward to find that due to the large IR corrections the difference between $n_p(\eta), \kappa_p(\eta)$ and $n_p^{(0)} = n_p(\eta_*), \kappa_p^{(0)} = \kappa_p(\eta_*)$ is proportional to the proper time elapsed from $\eta_*$ to $\eta$. Also one can approximate:

$$\frac{n_p(\eta) - n_p(\eta_*)}{\log(\eta) - \log(\eta_*)} \to \frac{d n_p(\eta)}{d \log(p\eta)}, \quad {\rm and} \quad \frac{\kappa_p(\eta) - \kappa_p(\eta_*)}{\log(\eta) - \log(\eta_*)} \to \frac{d \kappa_p(\eta)}{d \log(p\eta)}.$$
The coefficients of the proportionality between the elapsed time and the change of $n_p$ and $\kappa_p$ are the so-called collision integrals. With the use of the following matrixes:

\bqa
N_p(\eta_1,\, \eta_2) = \eta_2^{\frac{D-1}{2}} \, \left(
  \begin{array}{cc}
    n_p(\eta_1) \, h^*(p\eta_2) & \kappa_p(\eta_1) \, h(p\eta_2) \\
    \kappa_p(\eta_1) \, h(p\eta_2) & n_p(\eta_1) \, h^*(p\eta_2) \\
  \end{array}
\right), \quad P = \left(
  \begin{array}{cc}
    0 & 1 \\
    1 & 0 \\
  \end{array}
\right)
\eqa
the collision integrals for $n_p$ and $\kappa_p$ can be written compactly. The real one for $n_p$ has the form:

\bqa\label{callintPP0}
\frac{d n_p(\eta)}{d\log(p\eta)} = 2\,\lambda^2 \, \int \frac{d^{D-1}q}{(2\pi)^{(D-1)}} \, \int_{0}^{\infty} \frac{\eta\, d\eta'}{\left(\eta\, \eta'\right)^D} \, \, {\rm Re}\, \, Tr \nonumber \\ \left\{C_{p,q,p-q}(\eta)\, \left[\left(1+N^*_p\right)\, N_{q}\, N_{p-q}^{\phantom{\frac12}} - \,\, N^*_p \, \left(1+N_{q}\right)\, \left(1+N_{p-q}\right)\right](\eta,\eta') +  \right. \nonumber \\ + 2 \, C_{q,q-p,p}(\eta)\, \left[ N^*_{q}\, \left(1+N_{q-p}\right)\,\left(1+N_p\right)^{\phantom{\frac12}} - \,\,\left(1+N^*_{q}\right)\, N_{q-p}\, N_p \right](\eta,\eta') + \nonumber \\ \left.
+ D_{p,q,p+q}(\eta)\,\left[ \left(1+N_{p}\right)\, \left(1+N_{q}\right) \,\left(1+N_{p+q}\right)^{\phantom{\frac12}} - \,\, N_{p}\, N_{q}\, N_{p+q} \right](\eta,\eta')^{\phantom{\frac12}}\right\} + \left[N\to P\,N\right].
\eqa
At the same time the complex collision integral for $\kappa_p$ is as follows:

\bqa\label{callintPP00}
\frac{d \kappa_p(\eta)}{d\log(p\eta)} = - 2\,\lambda^2 \, \int \frac{d^{D-1}q}{(2\pi)^{(D-1)}} \, \int_{0}^{\eta} \frac{\eta \, d\eta'}{\left(\eta\, \eta'\right)^D} \, \, \left(\vec{p}\to - \vec{p}\right)\, \, Tr \nonumber \\ \left\{C_{p,q,p-q}(\eta)\,\left[ \left(1+N_{p}\right)\, \left(1+N_{q}\right) \,\left(1+N_{p-q}\right)^{\phantom{\frac12}} - \,\, N_{p}\, N_{q}\, N_{p-q} \right](\eta,\eta') +  \right. \nonumber \\ + 2\,C^*_{q,q-p,p}(\eta)\, \left[ N^*_{q}\, \left(1+N_{q-p}\right)\,\left(1+N_p\right)^{\phantom{\frac12}} - \,\,\left(1+N^*_{q}\right)\, N_{q-p}\, N_p \right](\eta,\eta') + \nonumber \\ \left.
+ D^*_{p,q,p+q}(\eta)\, \left[\left(1+N_p\right)\, N^*_{q}\, N_{p+q}^{*\phantom{\frac12}} - \,\, N_p \, \left(1+N^*_{q}\right)\, \left(1+N^*_{p+q}\right)\right](\eta,\eta')^{\phantom{\frac12}}\right\} - \left[N\to P\,N\right].
\eqa
The term $\left[N\to P\,N\right]$ means that we have to add to the explicitly written expressions in (\ref{callintPP0}) and (\ref{callintPP00}) the same quantities with every $N$ substituted by the product $P\,N$; Re $Tr$ means that one has to take the real part and the trace of the expression that stands after this sign; $\left(\vec{p}\to - \vec{p}\right)\,Tr$ means that one has to take the trace and add to the expression standing after this sign the same term with the exchange $\vec{p}\to - \vec{p}$. Finally, to simplify the equations we use the notations: $C_{k_1k_2k_3}(\eta) = \eta^{\frac{3\, (D-1)}{2}}\,h^*(k_1\eta)\,h(k_2\eta)\, h(k_3\eta)$ and $D_{k_1k_2k_3}(\eta) = \eta^{\frac{3(D-1)}{2}}\,h(k_1\eta)\,h(k_2\eta)\, h(k_3\eta)$.

Similarly to Eq. (\ref{EPP}), the system of equations (\ref{callintPP0}), (\ref{callintPP00}) represents just a preliminary version of the kinetic equations. It is not yet suitable to sum only the leading IR contributions in all loops because on top of the latter it also accounts for some of the subleading corrections. In particular, one should restrict the $dq$ integrals in (\ref{callintPP0}) and (\ref{callintPP00}) to the region $q \gg p$ and Taylor expand $h(p\eta)$ and $h(p\eta')$ around zero. As we discussed above, it is this region where the leading IR contributions come from.

\subsubsection{Physical meaning of infrared effects}

Before going further let us explain the meaning of the system of equations (\ref{callintPP0}) and (\ref{callintPP00}).
To do that let us forget for the moment about the presence of $\kappa_p$ and consider the situation in flat space. If the time of the turning on self-interactions is $t_0$ and the time of the observation is $t$, then the kinetic equation is:

\bqa\label{coll}
\frac{dn_p}{dt} \propto - \lambda^2 \, \int\frac{d^{D-1}q}{\epsilon(p)\,\epsilon(q)\, \epsilon(p-q)} \times \nonumber \\  \left\{ \int_{t_0}^t dt' \cos\left[\left(-\epsilon(p) + \epsilon(q) + \epsilon(p-q)\right)\, (t-t')\right] \, \left[(1+n_p)\, n_{q}\, n_{|p-q|}^{\phantom{\frac12}} - n_p \, (1+n_{q})\, (1+n_{|p-q|})\right](t) \right. \nonumber \\ + 2\, \int_{t_0}^t dt' \cos\left[\left(- \epsilon(q) + \epsilon(q-p) + \epsilon(p)\right)\, (t-t')\right] \, \left[n_{q}\,(1 + n_{|q-p|})\, (1 + n_p)^{\phantom{\frac12}} - (1+n_{q})\, n_{|q-p|} \, n_p\right](t) \nonumber \\ +
\left. \int_{t_0}^t dt' \cos\left[\left(\epsilon(q) + \epsilon(q+p) + \epsilon(p)\right)\, (t-t') \right] \, \left[(1 + n_{q})\,(1 + n_{q+p})\, (1 + n_p)^{\phantom{\frac12}} - n_{q}\, n_{q+p} \, n_p\right](t) \right\},
\eqa
where $\epsilon(p)$ is the energy of the particle with the momentum $\vec{p}$. There are many ways to derive this kinetic equation (see, e.g., appendix of \cite{AkhmedovKEQ}). One of them is similar to that method which was presented above for dS space.

In the limit $(t-t_0)\to \infty$ the $dt'$ integrals in (\ref{coll}) are converted into (minus) $\delta$-functions ensuring energy conservation in some processes, which we will define now.
The expression in the second line of Eq. (\ref{coll}), which is multiplying $dt'$ integral, describes the competition between the following two processes. One of them corresponds to the term $n_p \, (1+n_{q})\, (1+n_{|p-q|})$ and is such a process in which a particle with momentum $\vec{p}$ decays into two excitations --- $\vec{q}$ and $\vec{p}- \vec{q}$. This term appears with the minus sign in the collision integral because it describes the loss of an excitation with the momentum $\vec{p}$. The inverse gain process, corresponding to the term $(1+n_p)\, n_{q}\, n_{|p-q|}$ with the plus sign, is such that two particles, $\vec{q}$ and $\vec{p}-\vec{q}$, are merged together to create an excitation with momentum $\vec{p}$.

The third line in (\ref{coll}) also describes two competing processes. The first of them is such that a particle with momentum $\vec{p}$ joins together with another one, $\vec{q}-\vec{p}$, to create an excitation with momentum $\vec{q}$. This is the loss process. The inverse gain process is the one in which a particle with momentum $\vec{q}$ decays into two, one of which is with momentum $\vec{p}$. The coefficient 2 in front of this term is just the combinatoric factor.

Similarly the fourth line describes two processes.
The gain process is when three particles, one of which is with momentum $\vec{p}$,
are created by an external field, if any. The loss process is when three such excitations are annihilated into vacuum.

All these six types of processes are not allowed by the energy--momentum conservation for {\it massive}
fields with $\phi^3$ self-interaction in {\it flat} space-time\footnote{For the massless flat space $\phi^3$ theory the collision integral in (\ref{coll}) is not zero. The second and the third lines in (\ref{coll}) describe nontrivial decay processes for the collinear momenta of the products of the reactions. In this case (\ref{coll}) will describe, e.g., phonons in solid bodies if instead of the speed of light one would use that of sound.}. Hence, the collision integral is vanishing as $(t-t_0)\to\infty$. However, in dS space these processes are allowed \cite{Myrhvold}--\cite{Akhmedov:2009vh} because there is no energy conservation.

One can obtain from (\ref{coll}) the collision integral of Eq. (\ref{callintPP0}) with $\kappa_p$ set to zero. To do that it is necessary to exchanges $dt'$ for $d\eta'$, multiply it by the proper weight, $\sqrt{g}$, following from the nontrivial metric, and to use the dS harmonics, $g_p(\eta) = \eta^{\frac{D-1}{2}}\, h(p\eta)$, instead of the plane waves, $e^{-i\, \epsilon(p) \, t}/\sqrt{\epsilon(p)}$. Then, instead of the above mentioned $\delta$-functions, which ensure energy conservation, there are some expressions, which define the differential rates (per given range of $\vec{q}$) of the processes described in the above three paragraphs.

Note that in the free theory ($\lambda = 0$) $n_p$ remains constant. This is an indirect argument favoring that $n_p$ is the particle density per comoving volume. (The direct argument follows from the calculation of the density from the Wightman function.) Furthermore, due to the presence of $\kappa_p$, in (\ref{callintPP0}) we have extra terms on top of those which are shown in (\ref{coll}). All of them can be obtained from (\ref{coll}) via the simultaneous substitutions of some of $(1+n_{k})$'s and $n_{k}$'s by $\kappa_{k}$'s or $\kappa^*_{k}$'s. E.g. in (\ref{callintPP0}) we encounter terms of the type:

\bqa\label{meaning}
\left[(1+n_p)\, \kappa_{p}\, n_{|p-q|} - n_p \, \kappa_{q}\, (1+n_{|p-q|})\right].
\eqa
The meaning of (\ref{meaning}) is that it describes the following two competing processes --- a particle with momentum $\vec{p}$ is lost (gained) in such a situation, in which instead of the creation (annihilation) of two particles, with momenta $\vec{q}$ and $\vec{p}-\vec{q}$, we obtain a single excitation $\vec{p}-\vec{q}$ and missing momentum $\vec{q}$ is gone into (taken from) the background quantum state of the theory.

Similarly we encounter terms of the type

\bqa
\left[(1+n_p)\, \kappa_{q}\, \kappa_{|p-q|} - n_p \, \kappa_{q}\, \kappa_{|p-q|}\right]
\eqa
which describe the processes in which both momenta $\vec{q}$ and $\vec{p}-\vec{q}$ are coming from (going to) the background state. These observations, in particular, justify the interpretation of $\kappa_p$ as the measure of the strength of the backreaction on the background state of the theory.

Now put both $n_p$ and $\kappa_p$ to zero inside collision integral (\ref{callintPP0}). The only term which survives then is present in the last line of (\ref{callintPP0}) or (\ref{coll}). It describes the particle creation process by the dS gravitational field. If one takes the integral of this term over the conformal time from $\eta_0$ to $\eta$ he obtains exactly the expression for $n_p(\eta)$ from (\ref{EPP}).
Using the collision integral for $\kappa_p$, (\ref{callintPP00}), one can make similar observation about the origin of the corresponding expression in (\ref{EPP}).

Thus, in the nonstationary situations, i.e., when the collision integral is not zero, it is natural to expect a linear time-divergence, $\propto (t-t_0)$, in $n_p$ and $\kappa_p$. That indeed is true for the homogeneous in time background fields (such as, e.g., constant electric fields in QED), when particle production rates are constant.
But, in the time dependent metric of dS space one encounters a bit different situation.
In particular, in the EPP the largest IR contributions have the form of $\lambda^2 \,\log\left(\mu/p\eta\right)$ and there is no IR divergence as one takes the moment of turning on self-interactions to past infinity, $\eta_0 \to + \infty$. That happens because the creation of particles with comoving momentum $p$ effectively starts at $\eta_* \sim \mu/p$ rather than at the moment when we turn on self-interactions, $\eta_0\to +\infty$. For the same reason in the CPP the particle creation process goes on until $\eta_* \sim \mu/p$ and then stops, as we have seen from the one-loop calculation.

We encounter such a situation because,
first of all, the collision integral is not just a constant and depends on time. Second, the past (future) infinity of the EPP (CPP) corresponds to the UV limit of the physical momentum. In this limit $g_p(\eta)$ harmonics behave as plane waves. As a result, all the rates inside the collision integral can be well approximated by $\delta$-functions ensuring energy conservation \cite{AkhmedovKEQ}, \cite{AkhmedovBurda}, \cite{AkhmedovGlobal}. Hence, the collision integral becomes negligible as $p\eta \to \infty$ both in the EPP and CPP. That is the reason why in the EPP, independently of the type of the harmonics, one can put the moment of turning on self-interactions, $\eta_0$, to past infinity. Similarly these observations explain the time-independence of the leading IR contributions in the CPP, when $\eta > \mu/p$.

\subsubsection{The kinetic equation}

The DS equation (\ref{DSDK2}) is covariant under the simultaneous Bogolyubov rotation of the harmonics and of $n_p$ and $\kappa_p$. Hence, in principle a solution of (\ref{DSDK2}) in terms of one type of harmonics can be mapped to another type.

However, we would like to choose such harmonics, $h(p\eta)$, for which there is a solution of (\ref{callintPP0}), (\ref{callintPP00}) with $\kappa_p$ being zero. From the one-loop calculations it is not hard to find that most of $\alpha$-modes, including the BD ones, do not provide such a solution. That means that the backreaction on the corresponding ground states is strong. Only for the out-harmonics, $h(x) \propto J_{i\mu}(x)$, the situation is different. In particular, if one puts $\kappa_p(\eta)$ to zero it is not generated back in (\ref{np}), (\ref{kappa}). Or more precisely, contribution to it behaves as $\lambda^2$, i.e., is negligible in comparison with $\lambda^2\,\log(p\eta)$.

All in all, out-harmonics represent the proper quasi-particle states in future infinity. Which means that for out-modes the ansatz (\ref{ansatz}) solves the DS equation with $\kappa_p(\eta) \equiv 0$. This is the argument which favors the interpretation that independently of the initial state at the past infinity of the EPP the field theory state flows in future infinity to the out-vacuum with some density of particles on top of it \cite{AkhmedovKEQ}. To support such a conclusion, we will show in a moment that for the out-harmonics $\kappa_p(\eta)$ indeed flows to zero in future infinity, even if originally it had some small non-zero value.

The kinetic equation for out-modes is obtained from (\ref{np}), (\ref{kappa}), with $\kappa_p(\eta)$ set to zero, via the same ``renormalization group'' procedure as was used above. The result is:

\bqa\label{cint}
\frac{d n(x)}{d\log (x)} = - \frac{\lambda^2\, S_{D-2} \, |A|^2}{(2\,\pi)^{D-1}} \, \int^0_{\infty} dx_3 \,x_3^{\frac{D-3}{2}} \, \int_{\infty}^{0} dx_4 \, x_4^{\frac{D-3}{2}} \times \nonumber \\ \times \left\{ {\rm Re} \left[x_3^{-i\mu} \, V(x_3) \,x_4^{i\mu}\, V^*(x_4)\right] \, \left\{[1+n(x)]\, n(x_3)^{2\phantom{\frac12}} - \,\, n(x) \, [1+n(x_3)]^2\right\} + \right. \nonumber \\
+ 2\,{\rm Re} \left[x_3^{i\mu} \, W(x_3) \, x_4^{- i\mu} \, W(x_4) \right]\, \left\{ n(x_3)\, [1+n(x_3)]\,[1+n(x)]^{\phantom{\frac12}} - \,\,[1+n(x_3)]\, n(x_3)\, n(x) \right\} + \nonumber \\
+ \left. {\rm Re} \left[x_3^{i\mu} \, V(x_3) \, x_4^{- i\mu} \, V^*(x_4)\right] \, \left\{ [1+n(x_3)]^2 \,[1+n(x)]^{\phantom{\frac12}} - \,\, n(x_3)^2\, n(x) \right\}^{\phantom{2}}\right\},
\eqa
where $x=p\eta$, $V(x) = \left[h^2\left(x\right) - \frac{A_1\, A_2}{|x|} - \dots\right]$, $W(x) = \left[\left|h\left(x\right)\right|^2- \frac{A_1 \, A_2}{\left|x\right|} - \dots \right]$ and $h(x) = \sqrt{\frac{\pi}{\sinh(\pi\mu)}}\, J_{i\mu}(x)$. We have done here the same approximations as in the one-loop calculation.

Now we have to check the stability of (\ref{cint}) under the linearized perturbations of $\kappa_p(\eta)$. If one keeps only the linear in $\kappa_p$ terms, the integrodifferential form of the equation, which follows from (\ref{kappa}), degenerates to:

\bqa
\frac{d\kappa_p(\eta)}{d\log(p\eta)} = \Gamma_3 \, \kappa_p(\eta), \nn \\
\Gamma_3 = \frac{4\, i\, \lambda^2 \, S_{D-2} \, \left|A\right|^2}{(2\pi)^{D-1}}\, \iint^0_\infty dx_3 \, dx_4 \, x_3^{\frac{D-3}{2} - i\mu}\, x_4^{\frac{D-3}{2} + i\mu}\, \theta(x_4-x_3) \, {\rm Im}\left[V(x_3)V^*(x_4)\right].
\eqa
Below we will show that Re$\Gamma_3 > 0$ and, hence, the solution of this equation, $\kappa_p(\eta) \propto \left(p\eta\right)^{\Gamma_3}$, reduces back to zero, as $\eta\to 0$.

\subsection{Solution of the kinetic equation in the Poincar\'{e} patches}

To find a solution of the kinetic equation (\ref{cint}) it is instructive to consider simultaneously the situation in the EPP and CPP. Spatially homogeneous kinetic equations in EPP and CPP can be trivially related to each other, because to perform the map from EPP to CPP one just has to flip the limits of $d\eta'$ integration inside the collision integrals in (\ref{callintPP0}) and (\ref{callintPP00}). On top of that it is necessary to make the change of $h(x)$ to $h^*(x)$, because the positive energy harmonics are exchanged with the negative ones under the flip of the time direction. As a result, solutions in both patches also can be mapped to each other. That is physically meaningful because the system of kinetic equations (\ref{callintPP0}) and (\ref{callintPP00}) is valid for the spatially homogeneous states and is imposed on the quantities which are attributed to the comoving volume. It is worth mentioning that spatially homogenous situation in the CPP is unstable under small inhomogeneous density perturbations. However, it is still instructive to consider such an ideal situation in the CPP\footnote{Actually it is not very hard to find the inhomogeneous extension of the kinetic equation (\ref{cint}). In the case in which the particle density starts to depend also on the spatial position $n_p = n_p(\eta, \vec{x})$, one has to substitute $\eta\, d/d\eta$ by $\eta \, \partial_\eta + \eta^2 \vec{p} \, \vec{\partial}_x$ on the left hand side of (\ref{cint}). The extra $\eta^2 \vec{p} \vec{\partial}_x$ term is the simplest expression that is invariant under the symmetries of the EPP and CPP metrics: such as, e.g., simultaneous scaling of all coordinates.}.

Thus, all the observations made for the EPP can be extended to the CPP. For the Bessel (in-)harmonics in the CPP the corresponding system (\ref{callintPP0}) and (\ref{callintPP00}) can be reduced to the single equation:

\bqa\label{callintPP2}
\frac{d n_p(\eta)}{d\log \left(\eta/\eta_0\right)} = \frac{\lambda^2 \, S_{D-2} \, |A|^2}{(2\pi)^{D-1}} \, \int_0^{\infty} dq \, \eta \,\left(q \eta\right)^{\frac{D-1}{2}} \, \int^{\infty}_{0} d\eta' \, q \, \left(q \eta'\right)^{\frac{D-1}{2}} \times \nonumber \\ \times \Biggl\{ {\rm Re} \left[(q\eta)^{-i\mu} \, V(q\eta) \,\left(q\eta'\right)^{i\mu}\, V^*(q\eta')\right] \, \left\{[1+n_p]\, n_q^{2\phantom{\frac12}} - \,\, n_p \, [1+n_q]^2\right\}(\eta) \Biggr. + \nonumber \\
+ 2\,{\rm Re} \left[(q\eta)^{i\mu} \, W(q\eta) \, \left(q\eta'\right)^{- i\mu} \, W(q\eta')\right] \, \left\{ n_q\, [1+n_q]\,[1+n_p]^{\phantom{\frac12}} - \,\,[1+n_q]\, n_q\, n_p \right\}(\eta) + \nonumber \\
+ \Biggl. {\rm Re} \left[(q\eta)^{i\mu} \, V(q\eta) \, \left(q\eta'\right)^{- i\mu} \, V^*(q\eta')\right] \, \left\{ [1+n_q]^2 \,[1+n_p]^{\phantom{\frac12}} - \,\, n_q^2\, n_p \right\}^{\phantom{2}}(\eta)\Biggr\}.
\eqa
Neither (\ref{cint}) nor (\ref{callintPP2}) possess the Plankian distribution as a stationary solution, because there is no energy conservation. However, one can find a distribution which annihilates the collision integral of (\ref{cint}) or (\ref{callintPP2}) in a different manner: while Plankian distribution in ordinary kinetic equation annihilates each term, describing aforementioned couples of competing processes, separately, in the present case we will see the cancelation between different terms. The situation is somewhat similar to the one in which appears the turbulent stationary Kolmogorov like scaling. That is natural to expect in a system with energy pumping in at one scale and its loss at another one.

Thus, suppose we have started at the past infinity of the EPP with some very mild density perturbation over the BD state. After the Bogolyubov rotation to the out-harmonics, one has some initial values of $n_p$ and $\kappa_p$. As can be understood from the discussion above, for the given $p$, the density $n_p$ and the anomalous average $\kappa_p$ practically do not change before $\eta_* \sim \mu/p$. After this moment they begin to evolve according to (\ref{callintPP0}), (\ref{callintPP00}). If $\kappa_p$ is sufficiently small, it relaxes to zero, as we have shown above, and the problem is reduced to the solution of (\ref{cint}).

Now, if the initial value of $n_p$ after the rotation to the out-harmonics is much smaller than one, we can use the following approximations:

\bqa
[1+n(x)]\, n(x_3)^2 - n(x) \, [1+n(x_3)]^2 \approx - n(x),\nonumber \\
n(x_3)\, [1+n(x_3)]\,[1+n(x)] - [1+n(x_3)]\, n(x_3)\, n(x) \approx n(x_3), \nonumber
\eqa
\bqa\label{natur}
[1 + n(x_3)]^2 \,[1+n(x)] - n(x_3)^2 \, n(x) \approx 1.
\eqa
Because of the rapid oscillations of $h(x)$ as $x\to\infty$, the integrals on the right hand side of (\ref{cint}) are saturated at $x_{3,4} \sim \mu$. At the same time $p\eta \ll \mu$ and it is natural to assume that $n(x_{3,4} \sim 1) \ll n(p\eta)$ in the situation of the very small initial density perturbation and the vanishing production of the high momentum modes. Hence, we can neglect the second contribution ($\sim n(x_3)$) on the right hand side of (\ref{cint}), (\ref{natur}) in comparison with the first ($\sim n(x)$) and the third ($\sim 1$) ones.

As a result, Eq. (\ref{cint}) is reduced to

\bqa\label{solution1}
\frac{d n(x)}{d\log (x)} \approx \Gamma_1 \, n(x) - \Gamma_2, \nonumber \\
\Gamma_1 = \frac{\lambda^2 \, S_{D-2} \, \left|A\right|^2}{(2\pi)^{D-1}} \, \left|\int_0^{\infty} dy \,y^{\frac{D-3}{2} - i\, \mu} \, \left[h^2\left(y\right) - \frac{A_1 \, A_2}{|y|}\right] \right|^2, \nonumber \\ \Gamma_2 = \frac{\lambda^2 \, S_{D-2} \, \left|A\right|^2}{(2\,\pi)^{D-1}} \, \left|\int_0^{\infty} dy \,y^{\frac{D-3}{2} + i\,\mu} \, \left[h^2\left(y\right) - \frac{A_1\, A_2}{|y|}\right]\right|^2
\eqa
Here, $\Gamma_1$ and $\Gamma_2$ are the particle decay and production rates, correspondingly. Note that $x=p\eta$ is reducing to zero during the evolution towards the future infinity.

The obtained equation (\ref{solution1}) has the solution with the flat stationary point distribution $n(p\eta) = \Gamma_2/\Gamma_1$, which corresponds to the situation in which the production (gain) of particles on the level $p\eta$ is equilibrated by the particle decay (loss) from the same level. The obtained solution is self-consistent for the large enough $\mu$ because then $\Gamma_2/\Gamma_1 \approx e^{-2\,\pi\mu} \ll 1$, which is not hard to show using the following property of the Hankel functions:

\bqa
\int_0^{\infty} dx \, x^{\frac{D-3}{2} + i\mu} \, \left(H^{(1)}_{i\mu}(x)\right)^2 = e^{-2\pi\mu}\,\int_0^{\infty} dx \, x^{\frac{D-3}{2} - i\mu} \, \left(H^{(2)}_{i\mu}(x)\right)^2.
\eqa
(Also it is not hard to show that the introduced in the previous subsection $\Gamma_3$ obeys Re$\Gamma_3 = \Gamma_1 - \Gamma_2 \approx \left(1 - e^{-2\pi\mu}\right)\, \Gamma_1 > 0$.)

So the result of the summation of the large IR contributions may lead to the well behaving exact two-point function. As a result, the dS isometry, which was broken by the initial perturbation, is asymptotically restored at future infinity. But what if the evolution had started with some quite strong density perturbation (which is, however, still much smaller than the vacuum energy) over the BD vacuum state? Now, we are going to show that there is another very peculiar solution of the kinetic equation under consideration.

Suppose that, due to the particle self-reproduction, the density per comoving volume on the given level with $p\eta \ll \mu$ became big in comparison with one. Taking into account the flatness of the spectrum in dS space, it is natural to expect that, for the harmonics with the low physical momenta, the density very slowly depends on its argument. Hence, we can assume that $n(p\eta) \approx n(q \eta)$ when both $p\eta \ll \mu$ and $q \eta \ll \mu$.

Then, we can make the following approximations:

\bqa
[1+n(p\eta)]\, n^2(x_3) - n(p\eta) \, [1+n(x_3)]^2 \approx - n^2(p\eta), \nonumber \\
n(x_3)\, [1+n(x_3)]\,[1+n(p\eta)] - [1+n(x_3)]\, n(x_3)\, n(p\eta) \approx n^2(p\eta), \nonumber
\eqa
\bqa
[1+n(x_3)]^2 \, [1+n(p\eta)] - n(x_3)\, n(p\eta) \approx n^2(p\eta)
\eqa
and accept that, on the right hand side of (\ref{cint}), the main contribution to the $x_{3,4}$ integrals comes form the region in which $x_{3,4} \ll \mu$ because  $n(x) \gg n(y)$ if $x \ll \mu$ and $y \gg \mu$. As a result, the kinetic equation reduces to

\bqa
\frac{dn_p(\eta)}{d\log(\eta)} = \bar{\Gamma} \, n_p^2(\eta), \quad {\rm where} \nonumber \\
\bar{\Gamma} \approx \frac{\lambda^2}{(2\,\pi)^{D-1}\, \mu^3} \, \int_0^{\mu} dy \,(y)^{\frac{D-3}{2}} \, \left\{ -{\rm Re} \left[y^{-i\, \mu} \int^{\mu}_{0} dy' \,(y')^{\frac{D-3}{2} + i\, \mu}\right] \right. + \nonumber \\
+ 2\,{\rm Re} \left[y^{-i\,\mu}\int^{\mu}_{0} dy' \, (y')^{\frac{D-3}{2} + i\, \mu}\right]\, + \nonumber \\
+ \left. 3\,{\rm Re} \left[y^{-3\,i\, \mu}\int^{\mu}_{0} dy'\, (y')^{\frac{D-3}{2} + 3\, i\, \mu} \right] \right\} > 0.
\eqa
Note that $\bar{\Gamma}$ is independent of $p$. This equation has the following solution:

\bqa\label{solu2}
n_p(\eta) \approx \frac{1}{\bar{\Gamma} \, \log\left(\eta/\eta_{\star}\right)},
\eqa
where $\eta_{\star} = \frac{\mu}{p} \, e^{-\frac{C}{\bar{\Gamma}}}$ and $C$ is an integration constant, which depends on initial conditions. The obtained solution is valid if $\mu/p = \eta_* > \eta > \eta_{\star}$.

Thus, we see that there is a singular solution of the kinetic equation under consideration, which corresponds to the explosion of the particle number density per comoving volume within a finite proper time. Of course, such an explosion wins against the expansion of the EPP because $D^K \propto \frac{\eta^{D-1}}{\log{\eta/\eta_{\star}}}$. Hence, there are contributions to the energy-momentum tensor of the created particles which become huge, and the backreaction has to be taken into account. As a result, the dS space gets modified. But that is the problem for a separate study. At this point, we just would like to stress that we see catastrophic IR effects even for the massive fields. It is natural to expect that, for the light fields, the situation will be even more dramatic.

In the CPP, we can find similar solutions of the kinetic equation (\ref{callintPP2}). For example, (\ref{solution1}) is mapped to

\bqa\label{solu4}
\frac{d n_p(\eta)}{d \log(\eta/\eta_0)} \approx - \Gamma_1 \, n_p(\eta) + \Gamma_2,
\eqa
which shows a very peculiar phenomenon that, although in the first loop $\eta_0$ cannot be taken to the past infinity, after the summation of all loops, we can find the theory at the stationary point state, $n_p = \Gamma_2/\Gamma_1$, which allows one to remove the IR cutoff, $\eta_0$, and restore the dS isometry.

At the same time, the solution (\ref{solu2}) is mapped to

\bqa\label{solu3}
n_p(\eta) \approx \frac{1}{\bar{\Gamma} \, \log\left(\eta_{\star}/\eta\right)},
\eqa
where $\eta_{\star} = \eta_0 \, e^{C/\bar{\Gamma}} < \mu/p$. Of course, whether the field theory state goes into (\ref{solu3}) or to (\ref{solu4}) depends on initial conditions.

\subsection{Kinetic equation in global de Sitter space}

Having in mind the discussion of the EPP and CPP above, it is not hard to derive the system of kinetic equations in global dS:

\bqa\label{global1}
\frac{dn_j}{dt} = [N\to PN] + 2\,\lambda^2 \sum_{j_1,\vec{m}_1;j_2,\vec{m}_2} {\rm Re} Tr \Biggl\{J_{j,\vec{m};j_1,\vec{m}_1;j_2\vec{m}_2} \, \int_{-\infty}^{+\infty} dt' \, C^*_{j,j_1,j_2}(t) \times \nn \\ \times \left[\left(1+N^*_j\right)\, N_{j_1} \, N_{j_2}^{\phantom{\frac12}} - N^*_j\, \left(1 + N_{j_1}\right)\, \left(1 + N_{j_2}\right)\right](t,t')\Biggr. + \nn \\
+ 2\, J_{j_1,\vec{m}_1;j_2,\vec{m}_2;j\vec{m}} \, \int_{-\infty}^{+\infty} dt'\, C^*_{j_1,j_2,j}(t)\, \left[N^*_{j_1}\, \left(1+N_{j_2}\right) \, \left(1+N_{j}\right)^{\phantom{\frac12}} - \left(1+N^*_{j_1}\right)\, N_{j_2}\, N_{j}\right](t,t') + \nn \\
\Biggl. I_{j,\vec{m};j_1,\vec{m}_1;j_2\vec{m}_2} \, \int_{-\infty}^{+\infty} dt' \, D^*_{j,j_1,j_2}(t)\, \left[\left(1+N_j\right)\, \left(1+N_{j_1}\right) \, \left(1+N_{j_2}\right)^{\phantom{\frac12}} - N_j\, N_{j_1} \, N_{j_2}\right](t,t')\Biggr\}
\eqa
and

\bqa\label{global2}
\frac{d\kappa_j}{dt} = - [N\to PN] - 2\,\lambda^2 \sum_{j_1,\vec{m}_1;j_2,\vec{m}_2} (\vec{m} \to - \vec{m}) Tr \Biggl\{J_{j,\vec{m};j_1,\vec{m}_1;j_2\vec{m}_2} \, \int_{-\infty}^{t} dt' \, C^*_{j,j_1,j_2}(t)\times \nn \\ \times \left[\left(1+N_j\right)\, \left(1+N_{j_1}\right) \, \left(1+ N_{j_2}\right)^{\phantom{\frac12}} - N_j\, N_{j_1}\, N_{j_2}\right](t,t')\Biggr. + \nn \\
+ 2\, J_{j_1,\vec{m}_1;j_2,\vec{m}_2;j\vec{m}} \, \int_{-\infty}^{t} dt' \
C^*_{j_1,j_2,j}(t) \, \left[N^*_{j_1}\, \left(1+N_{j_2}\right) \, \left(1+N_{j}\right)^{\phantom{\frac12}} - \left(1+N^*_{j_1}\right)\, N_{j_2}\, N_{j}\right](t,t') + \nn \\ +
\Biggl. I_{j,\vec{m};j_1,\vec{m}_1;j_2\vec{m}_2} \, \int_{-\infty}^{t} dt' \, D^*_{j,j_1,j_2}(t) \, \left[\left(1+N_j\right)^{\phantom{\frac12}} N^*_{j_1} \, N^*_{j_2} - N_j\, \left(1+N^*_{j_1}\right) \, \left(1+N^*_{j_2}\right)\right](t,t')\Biggr\}
\eqa
The notations $[N\to PN]$ and $(\vec{m} \to - \vec{m})$ have been introduced above.
In (\ref{global1}) and (\ref{global2}) instead of the usual flat space $\delta$-functions, which ensure energy-momentum conservation, we have

\bqa
J_{j_1\vec{m}_1;j_2\vec{m}_2;j_3\vec{m}_3} = \int d\Omega Y^*_{j_1\vec{m}_1}\left(\Omega\right)\, Y_{j_2\vec{m}_2}\left(\Omega\right)\, Y_{j_3\vec{m}_3}\left(\Omega\right), \quad C_{j_1,j_2,j_3} = \cosh^{D-1}\left(t\right)\, g^*_{j_1}\, g_{j_2}\, g_{j_3}, \nonumber \\
I_{j_1\vec{m}_1;j_2\vec{m}_2;j_3\vec{m}_3} = \int d\Omega Y_{j_1\vec{m}_1}\left(\Omega\right)\, Y_{j_2\vec{m}_2}\left(\Omega\right)\, Y_{j_3\vec{m}_3}\left(\Omega\right), \quad D_{j_1,j_2,j_3} = \cosh^{D-1}\left(t\right)\, g_{j_1}\, g_{j_2}\, g_{j_3}.
\eqa
Unlike the case of EPP and CPP, we do not yet understand what is the appropriate choice of the harmonics for this system of kinetic equations. In fact, if one considers the entire dS space, then $\kappa_p$ is divergent either when $t_0\to -\infty$ or when $t\to \infty$, or in both cases simultaneously. In this case none of the harmonics provides a proper quasi-particle description. However, if one will consider that half of global dS space, which is above its neck, then the situation would be similar to that in the EPP. I.e., in this case one can find the same solutions as we have found above.

\section{Sketch of the results for $\lambda \phi^4$ theory (instead of conclusions)}

Instead of conclusions let us present here the results for $\lambda \phi^4$ theory both for the principal and complementary series (see \cite{AkhmedovPopovSlepukhin} for grater details).
It is straightforward to show that $\phi^4$ theory, unlike $\phi^3$ one, does not possess any large IR contributions to the Keldysh propagator at the first-loop level ($\sim \lambda$). However, in the second-loop ($\sim \lambda^2$), there are large IR contributions to $D^K$, which are of interest for us. They come from the so-called sunset diagrams. There are no IR corrections to $D^{R,A}$ propagators and to the vertices.

The kinetic equation for the principal series in the EPP can be derived in the same manner as in $\phi^3$ theory. For the out-modes it is as follows:

\bqa
\frac{d n_{p\eta}}{d \log(p\eta)} = - \frac{\lambda^2 \, |A|^2}{6} \, \int \frac{d^{D-1} l_1}{(2\pi)^{D-1}} \frac{d^{D-1} l_2}{(2 \pi)^{D-1}} \int_\infty^0 dv \, v^{D-2}\, \nonumber \\
\left\{3 {\rm Re} \left [v^{i\mu\phantom{\frac{a}{b}}} \,h^*(l_1) h^*(l_2) h\left(\left|\vec{l}_1 + \vec{l}_2\right|\right) h(l_1 v) h(l_2 v) h^*\left(\left|\vec{l}_1 + \vec{l}_2\right| v\right) \right   ] \right. \times \nonumber \\ \times \left[(1+n_{p\eta}) n_{l_1} n_{l_2} (1+n_{|\vec{l}_1 + \vec{l}_2|}) \,\, - \phantom{\frac{a}{b}} n_{p\eta} (1+n_{l_1}) (1+n_{l_2}) n_{|\vec{l}_1 + \vec{l}_2|} \right]
\nonumber \\+ 3  {\rm Re} \left [v^{i\mu\phantom{\frac{a}{b}}} \,h^*(l_1) h(l_2) h\left(\left|\vec{l}_1 - \vec{l}_2\right|\right) h(l_1 v) h^*(l_2 v) h^*\left(\left|\vec{l}_1 - \vec{l}_2\right| v\right) \right] \times \nonumber \\ \times \left  [(1+n_{p\eta}) n_{l_1} (1+n_{l_2}) (1+n_{|\vec{l}_1 - \vec{l}_2|}) \,\, - \phantom{\frac{a}{b}} n_{p\eta} (1+n_{l_1}) n_{l_2} n_{|\vec{l}_1 - \vec{l}_2|}\right ] \nonumber \\ +
 {\rm Re} \left [v^{i\mu \phantom{\frac{a}{b}}} \, h^*(l_1) h^*(l_2) h^*\left(\left|\vec{l}_1 + \vec{l}_2\right|\right) h(l_1 v) h(l_2 v) h\left(\left|\vec{l}_1+\vec{l}_2\right| v\right) \right   ]\times \nonumber \\ \times \left  [(1+n_{p\eta}) n_{l_1} n_{l_2} n_{|\vec{l}_1 + \vec{l}_2|} \,\, -  \phantom{\frac{a}{b}} n_{p\eta} (1+n_{l_1}) (1+n_{l_2}) (1+n_{|\vec{l}_1 + \vec{l}_2|})  \right ] \nonumber \\ +
 {\rm Re}  \left [v^{i\mu \phantom{\frac{a}{b}}} \, h(l_1) h(l_2) h\left(\left|\vec{l}_1 + \vec{l}_2\right|\right) h^*(l_1 v) h^*(l_2 v) h^*\left(\left|\vec{l}_1+\vec{l}_2\right| v\right) \right   ]\times \nonumber \\ \times \left. \left  [ (1+n_{p\eta}) (1+n_{l_1}) (1+n_{l_2}) (1+n_{|\vec{l}_1 + \vec{l}_2|}) \,\, - \phantom{\frac{a}{b}} n_{p\eta} n_{l_1} n_{l_2} n_{|\vec{l}_1 + \vec{l}_2|} \right ] \right\}.
\eqa
One can find solutions to this kinetic equation which are similar to those which we have found above.

At the same time the two-loop corrections for the complementary series are as follows. This series corresponds to the imaginary $\mu$, i.e. to the real index of the solution of the Bessel equation $h(p\eta)$. Below we use the notation $\nu = - i\mu$. Then, for the in harmonics we have that $h = J_\nu + iY_\nu$ and both Bessel functions $J_\nu$ and $Y_\nu$ are real in the case of the real index $\nu$. Expanding them near zero, we get $Y_\nu(x) \approx D_- x^{-\nu} + B x^{ -\nu + 2}$ and $J_\nu(x) \approx D_+ x^{\nu}$.
Because of the possible differences between the behavior of $h$ and $h^*$ near zero, we have to pay attention separately to $\kappa_p$ and $\kappa_p^*$.

The contributions to $n_p$ and $\kappa_p$, $\kappa^*_p$ can be expressed as

\bqa\label{66}
n_p(\eta) \approx - \frac{2 \lambda^2}{3\,(2\,\pi)^{2\, (D-1)}} \, \int^0_\infty \frac{du}{u} \, \int^0_\infty \frac{dv}{v^{2D - 1}} \, F[v] \, h(u v)  h^*\left(\frac{u}{v}\right) \,
\theta[uv - p \eta] \, \theta\left[\frac{u}{v}  - p \eta\right],
\nonumber \\
\kappa_p(\eta) \approx \frac{4\, \lambda^2}{3\,(2\pi)^{2(D-1)}} \, \int^0_\infty \frac{du}{u} \int_\infty^0 \frac{dv}{v^{2D - 1}} \, F[v] \, h^*(u v)  h^*\left(\frac{u}{v}\right) \, \theta[uv - p \eta] \, \theta\left[\frac{u}{v}  -  u v\right],
\nonumber \\
\kappa^*_p(\eta) \approx \frac{4\, \lambda^2}{3\,(2\pi)^{2(D-1)}} \, \int^0_\infty \frac{du}{u} \int_\infty^0 \frac{dv}{v^{2D - 1}} \, F[v] \, h(u v)  h\left(\frac{u}{v}\right) \, \theta[uv - p \eta] \, \theta\left[\frac{u}{v}  -  u v\right],
\eqa
with the use of the following notations

\bqa
F(\eta_3, \eta_4) = \int \left[\prod_{i=1}^3 d^{D-1} q_i h(q_i\eta_3) h^*(q_i \eta_4)\right]\, (\eta_3 \eta_4)^{D-2} \, \delta^{(D-1)}\left(\vec{p} - \vec{q}_1 - \vec{q}_2 - \vec{q}_3\right),
\eqa
which, after the change of integration variables $u=p \sqrt{\eta_3 \eta_4}$, $v=\sqrt{\frac{\eta_3}{\eta_4}}$,
can be expressed as $F(\eta_3, \eta_4) = \frac{p^2}{(uv)^{2 \, (D - 1)}} \, u^{2 \, (D - 2)} \, F[v]$, where $F[v] = F^*\left[1/v\right]$ is some function of one variable $v$.

The leading corrections to $n_p$ and $\kappa_p$, $\kappa^*_p$ are given by Eq. (\ref{66}), where from the Hankel functions, $h(uv)$ and $h(u/v)$, we single out only $Y$'s. Such contributions give for $n_p$ and $\kappa_p$, $\kappa^*_p$ the inverse powerlike behavior in $p\eta$, which, however, cancels out after the substitution into $D^K$ because $Y$ is real. The next order
is obtained as follows. One also has to express $h(p\eta_{1,2})$ through $J_\nu$ and $Y_\nu$ in the full propagator $D^K$. Then, from one of the four $h$'s $[h(p\eta_{1,2})$ and $h(uv)$, $h(u/v)]$, we have to single out $J_\nu$, while from the other three --- $Y_\nu$'s.
This expression does not cancel out and provides the leading IR contribution to $D^K$:

\bqa
D^K(\eta_1, \eta_2, p) = \frac{8\,\lambda^2 \, D_-^3 \, D_+ \, \eta^{D-1}}{3\,(2\pi)^{2\, (D-1)}} \,
\int \limits_\infty^0 du \, u^{-1-2 \nu} \int \limits_\infty^0 dv \, v^{1-2\,D} F[v] \nonumber \\
\left\{\left[-1+\left(\frac{u}{p\eta \, v}\right)^{2 \nu}\right] \theta \left[uv - \frac{u}{v}\right]\, \theta\left[-p\eta + uv\right] + \left[1-\left(\frac{u
v}{p\eta}\right)^{2 \nu }\right] \, \theta\left[- u v + \frac{u}{v}\right] \, \theta\left[-p\eta + \frac{u}{v}\right] \right\}.
\eqa
After the straightforward manipulations, the obtained expression can be reduced to

\bqa\label{9}
D^K(\eta_1,\eta_2, p) = - \frac{8\,\lambda^2 \, D_-^3\,D_+ \,\eta^{D-1} \, \log(p\eta)}{3\,(2\pi)^{2\, (D-1)} \, (p\eta)^{2 \nu}}\, \nonumber \\
\left\{\int \limits^\infty_1 dv \, v^{1 - 2D}\, F[v] \, \left[-\frac{1}{2 \nu} v^{2 \nu} + \left(\frac{1}{v}\right)^{2 \nu}\right] - \int \limits^1_0 dv \, v^{1 - 2D} \, F[v]\,  \left[\frac{1}{2 \nu} \, ( v)^{-2 \nu} +
v^{2 \nu}\right]\right\}.
\eqa
The integral over $v$ is convergent in the IR limit (as $v\to \infty$) if $D > 1 + 4 \nu$. In the UV limit $(v\to 0)$, it is convergent in the sense of generalized functions. Thus, for the complementary series, the loop contributions seem to be powerlike, i.e., stronger than for the case of the principal series.

For the out-harmonics, the situation is a bit different. In this case, $h = J_\nu$, $h^* = Y_\nu$. The straightforward calculation shows that

\bqa
n_p(\eta) \propto \lambda^2 \, D_- \, D_+ \, \log(p \eta) \, \int_0^1 dv \, F[v] v^{2\nu + 1 - 2D},
\nonumber \\
\kappa_p(\eta) \propto \lambda^2 \, D_-^2 \, (p \eta)^{-2 \nu} \, \int^0_1 dv \, F[v] \, v^{2\nu + 1 - 2D}, \nonumber \\
\kappa^*_p(\eta) \propto \lambda^2 \, D_+^2 \, (p \eta)^{2 \, \nu} \int_0^1 dv \, F[v] \, v^{-2\nu + 1 - 2D}.
\eqa
After the substitution into the Keldysh propagator, the leading contribution comes from $n_p$ and is logarithmic.
However, because of the character of these IR contributions for the light fields, we do not yet understand their physical meaning and think that the kinetic equations obtained above are not applicable for the fields from the complementary series. In the case of the complementary series, we do not yet know how to perform the summation of the leading IR contributions from all loops. But on general physical grounds and from the two-loop result, we expect that light fields from the complementary series will show stronger IR effects than the heavy ones from the principal series.

\section{Acknowledgements}

This work is dedicated to the memory of M.I.Polykarpov.

These notes are based on the lectures given at Scuola Normale Superiore, Pisa, Italy and Max-Plank-Institute for the Gravitational Physics (Albert-Einstein-Institute), Golm, Germany. I would like to thank D.Francia, A.Sagnotty, H.Nicolai, S.Theisen and S.Fredenhagen for the hospitality during my stay in these institutes.

I would like to thank P.Burda, A.Sadofyev, V.Slepukhin and F.Popov for the collaboration on this project.
I would like to acknowledge valuable discussions with A.Roura, D.Marolf, I.Morrison, A.Morozov, A.Mironov, V.Losyakov, S.Apenko, D.Francia, S.Theisen and A.Polyakov. This work was partially supported by the Ministry of Education and Science of Russian Federation under the contract 8207, by the grant "Leading Scientific Schools" No. NSh-6260.2010.2 and by the grant RFBR-11-02-01227-a.


\begin{thebibliography}{99}


\bibitem{Schrodinger:2010zz}
  E.~Schrodinger,
  Cambridge, UK: Univ. Pr. (2010) 93 p

\bibitem{Allen:1985ux}
  B.~Allen,
  ``Vacuum States In De Sitter Space,''
  Phys.\ Rev.\  D {\bf 32}, 3136 (1985).

\bibitem{Allen:1987tz}
  B.~Allen and A.~Folacci,
  Phys.\ Rev.\ D {\bf 35}, 3771 (1987).

\bibitem{Miao:2013isa}
  S.~P.~Miao, P.~J.~Mora, N.~C.~Tsamis and R.~P.~Woodard,
  ``The Perils of Analytic Continuation,''
  arXiv:1306.5410 [gr-qc].

\bibitem{Tsamis:1993ub}
  N.~C.~Tsamis and R.~P.~Woodard,
  Class.\ Quant.\ Grav.\  {\bf 11}, 2969 (1994).

\bibitem{Tsamis:1992zt}
  N.~C.~Tsamis and R.~P.~Woodard,
  Phys.\ Lett.\  B {\bf 292}, 269 (1992).

\bibitem{Tsamis:1992xa}
  N.~C.~Tsamis and R.~P.~Woodard,
  Commun.\ Math.\ Phys.\  {\bf 162}, 217 (1994).

\bibitem{Tsamis:1994ca}
  N.~C.~Tsamis and R.~P.~Woodard,
  Annals Phys.\  {\bf 238}, 1 (1995).

\bibitem{Tsamis:1996qq}
  N.~C.~Tsamis and R.~P.~Woodard,
  Nucl.\ Phys.\  B {\bf 474}, 235 (1996)
  [arXiv:hep-ph/9602315].

\bibitem{Tsamis:1996qk}
  N.~C.~Tsamis and R.~P.~Woodard,
  Phys.\ Rev.\  D {\bf 54}, 2621 (1996)
  [arXiv:hep-ph/9602317].

\bibitem{Tsamis:1996qm}
  N.~C.~Tsamis and R.~P.~Woodard,
  Annals Phys.\  {\bf 253}, 1 (1997)
  [arXiv:hep-ph/9602316].

\bibitem{Garriga:2007zk}
  J.~Garriga and T.~Tanaka,
  Phys.\ Rev.\  D {\bf 77}, 024021 (2008)
  [arXiv:hep-th/0706.0295].

\bibitem{Higuchi:2011vw}
  A.~Higuchi, D.~Marolf and I.~A.~Morrison,
  Class.\ Quant.\ Grav.\  {\bf 28}, 245012 (2011)
  [arXiv:1107.2712 [hep-th]].

\bibitem{Morrison:2013rqa}
  I.~A.~Morrison,
  ``On cosmic hair and "de Sitter breaking" in linearized quantum gravity,''
  arXiv:1302.1860 [gr-qc].

\bibitem{PerezNadal:2008ju}
  G.~P\'erez-Nadal, A.~Roura and E.~Verdaguer,
  Class.\ Quant.\ Grav.\  {\bf 25}, 154013 (2008)
  [arXiv:gr-qc/0806.2634].

\bibitem{PerezNadal:2007ez}
  G.~Perez-Nadal, A.~Roura and E.~Verdaguer,
  Phys.\ Rev.\ D {\bf 77}, 124033 (2008)
  [arXiv:0712.2282 [gr-qc]].

\bibitem{PerezNadal:2009hr}
  G.~P\'erez-Nadal, A.~Roura and E.~Verdaguer,
  JCAP {\bf 1005}, 036 (2010)
  [arXiv:0911.4870 [gr-qc]].

\bibitem{Roura:1999fq}
  A.~Roura and E.~Verdaguer,
  Int.\ J.\ Theor.\ Phys.\  {\bf 38}, 3123 (1999)
  [arXiv:gr-qc/9904039].

\bibitem{Mottola:1984ar}
  E.~Mottola,
  Phys.\ Rev.\  D {\bf 31}, 754 (1985).

\bibitem{MarolfMorrison} 
  D.~Marolf, I.~A.~Morrison and M.~Srednicki,
  ``Perturbative S-matrix for massive scalar fields in global de Sitter space,''
  arXiv:1209.6039 [hep-th].

\bibitem{Marolf:2010zp}
  D.~Marolf and I.~A.~Morrison,
  Phys.\ Rev.\ D {\bf 82}, 105032 (2010)
  [arXiv:1006.0035 [gr-qc]].

\bibitem{Marolf:2011sh}
  D.~Marolf and I.~A.~Morrison,
  Gen.\ Rel.\ Grav.\  {\bf 43}, 3497 (2011)
  [arXiv:1104.4343 [gr-qc]].

\bibitem{Higuchi:2010xt}
  A.~Higuchi, D.~Marolf, I.~A.~Morrison,
  Phys.\ Rev.\  {\bf D83}, 084029 (2011).
  [arXiv:1012.3415 [gr-qc]].

\bibitem{Marolf:2010nz}
  D.~Marolf and I.~A.~Morrison,
  Phys.\ Rev.\ D {\bf 84}, 044040 (2011)
  [arXiv:1010.5327 [gr-qc]].

\bibitem{Hollands:2010pr}
  S.~Hollands,
  Commun.\ Math.\ Phys.\  {\bf 319}, 1 (2013)
  [arXiv:1010.5367 [gr-qc]].

\bibitem{Hollands:2011we}
  S.~Hollands,
  Annales Henri Poincare {\bf 13}, 1039 (2012)
  [arXiv:1105.1996 [gr-qc]].

\bibitem{Antoniadis:2006wq}
  I.~Antoniadis, P.~O.~Mazur and E.~Mottola,
  New J.\ Phys.\  {\bf 9}, 11 (2007)
  [arXiv:gr-qc/0612068].

\bibitem{Mottola:1985qt}
  E.~Mottola,
  Phys.\ Rev.\  D {\bf 33}, 1616 (1986).

\bibitem{Mottola:1985ee}
  E.~Mottola,
  Phys.\ Rev.\  D {\bf 33}, 2136 (1986).

\bibitem{Mazur:1986et}
  P.~Mazur and E.~Mottola,
  Nucl.\ Phys.\  B {\bf 278}, 694 (1986).

\bibitem{Anderson:2009ci}
  P.~R.~Anderson, C.~Molina-Paris, E.~Mottola,
  Phys.\ Rev.\  {\bf D80}, 084005 (2009).
  [arXiv:0907.0823 [gr-qc]].

\bibitem{Mottola:2010gp}
  E.~Mottola,
  [arXiv:1008.5006 [gr-qc]].

\bibitem{Dolgov:1994cq}
  A.~D.~Dolgov, M.~B.~Einhorn and V.~I.~Zakharov,
  Phys.\ Rev.\  D {\bf 52}, 717 (1995)
  [arXiv:gr-qc/9403056].

\bibitem{Dolgov:1994ra}
  Acta Phys.\ Polon.\  B {\bf 26}, 65 (1995)
  [arXiv:gr-qc/9405026].

\bibitem{Polyakov:2007mm}
  A.~M.~Polyakov,
  Nucl.\ Phys.\  B {\bf 797}, 199 (2008)
  [arXiv:0709.2899 [hep-th]].

\bibitem{Polyakov:2009nq}
  A.~M.~Polyakov,
  Nucl.\ Phys.\  B {\bf 834}, 316 (2010)
  [arXiv:0912.5503 [hep-th]].

\bibitem{Higuchi:2008tn}
  A.~Higuchi,
  Class.\ Quant.\ Grav.\  {\bf 26}, 072001 (2009)
  [arXiv:0809.1255 [gr-qc]].

\bibitem{Higuchi:2009ew}
  A.~Higuchi and Y.~C.~Lee,
  Class.\ Quant.\ Grav.\  {\bf 26}, 135019 (2009)
  [arXiv:0903.3881 [gr-qc]].

\bibitem{Alvarez:2009kq}
  E.~Alvarez and R.~Vidal,
  JHEP {\bf 0910}, 045 (2009)
  [arXiv:0907.2375 [hep-th]].

\bibitem{Alvarez:2010te}
  E.~Alvarez and R.~Vidal,
  JCAP {\bf 1011}, 043 (2010)
  [arXiv:1004.4867 [hep-th]].

\bibitem{Burgess:2010dd}
  C.~P.~Burgess, R.~Holman, L.~Leblond and S.~Shandera,
  JCAP {\bf 1010}, 017 (2010)
  [arXiv:1005.3551 [hep-th]].

\bibitem{Burgess:2009bs}
  C.~P.~Burgess, L.~Leblond, R.~Holman and S.~Shandera,
  JCAP {\bf 1003}, 033 (2010)
  [arXiv:0912.1608 [hep-th]].

\bibitem{Giddings:2010ui}
  S.~B.~Giddings, M.~S.~Sloth,
  JCAP {\bf 1007}, 015 (2010).
  [arXiv:1005.3287 [hep-th]].

\bibitem{Giddings:2010nc}
  S.~B.~Giddings, M.~S.~Sloth,
  JCAP {\bf 1101}, 023 (2011).
  [arXiv:1005.1056 [hep-th]].

\bibitem{Riotto:2008mv}
  A.~Riotto, M.~S.~Sloth,
  JCAP {\bf 0804}, 030 (2008).
  [arXiv:0801.1845 [hep-ph]].

\bibitem{Onemli:2002hr}
  V.~K.~Onemli, R.~P.~Woodard,
  Class.\ Quant.\ Grav.\  {\bf 19}, 4607 (2002).
  [gr-qc/0204065].

\bibitem{Onemli:2004mb}
  V.~K.~Onemli, R.~P.~Woodard,
  Phys.\ Rev.\  {\bf D70}, 107301 (2004).
  [gr-qc/0406098].

\bibitem{Kahya:2006hc}
  E.~O.~Kahya, V.~K.~Onemli,
  Phys.\ Rev.\  {\bf D76}, 043512 (2007).
  [gr-qc/0612026].

\bibitem{Kahya:2009sz}
  E.~O.~Kahya, V.~K.~Onemli, R.~P.~Woodard,
  Phys.\ Rev.\  {\bf D81}, 023508 (2010).
  [arXiv:0904.4811 [gr-qc]].

\bibitem{Nacir:2013xca}
  D.~L.~Lуp.~Nacir, F.~D.~Mazzitelli and L.~G.~Trombetta,
  ``The Hartree approximation in curved spacetimes revisited I: the effective potential in de Sitter,''
  arXiv:1309.0864 [hep-th].

\bibitem{Xue:2011hm}
  W.~Xue, K.~Dasgupta, R.~Brandenberger,
  Phys.\ Rev.\  {\bf D83}, 083520 (2011).
  [arXiv:1103.0285 [hep-th]].

\bibitem{Polyakov:2012uc}
  A.~M.~Polyakov,
  ``Infrared instability of the de Sitter space,''
  arXiv:1209.4135 [hep-th].

\bibitem{Bunch:1978yq}
  T.~S.~Bunch and P.~C.~W.~Davies,
  Proc.\ Roy.\ Soc.\ Lond.\  A {\bf 360}, 117 (1978).

\bibitem{Chernikov:1968zm}
  N.~A.~Chernikov and E.~A.~Tagirov,
 Ann.\ Inst.\ Henri Poincar\'e  A {\bf 9}, 109 (1968).

\bibitem{AkhmedovSadofyev}
  E.~T.~Akhmedov, A.~V.~Sadofyev,
  Phys.\ Lett.\ B {\bf 712}, 138 (2012)
  [arXiv:1201.3471 [hep-th]].

\bibitem{PolyakovKrotov}
  D.~Krotov, A.~M.~Polyakov,
  Nucl.\ Phys.\  {\bf B849}, 410-432 (2011).
  [arXiv:1012.2107 [hep-th]].

\bibitem{AkhmedovKEQ}
  E.~T.~Akhmedov,
  JHEP {\bf 1201}, 066 (2012)
  [arXiv:1110.2257 [hep-th]].

\bibitem{AkhmedovPopovSlepukhin}
  E.~T.~Akhmedov, F.~K.~Popov and V.~M.~Slepukhin,
  Phys.\ Rev.\ D {\bf 88}, 024021 (2013)
  [arXiv:1303.1068 [hep-th]].

\bibitem{LL}
L.~D.~Landau and E.~M.~Lifshitz, Vol. 10 (Pergamon Press, Oxford, 1975).

\bibitem{Kamenev} A.Kamenev, ``Many-body theory of non-equilibrium systems'',  arXiv:cond-mat/0412296;
Bibliographic Code:	2004cond.mat.12296K.

\bibitem{AkhmedovBurda}
  E.~T.~Akhmedov and P.~.Burda,
  Phys.\ Rev.\ D {\bf 86}, 044031 (2012)
  [arXiv:1202.1202 [hep-th]].

\bibitem{AkhmedovGlobal}
  E.~T.~Akhmedov,
  Phys.\ Rev.\ D {\bf 87}, 044049 (2013)
  [arXiv:1209.4448 [hep-th]].

\bibitem{Misner:1974qy}
Misner C. W., Wheeler J. A. Gravitation: Charles W. Misner, Kip S. Thorne, John Archibald Wheeler. ``Gravitation'', Macmillan, San Francisco 1973, 1279p; ISBN-9780716703440

\bibitem{Unruh:1976db}
  W.~G.~Unruh,
  Phys.\ Rev.\ D {\bf 14}, 870 (1976).

\bibitem{AkhmedovUnruh}
  E.~T.~Akhmedov and D.~Singleton,
  Int.\ J.\ Mod.\ Phys.\ A {\bf 22}, 4797 (2007)
  [hep-ph/0610391].

\bibitem{Akhmedov:2007xu}
  E.~T.~Akhmedov and D.~Singleton,
  Pisma Zh.\ Eksp.\ Teor.\ Fiz.\  {\bf 86}, 702 (2007)
  [arXiv:0705.2525 [hep-th]].

\bibitem{BoussoStrominger}
 R.~Bousso, A.~Maloney and A.~Strominger,
  Phys.\ Rev.\  D {\bf 65}, 104039 (2002)
  [arXiv:hep-th/0112218].

\bibitem{GribMamaevMostepanenko} Grib A. A., Mamaev S. G., Mostepanenko V. M. ``Quantum effects in strong external fields'', Atomizdat, Moscow 1980, 296.\\ Grib A. A., Mamayev S. G., Mostepanenko V. M. Vacuum quantum effects in strong fields. – St. Petersburg : Friedmann Laboratory, 1994.

\bibitem{vanderMeulen:2007ah}
  M.~van der Meulen, J.~Smit,
  JCAP {\bf 0711}, 023 (2007).
  [arXiv:0707.0842 [hep-th]].

\bibitem{GradshteinRizhik} Gradshteyn I. S., Ryzhik I. M., ``Table of integrals'', Series, and Products (Academic, New York, 1980), 1.421.

\bibitem{Leblond}
  D.~P.~Jatkar, L.~Leblond and A.~Rajaraman,
  Phys.\ Rev.\ D {\bf 85}, 024047 (2012)
  [arXiv:1107.3513 [hep-th]].

\bibitem{Myrhvold} N. Myrhvold, Phys.\ Rev.\ D {\bf 28} 2439 (1983).

\bibitem{Boyanovsky:2004ph}
  D.~Boyanovsky, H.~J.~de Vega and N.~G.~Sanchez,
  Phys.\ Rev.\  D {\bf 71}, 023509 (2005)
  [arXiv:astro-ph/0409406].

\bibitem{Boyanovsky:2004gq}
  D.~Boyanovsky and H.~J.~de Vega,
  Phys.\ Rev.\  D {\bf 70}, 063508 (2004)
  [arXiv:astro-ph/0406287].

\bibitem{Boyanovsky:2012qs}
  D.~Boyanovsky,
  Phys.\ Rev.\ D {\bf 85}, 123525 (2012)
  [arXiv:1203.3903 [hep-ph]].

\bibitem{Boyanovsky:2012nd}
  D.~Boyanovsky,
  Phys.\ Rev.\ D {\bf 86}, 023509 (2012)
  [arXiv:1205.3761 [astro-ph.CO]].

\bibitem{Bros:2009bz}
  J.~Bros, H.~Epstein, M.~Gaudin, U.~Moschella and V.~Pasquier,
  ``Triangular invariants, three-point functions and particle stability on the
  de Sitter universe,''
  Commun.\ Math.\ Phys.\  {\bf 295}, 261 (2010)
  [arXiv:0901.4223 [hep-th]].

\bibitem{Bros:2008sq}
  J.~Bros, H.~Epstein and U.~Moschella,
  Annales Henri Poincare {\bf 11}, 611 (2010)
  [arXiv:0812.3513 [hep-th]].

\bibitem{Bros:2006gs}
  J.~Bros, H.~Epstein and U.~Moschella,
  ``Lifetime of a massive particle in a de Sitter universe,''
  JCAP {\bf 0802}, 003 (2008)
  [arXiv:hep-th/0612184].

\bibitem{Volovik:2008ww}
  G.~E.~Volovik,
  ``On de Sitter radiation via quantum tunneling,''
  [arXiv:gr-qc/0803.3367].

  JETP Lett.\  {\bf 90}, 1 (2009)
  [arXiv:gr-qc/0905.4639].

\bibitem{Akhmedov:2009be}
  E.~T.~Akhmedov, P.~V.~Buividovich and D.~A.~Singleton,
  Phys.\ Atom.\ Nucl.\  {\bf 75}, 525 (2012)
  [arXiv:0905.2742 [gr-qc]].

\bibitem{AkRoSa}
  E.~T.~Akhmedov, A.~Roura and A.~Sadofyev,
  Phys.\ Rev.\  D {\bf 82}, 044035 (2010)
  [arXiv:1006.3274 [gr-qc]].

\bibitem{Akhmedov:2008pu}
 E.~T.~Akhmedov and P.~V.~Buividovich,
  ``Interacting Field Theories in de Sitter Space are Non-Unitary,''
 Phys.\ Rev.\  D {\bf 78}, 104005 (2008)
  [arXiv:0808.4106 [hep-th]].

\bibitem{Akhmedov:2009vs}
  E.~T.~Akhmedov and P.~Burda,
  Phys.\ Lett.\ B {\bf 687}, 267 (2010)
  [arXiv:0912.3435 [hep-th]].

\bibitem{Akhmedov:2009ta}
  E.~T.~Akhmedov,
  Mod.\ Phys.\ Lett.\ A {\bf 25}, 2815 (2010)
  [arXiv:0909.3722 [hep-th]].

\bibitem{Akhmedov:2009vh}
  E.~T.~Akhmedov and E.~T.~Musaev,
  ``Comments on QED with background electric fields,''
  New J.\ Phys.\  {\bf 11}, 103048 (2009)
  [arXiv:0901.0424 [hep-ph]].

\bibitem{Youssef:2013by}
  A.~Youssef and D.~Kreimer,
  ``Resummation of infrared logarithms in de Sitter space via Dyson-Schwinger equations: the ladder-rainbow approximation,''
  arXiv:1301.3205 [gr-qc].

\bibitem{Gautier:2013aoa}
  F.~Gautier and J.~Serreau,
  ``Solving the Schwinger-Dyson equation for a scalar field in de Sitter space,''
  arXiv:1305.5705 [hep-th].

\end{thebibliography}
\end{document}